\documentclass[10pt]{article}

\usepackage{authblk}
\usepackage{natbib}
\usepackage{mathtools}
\usepackage{amsmath}
\usepackage{amsthm}
\usepackage{amssymb}
\usepackage{amsbsy}
\usepackage{amsfonts}
\usepackage{amscd}
\usepackage{mathrsfs}
\usepackage{bm}
\usepackage{bbm}
\usepackage{breqn}
\usepackage{color, soul}
\usepackage{enumerate}
\usepackage{empheq}
\usepackage{rotating}
\usepackage{ftnxtra}
\usepackage{fnpos}
\usepackage[titletoc,toc,title]{appendix}
\usepackage{euscript}
\usepackage{graphicx}
\usepackage{epsfig}
\usepackage{epstopdf}
\DeclareGraphicsExtensions{.pdf,.png,.jpg,.eps}
\usepackage{pstool}
\usepackage{upgreek}
\usepackage{mathrsfs}
\usepackage{textcomp}

\usepackage{psfrag}

\numberwithin{equation}{section}

\theoremstyle{plain}	
\newtheorem{thm}{Theorem}[section]

\newtheorem*{prop*}{Proposition}
\theoremstyle{definition}	

\newtheorem{remark}[thm]{Remark}

\usepackage{caption}
\usepackage{subcaption}

\usepackage{hyperref}
\hypersetup{colorlinks=true, linkcolor=blue}
\hypersetup{colorlinks=true,citecolor=blue}

\DeclareMathAlphabet{\mathpzc}{OT1}{pzc}{m}{it}

\usepackage{amsmath, amsthm, amssymb}

\usepackage{cleveref}

\DeclarePairedDelimiter\abs{\lvert}{\rvert}

\makeatletter
\newsavebox{\@brx}
\newcommand{\llangle}[1][]{\savebox{\@brx}{\(\m@th{#1\langle}\)}%
  \mathopen{\copy\@brx\mkern2mu\kern-0.9\wd\@brx\usebox{\@brx}}}
\newcommand{\rrangle}[1][]{\savebox{\@brx}{\(\m@th{#1\rangle}\)}%
  \mathclose{\copy\@brx\mkern2mu\kern-0.9\wd\@brx\usebox{\@brx}}}%
\let\oldabs\abs
\def\abs{\@ifstar{\oldabs}{\oldabs*}}
\makeatother

\usepackage{accents}

\begin{document}

\title{\textbf{Transformation Cloaking in Elastic Plates}
}
	
\author[1,2]{Ashkan Golgoon}
\author[2,3]{Arash Yavari\thanks{Corresponding author, e-mail: arash.yavari@ce.gatech.edu}}
\affil[1]{\small \textit{Department of Mechanical Engineering, Northwestern University, Evanston, IL 60208, USA}}
\affil[2]{\small \textit{School of Civil and Environmental Engineering, Georgia Institute of Technology, Atlanta, GA 30332, USA}}
\affil[3]{\small \textit{The George W. Woodruff School of Mechanical Engineering, Georgia Institute of Technology, Atlanta, GA 30332, USA}}

\maketitle

\begin{abstract} 
In this paper we formulate the problem of elastodynamic transformation cloaking for Kirchhoff-Love plates and elastic plates with both the in-plane and out-of-plane displacements. A cloaking transformation maps the boundary-value problem of an isotropic and homogeneous elastic plate (virtual problem) to that of an anisotropic and inhomogeneous elastic plate with a hole surrounded by a cloak that is to be designed (physical problem). For Kirchhoff-Love plates, the (out-of-plane) governing equations of the virtual plate is transformed to those of the physical plate up to an unknown scalar field. In doing so, one finds the initial stress and the initial tangential body force for the physical plate, along with a set of constraints that we call cloaking compatibility equations. These constraints involve the cloaking transformation, the unknown scalar field, and the elastic constants of the virtual plate. It is noted that the cloaking map needs to satisfy certain conditions on the outer boundary of the cloak and the surface of the hole. In particular, the cloaking map needs to fix the outer boundary of the cloak up to the third order. Assuming a generic radial cloaking map, we show that cloaking a circular hole in Kirchoff-Love plates is not possible; the cloaking compatibility equations and the boundary conditions that the cloaking map needs to satisfy are the obstruction to cloaking. Next, relaxing the pure bending assumption, the transformation cloaking problem of an elastic plate in the presence of in-plane and out-of-plane displacements is formulated. In this case, there are two sets of governing equations that need to be simultaneously transformed under the cloaking map. We show that cloaking a circular hole is not possible for a general radial cloaking map; similar to the case of Kirchoff-Love plates, the cloaking compatibility equations and the boundary conditions that the cloaking map needs to satisfy obstruct transformation cloaking. 
\end{abstract}

\begin{description}
\item[Keywords:] Cloaking, Nonlinear elasticity, Elastic plates, Elastic waves.
\end{description}
 
\tableofcontents

\section{Introduction}

Hiding objects from electromagnetic waves has been a subject of intense interest in recent years. \citet{pendry2006} and \citet{leonhardt2006} studied the possibility of electromagnetic transformation cloaking, which was later experimentally verified for microwave frequencies by \citet{Schurig2006} and for optical wavelengths ($1.4-2.7\,\mu m$) by \citet{Ergin2010}. The first ideas pertaining to cloaking in elasticity can be found in the works of \citet{Gurney1938} and \citet{Reissner1949} on reinforced holes in linear elastic sheets. \citet{Mansfield1953} systematically studied cloaking in the context of linear elasticity by introducing the concept of neutral holes. Mansfield considered a hole(s) in a sheet under a given far-field in-plane loading and showed that the hole(s) can be reinforced such that the stress field outside the hole(s) is identical to that of an uncut sheet under the same (far-field) loading. The shape of the boundary of the (neutral) hole and the characteristics of the reinforcement are determined based on the stress field of the uncut sheet, and thus, they explicitly depend on the applied far-filed loading. The main difference between electromagnetic and elastodynamic transformation cloaking is that unlike Maxwell's equations (with only one configuration, i.e., the ambient space), the governing equations of elasticity are written with respect to two inherently different configurations (frames): a reference and a current configuration. This, in turn, leads to two-point tensors in the governing equations. \citet{MaHu1983}, \citet{yavari2008covariance}, and \citet{Steigmann2007} showed that if formulated properly, the governing equations of nonlinear and linearized elasticity are covariant (invariant) under arbitrary time-dependent changes in the current configuration (frame). The governing equations of elasticity are invariant under any time-independent changes in the reference configuration (or referential coordinate transformations) as well \citep{MazzucatoRachele2006,YaMaOr2006}. Nevertheless, as was shown in \citep{yavari2018nonlinear}, referential or spatial covariance of the governing equations is not the direct underlying principle of transformation cloaking; a cloaking map is neither a referential nor a spatial change of coordinates. Rather, a cloaking transformation maps the boundary-value problem of an isotropic and homogeneous elastic body containing an infinitesimal hole (virtual problem) to that of a generally anisotropic and inhomogeneous elastic body with a finite hole surrounded by a cloak (physical problem). The cloak should be designed such that the physical and virtual problems have identical solutions (elastic measurements in the case of elastodynamics) outside the cloak.

The idea of cloaking has been thoroughly studied and is well understood in the context of conductivity \citep{Greenleaf2003a,Greenleaf2003b}, electrical impedance tomography and electromagnetism \citep{BryanLeise2010,Greenleaf2007,Greenleaf2009a,Greenleaf2009b}. The possibility of cloaking has been examined in many other fields of science and engineering, e.g., acoustics \citep{ChenChan2007,Farhat2008,Farhat2009a,Farhat2009b,Norris2008,Cummer2008,cummer2008material,Zhou2008}, optics \citep{leonhardt2012geometry}, thermodynamics (i.e., design of thermal cloaks) \citep{Guenneau2012,han2014full}, diffusion \citep{guenneau2013fick}, quantum mechanics \citep{Zhang2008cloaking}, thermoelasticity \citep{syvret2019thermoelastic,hostos2019metamaterial}, seismology \citep{Al-Attar2016,sklan2018seismic}, and elastodynamics \citep{Milton2006,Parnell2012} (see the recent reviews \citep{Kadic2013, kadic2015experiments,koschny2017metamaterials} for a discussion of these applications in some detail). Recently, we formulated both the nonlinear and linearized elastodynamic transformation cloaking problems (in $3$D) in a mathematically precise form \citep{yavari2018nonlinear}. In this paper, we provide a geometric formulation of transformation cloaking in elastic plates starting from nonlinear shell theory. In our opinion, none of the existing works in the literature has properly formulated the transformation cloaking problem in elastic plates. In particular, the boundary (and continuity) conditions on the hole surface and the outer surface of the cloak, and the restrictions they impose on cloaking transformations have not been discussed. Additionally, we derive several constraints that involve the cloaking transformation, along with the elastic parameters of the virtual plate. These constraints seem to have been ignored in the literature to this date.  

Many solid mechanics workers traditionally have used the classical formulation of linear elasticity. This is appropriate for many practical engineering applications. In the case of elastodynamic transformation cloaking, however, as observed in \citep{yavari2018nonlinear}, starting from linear elasticity is not appropriate. This is because linear elasticity does not distinguish between the reference and the current configurations and the corresponding changes of coordinates defined in these inherently different configurations. This has been a source of confusion in the recent literature of transformation cloaking in elastodynamics. Coordinate transformations in the reference and current configurations are physically very different: Local referential changes of frame are related to the local material symmetry group, whereas the global coordinate transformations in the ambient space are related to objectivity (or material frame indifference). This, in turn, implies that even in the case of small strains, any elastodynamic transformation cloaking study needs to be formulated in the nonlinear framework. 

Examples of improper formulations of transformation cloaking in elastic plates can be seen in almost all the existing works in the literature. \citet{colquitt2014tra} studied transformation cloaking in Kirchhoff-Love plates subjected to time-harmonic out-of-plane displacements in the setting of linear elasticity and started from a governing equation simplified for the case of an isotropic and homogeneous (Kirchhoff-Love) plate. In \S\ref{four4.1}, we show in detail that their transformation cloaking formulation is, unfortunately, incorrect. In particular, their transformed rigidity tensor is incorrect and does not agree with that of an isotropic and homogeneous elastic plate when the cloaking map is the identity, and is not positive definite. \citet{colquitt2014tra} and many other researchers, e.g., \citet{brun2014transformation,jones2015singular,misseroni2016cymatics,alamzareei17,darabi2018experimental,liu2019nonlinear} start from the following equation
\begin{equation}\label{key-wave1}
D^{(0)}\nabla_{\mathbf X}^4W{(\mathbf X)}-Ph\omega^2W{(\mathbf X)}=0\,,\quad\mathrm{or}\quad\left(\nabla_{\mathbf X}^4-\frac{Ph}{D^{(0)}}\omega^2\right)W{(\mathbf X)}=0\,,\quad\mathbf{X}\in\chi\subseteq\mathbb{R}^2\,,
\end{equation}
where $h$ is the plate thickness, $P$ is the mass density, $W(\mathbf X)$ is the amplitude and $\omega$ is the frequency of time-harmonic transverse waves, and $D^{(0)}=Eh^3/12(1-\nu^2)$ is the bending rigidity. They next transform the governing equation under an invertible transformation $\mathcal{F}:\chi\to\Omega$, where $\mathbf{x}=\mathcal{F}(\mathbf X)$,  $\mathbf{F}=\nabla_{\mathbf X}\mathbf{x}$, and $J=\det\mathbf{F}$, using \citep[Lemma 2.1]{Norris2008} twice and obtain
\begin{equation}\label{key-fl-tr1}
\left(\nabla\cdot J^{-1}\mathbf{F}\mathbf{F}^{\mathsf T}\nabla J\nabla\cdot J^{-1}\mathbf{F}\mathbf{F}^{\mathsf T}\nabla -\frac{Ph}{JD^{(0)}}\omega^2\right)W(\mathbf x)=0\,,\quad\mathbf{x}\in{\Omega}\,.
\end{equation}
We show that applying \citep[Lemma 2.1]{Norris2008} twice, one assumes that the gradients of (out-of-plane) displacements and the gradients of the Laplacian of displacements in the physical and virtual plates are related by the Piola transformation. Surprisingly, in none of the works that use this lemma to transform the biharmonic equation \eqref{key-wave1} is there any discussion of these strong assumptions and whether they are compatible with the fact that displacements in the physical and virtual plates are required to be equal. In particular, we see in \S\ref{four4.1} that taking these assumptions into account, the cloaking map is forced to isometrically transform the governing equations of the virtual plate to those of the physical plate. Therefore, this formulation of the cloaking problem does not result in any new information, and the physical and virtual plates are essentially the same elastic plate with the same mechanical response. Furthermore, we will explain that ignoring these assumptions and what constraints they impose on the cloaking map has resulted in obtaining incorrect transformed fields for the physical problem.


In another related work, \citet{Colquitt2013} presented a formulation for the cloaking of the out-of-plane shear waves for the Helmholtz equation. In Remark. \ref{key-helm-holtz}, we illustrate that their cloaking scheme is, unfortunately, incorrect. Similar to \citep{colquitt2014tra}, they use \citep[Lemma 2.1]{Norris2008} to transform the Helmholtz equation for the isotropic and homogeneous medium without considering the restrictions that are imposed on the cloaking map by this lemma. We show that their cloaking map does not satisfy these restrictions, and therefore, does not result in cloaking.

More recently, \citet{pomot2019form} pointed out some shortcomings of \citet{colquitt2014tra}'s formulation of flexural cloaking\footnote{They, however, do not provide a clear mathematical reasoning as to why \citet{colquitt2014tra}'s formulation is incorrect.} and attempted to formulate the flexural cloaking problem. However, their formulation is not consistent. In particular, it is not clear what type of coordinate transformation is being used as they do not distinguish between spatial and referential coordinate transformations. Furthermore, their formulation is missing the important constraint \eqref{Plate-Constraint} that puts severe restrictions on the cloaking map. In particular, their claim that ``the transformed equation does not verify in general the equilibrium equation $N_{IJ,J}+S_{I}=0$." is incorrect. 
We show in \S\ref{four4.1} and \S\ref{linear-elas} that the equilibrium equations for a finitely-deformed plate for a generic radial cloaking transformation do not put any extra constraints on the cloaking map as long as the cloaking map satisfies the \emph{cloaking compatibility equations} \eqref{key-constraint}. Finally, there is no discussion on the boundary conditions and what restrictions they impose on the cloaking map. In particular, their cloaking map is inadmissible, because its derivative is not the identity on the outer boundary of the cloak. \citet{pomot2019form} consider two spaces $E$ and $e$ related by the mapping $\Phi$ (which they call the geometrical transformation) such that $F=\nabla_xX$, denotes the Jacobian of $\Phi$, $J=\det\Phi$, and $\Gamma_{Iij}=\frac{\partial^2 X_{I}}{\partial x_{i}\partial x_j}$, which they call the Hessian matrix. The flexural rigidities, mass densities, and infinitesimal displacements in $E$ and $e$, are, denoted by $(D_{IJKL},P,U)$ and $(D_{ijkl},\rho,u)$, respectively.
They write the total energy density in the transformed space as $\mathcal E=\mathcal{W}+\mathcal{T}$, where $\mathcal{W}=\int_{e}\frac{1}{2}u_{,ij}D_{ijkl}(\mathbf{x})u_{,kl}\mathrm{d}\mathrm{v}$, and $\mathcal{T}=\int_{e}\frac{1}{2}\rho(\mathbf{x})h\dot{u}^2\mathrm{d}\mathrm{v}$, with $h$ being the plate thickness. 
In the initial space, they write the Hamiltonian as $\mathcal{E}_0=\mathcal{W}_0+\mathcal{T}_0$, where
\begin{equation}
\begin{split}
\mathcal{W}_0&=\int_{E}\frac{1}{2}\left(U_{,IJ}F_{Ii}F_{Jj}+U_{,I}\Gamma_{Iij}\right)D_{ijkl}(\mathbf{X})\left(U_{,KL}F_{Kk}F_{Ll}+U_{,K}\Gamma_{Kkl}\right)J^{-1}\mathrm{d}\mathrm{V}\,,\\
\mathcal{T}_{0}&=\int_{E}\frac{1}{2}P(\mathbf X)\dot{U}^2J^{-1}\mathrm{d}\mathrm{V}\,.
\end{split}
\end{equation}
Next, they obtain the equations of motion from the stationarity of the Lagrangian density $\mathcal{L}_{0}=\mathcal{W}_{0}-\mathcal{T}_{0}$. They write the energy variation as 
\begin{equation}\label{key-gar}
\begin{split}
\delta\mathcal{W}_{0}=&\int_{E}\delta U\left(J^{-1}D_{ijkl}F_{Ii}F_{Jj}F_{Kk}F_{Ll}U_{,KL}\right)_{,IJ}\mathrm{d}\mathrm{V}+\int_{E}\delta U\left(J^{-1}D_{ijkl}F_{Ii}F_{Jj}\Gamma_{Kkl}U_{,K}\right)_{,IJ}\mathrm{d}\mathrm{V}\\&-\int_{E}\delta U\left(J^{-1}D_{ijkl}\Gamma_{Iij}F_{Kk}F_{Ll}U_{,KL}\right)_{,I}\mathrm{d}\mathrm{V}-\int_{E}\delta U\left(J^{-1}D_{ijkl}\Gamma_{Iij}\Gamma_{Kkl}U_{,K}\right)_{,I}\mathrm{d}\mathrm{V}\,,
\end{split}
\end{equation}
that is rewritten as
\begin{equation}\label{key-gar1}
 \delta\mathcal{W}_0=\int_{E}\delta U\left(-M_{IJ,IJ}-N_{IJ}U_{,IJ}+S_{I}U_{,I}\right)\mathrm{d}\mathrm{V}\,. 
\end{equation} 
Using \eqref{key-gar}, in turn, they determine $D_{IJKL}$, $N_{IJ}$, and $S_I$ (their Eq. (14)).\footnote{Note the typo in the expression for $S_{I}$ in their Eq. (14).}
Two comments are in order here: (i) Rewriting 
\eqref{key-gar} as \eqref{key-gar1}, one needs to take the following constraint into account
\begin{equation} \label{Plate-Constraint}
	D_{ijkl}F_{Ii}F_{Jj}\Gamma_{Kkl}=D_{ijkl} F_{Kk}F_{Il}\Gamma_{Jij}\,,
\end{equation}
which is a consequence of the fact that the coefficient of the third gradient of $U$ must be identical in these expressions. This has been ignored in their work. We derive the constraints of this type in their general form for Kirchhoff-Love plates (see Eq. \eqref{key-constraint}) and for plates with both the in-plane and out-of-plane displacements (see Remark \ref{key-consq}). We call these constraints the cloaking compatibility equations. (ii) They suggested using a cloaking map for which $\Gamma=\mathbf{0}$, i.e., a cloaking map with a covariantly constant derivative map. However, as we explain in Remark \ref{key-pp}, such mappings are not admissible for cloaking in Kirchhoff-Love plates.

In this paper, in order to obtain the governing equations of an elastic plate, we first obtain those of a nonlinear elastic shell. This is crucial in order to properly account for the variations of the geometric objects of a surface, and thus, obtain the correct linearized governing equations of an elastic plate. An elastic shell is modeled by an orientable two-dimensional Riemannian submanifold embedded in the Euclidean space. The geometry of the shell, is thus, characterized by its first and second fundamental forms that, respectively, represent the intrinsic (in-plane) and extrinsic (out-of-plane) geometries of the surface. Utilizing a Lagrangian field theory, we derive the governing equations of motion. To account for the contribution of body forces and body moments associated with the variations of the position and the orientation (i.e., the normal vector) of the surface, we use the Lagrange--d'Alembert principle. The linearized governing equations for initially stress-free shells, and also for elastic shells with non-vanishing initial stress, couple-stress, and initial body forces and moments are also derived.

Next, the transformation cloaking problem for Kirchhoff-Love plates is formulated. We start with the balance of linear momentum for the virtual plate with uniform elastic properties in the absence of initial stress (and couple-stress) and initial body forces (and moments). The governing equation of the virtual plate is then transformed to that of the physical plate up to an unknown scalar field. The physical plate is subjected to initial stress and tangential body forces that are determined using the elastic constants of the virtual plate and the cloaking map. 
The transformed couple-stress is then determined. It is seen that the cloaking map needs to satisfy certain conditions on the boundaries of the cloak and the hole. In particular, we show that the cloaking transformation needs to fix the outer boundary of the cloak up to the third order. Assuming a generic radial cloaking map, we show that cloaking a circular hole in Kirchhoff-Love plates is not possible.

The pure bending assumption is then relaxed and the transformation cloaking problem of an elastic plate, for which both the in-plane and the out-of-plane displacements are allowed, is formulated. To our best knowledge, this has not been discussed in the literature. In this case, in addition to the flexural rigidity, one needs to consider the in-plane rigidity (stiffness), along with the tensor of elastic constants corresponding to the coupling between the in-plane and the out-of-plane deformations. The physical plate is subjected to initial stress, initial body forces (normal and tangential) and moments. Also, we allow the physical plate to undergo finite in-plane deformations while remaining flat. The virtual plate is assumed to have uniform elastic parameters with vanishing pre-stress and body forces (and moments). The governing equations of the virtual plate are then mapped to those of the physical plate. 
There are two sets of governing equations (i.e., in-plane and out-of-plane) that need to be simultaneously transformed under a cloaking map. The pre-stress, initial body forces and moments, along with the elastic parameters of the physical plate are then determined and the symmetries of the elastic constants in the physical problem are discussed. 
Calculating the stress and couple-stress in the physical problem, and determining the boundary conditions and the restrictions they impose on the cloaking map are discussed. We show that cloaking a circular hole in a plate with both the in-plane and out-of-plane displacements using a general radial cloaking map is not possible. Finally, we prove that if in the virtual plate the tensor of elastic constants corresponding to the coupling between the in-plane and the out-of-plane deformations is positive definite, then cloaking is not possible for any hole covered by a cloak with an arbitrary shape.  

This paper is structured as follows: In \S\ref{two2}, we tersely review some elements of the differential geometry and the kinematics of embedded hypersurfaces in three-dimensional Riemannian manifolds. The governing equations of nonlinear elastic shells are derived in \S\ref{three3}. We then obtain the equations of motion of linear initially stress-free and pre-stressed elastic shells (and thus, plates) by linearizing the nonlinear shell equations. In \S\ref{four4} the problem of transformation cloaking in elastic plates is formulated. We discuss how the geometry of the physical and virtual shells as well as the boundary conditions in the physical and virtual problems are related under a cloaking transformation. In \S\ref{four4.1}, we formulate the transformation cloaking problem for Kirchhoff-Love plates and study the example of a circular cloak assuming a generic radial cloaking map. Next, we relax the pure bending assumption and formulate the transformation cloaking problem for an elastic plate in the presence of both the in-plane and out-of-plane displacements in \S\ref{linear-elas}. We solve the example of a circular cloak using a radial cloaking map. Finally, we discuss the obstruction to transformation cloaking for a hole of arbitrary shape.
Conclusions are given in \S\ref{five5}.

\section{Differential geometry of surfaces}\label{two2}

In this section, we briefly review some concepts of the geometry of two-dimensional embedded surfaces in three-manifolds and the kinematics of elastic shells (see \citep{hicks1965notes,do1992riemannian,sadik2016shell} for more detailed discussions).

\subsection{Geometry of an embedded surface}
Consider an orientable Riemannian manifold $(\mathcal{B},\bar{\mathbf{G}})$, and let $(\mathcal{H},\mathbf{G})$ be an orientable two-dimensional Riemannian submanifold of $(\mathcal{B},\bar{\mathbf{G}})$ such that $\mathbf{G}$ is the induced metric on $\mathcal{H}$, i.e., $\mathbf{G}=\bar{\mathbf{G}}|_{\mathcal{H}}$. Let us denote the space of smooth vector fields on $\mathcal{H}$ and $\mathcal{B}$ by $\mathcal{X}(\mathcal{H})$ and $\mathcal{X}(\mathcal{B})$, respectively. First, we note that for any $X\in\mathcal{B}$,
\begin{equation}
T_X\mathcal{B}=T_X\mathcal{H}\otimes(T_X\mathcal{H})^{\perp}\,,
\end{equation}
that is any vector field $\mathbf{W}\in T_X\mathcal{B}$ may be uniquely written as sum of a vector $\mathbf{W}^\top\in T_X\mathcal{H}$ (which is tangent to $\mathcal H$) and a vector $\mathbf{W}^\perp:=\mathbf{W}-\mathbf{W}^\top$ (which is normal to $\mathcal{H}$, i.e., $\mathbf{W}^\perp\in (T_X\mathcal{H})^\perp$). Given the Levi-Civita connection $\bar{\nabla}$ of the Riemannian manifold $(\mathcal{B},\bar{\mathbf{G}})$, the induced Levi-Civita connection on $(\mathcal{H},\mathbf{G})$ is denoted by $\nabla$ and is given by 
\begin{equation}
\nabla_{\mathbf{X}}\mathbf{Y}=\bar{\nabla}_{\bar{\mathbf{X}}}\bar{\mathbf{Y}}-\bar{\mathbf{G}}(\bar{\nabla}_{\bar{\mathbf{X}}}\bar{\mathbf{Y}},\bar{\mathbf{N}})\mathbf{N}\,,\quad\forall ~ \mathbf{X}\,,\mathbf{Y}\in\mathcal{X}(\mathcal{H})\,, 
\end{equation}
where $\bar{\mathbf{X}},\bar{\mathbf{Y}}\in\mathcal{X}(\mathcal{B})$ are arbitrary local extensions of $\mathbf{X}$ and $\mathbf{Y}$ (i.e., $\bar{\mathbf{X}}(X)=\mathbf{X}(X),\bar{\mathbf{Y}}(X)=\mathbf{Y}(X)\,,\forall X\in\mathcal{H}$), and $\mathbf{N}\in \mathcal{X}(\mathcal{H})^\perp$ is the smooth unit normal vector field to $\mathcal{H}$ with $\bar{\mathbf{N}}$ being its local extension. The second fundamental form of the hypersurface $\mathcal{H}$ is a bilinear and symmetric mapping $\mathbf{B}:\mathcal{X}(\mathcal{H})\times\mathcal{X}(\mathcal{H})\to\mathcal{X}(\mathcal{H})^\perp$ given by
\begin{equation}\label{sec-fun-form}
\mathbf{B}(\mathbf{X},\mathbf{Y})=\bar{\nabla}_{\bar{\mathbf{X}}}\bar{\mathbf{Y}}-{\nabla}_{\mathbf{X}}\mathbf{Y}\,,\quad \forall~\mathbf{X},\mathbf{Y}\in\mathcal{X}(\mathcal{H})\,.
\end{equation}
The set of symmetric $0\choose2$-tensors on $\mathcal H$ is indicated by $\Gamma(S^2T^*\mathcal{H})$. The second fundamental form can be considered as the symmetric tensor $\mathbf{B}\in\Gamma(S^2T^*\mathcal{H})$ defined with a slight abuse of notation as\footnote{
A linear connection is said to be compatible with a metric $\bar{\mathbf{G}}$ on the manifold provided that
\begin{equation}\label{me-comp}
\bar{\nabla}_{\bar{\mathbf{X}}}\llangle \bar{\mathbf{Y}},\bar{\mathbf{Z}}\rrangle_{\bar{\mathbf{G}}}
=\llangle \bar{\nabla}_{\bar{\mathbf{X}}}\bar{\mathbf{Y}},\bar{\mathbf{Z}} \rrangle_{\bar{\mathbf{G}}}
+\llangle \bar{\mathbf{Y}},\bar{\nabla}_{\bar{\mathbf{X}}}\bar{\mathbf{Z}} \rrangle_{\bar{\mathbf{G}}}\,,
\end{equation}
where $\left\langle\!\left\langle .,. \right\rangle\!\right\rangle_{\bar{\mathbf{G}}}$ is the inner product induced by the metric $\bar{\mathbf{G}}$. It is straightforward to show that $\bar{\nabla}$ is compatible with $\bar{\mathbf{G}}$ if and only if $\bar{\nabla}\bar{\mathbf{G}}=\mathbf{0}$, or, in components
\begin{equation*}
\bar{G}_{AB|C}=\frac{\partial \bar{G}_{AB}}{\partial X^C}-\bar{\Gamma}^S{}_{CA}\bar{G}_{SB}-\bar{\Gamma}^S{}_{CB}\bar{G}_{AS}=0.
\end{equation*}
On any Riemannian manifold $(\mathcal{B},\bar{\mathbf{G}})$ the Levi-Civita connection is the unique linear connection $\bar{\nabla}^{\bar{\mathbf{G}}}$ that is compatible with $\bar{\mathbf{G}}$ and is symmetric (torsion-free).	
	Note that the metric compatibility of $\bar{\nabla}$ (see \eqref{me-comp}) and the fact that $\bar{\mathbf{G}}(\bar{\mathbf{N}},\bar{\mathbf{Y}})=0$, are used in deriving the second equality in \eqref{2nd-fun}.} 
\begin{equation}\label{2nd-fun}
\mathbf{B}(\mathbf{X},\mathbf{Y})=\bar{\mathbf{G}}(\bar{\nabla}_{\bar{\mathbf{X}}}\bar{\mathbf{Y}},\bar{\mathbf{N}})=-\bar{\mathbf{G}}(\bar{\nabla}_{\bar{\mathbf{X}}}\bar{\mathbf{N}},\bar{\mathbf{Y}})\,,\quad\forall\,\mathbf{X},\mathbf{Y}\in\mathcal{X}(\mathcal{H})\,,
\end{equation}
which is known as the \emph{Weingarten formula}. The covariant derivative extends to tensors in $S^2T^*\mathcal{H}$ in a natural way ($\mathbf{X}, \mathbf{Y}, \mathbf{Z}\in\mathcal{X}(\mathcal{H})$): 
\begin{equation}
(\nabla_{\mathbf{X}}\mathbf{A})(\mathbf{Y},\mathbf{Z})=\mathbf{X}(\mathbf{A}(\mathbf{Y},\mathbf{Z}))-\mathbf{A}({\nabla}_{\mathbf{X}}\mathbf{Y},\mathbf{Z})-\mathbf{A}(\mathbf{Y},{\nabla}_{\mathbf{X}}\mathbf{Z})\,,\quad\forall\mathbf{A}\in\Gamma(S^2T^*\mathcal{H})\,.
\end{equation}
The curvature tensor $\bar{\boldsymbol{\mathcal{R}}}$ associated with the Riemannian manifold $(\mathcal{B},\bar{\mathbf{G}})$ is defined as
\begin{equation}
\bar{\boldsymbol{\mathcal{R}}}(\mathbf{X},\mathbf{Y},\mathbf{Z},\mathbf{T})=\bar{\mathbf{G}}(\bar{\mathbf{R}}(\mathbf{X},\mathbf{Y})\mathbf{Z},\mathbf{T})\,, \quad \forall~\mathbf{X},\mathbf{Y},\mathbf{Z},\mathbf{T}\in\mathcal{X}(\mathcal{B})\,,
\end{equation}
where $\bar{\mathbf{R}}(\mathbf{X},\mathbf{Y}):\mathcal{X}(\mathcal{B})\to\mathcal{X}(\mathcal{B})$, the Riemann curvature tensor, is given by
\begin{equation}
\bar{\mathbf{R}}(\mathbf{X},\mathbf{Y})\mathbf{Z}={\bar{\nabla}}_{\mathbf{Y}}{\bar{\nabla}}_{\mathbf{X}}\mathbf{Z}-{\bar{\nabla}}_{\mathbf{X}}{\bar{\nabla}}_{\mathbf{Y}}\mathbf{Z}+{\bar{\nabla}}_{[\mathbf{X},\mathbf{Y}]}\mathbf{Z}\,,\quad \mathbf{Z}\in\mathcal{X}(\mathcal{B})\,,
\end{equation}
with $[\mathbf{X},\mathbf{Y}]$ indicating the Lie brackets of $\mathbf{X}$ and $\mathbf{Y}$, i.e., in components, $[\mathbf{X},\mathbf{Y}]^a=\frac{\partial Y^a}{\partial x^b}X^b-\frac{\partial X^a}{\partial x^b}Y^b$.
The Riemann curvature tensor has the following components
\begin{equation}
\bar{\mathcal{R}}_{ABCD}=\left(\bar{\Gamma}^K{}_{AC,B}-\bar{\Gamma}^K{}_{BC,A}+\bar{\Gamma}^L{}_{AC}\bar{\Gamma}^K{}_{BL}-\bar{\Gamma}^L{}_{BC}\bar{\Gamma}^K{}_{AL}\right)\bar{G}_{KD}\,.
\end{equation}
Note that the components of the Christoffel symbols of the connection $\bar{\nabla}$ read
\begin{equation}\label{key-me4}
\bar{\Gamma}^A{}_{BC}=\frac{1}{2}\bar{G}^{AK}(\bar{G}_{KB,C}+\bar{G}_{KC,B}-\bar{G}_{BC,K})\,,
\end{equation}
where a comma preceding a subscript denotes partial differentiation with respect to that subscript. The curvature tensor $\boldsymbol{\mathcal{R}}$ for the hyperplane is similarly defined by the induced metric $\mathbf{G}$ and connection $\nabla$ on $\mathcal{H}$. The Gauss equation reads
\begin{equation}\label{ga-f}
\bar{\boldsymbol{\mathcal{R}}}(\bar{\mathbf{X}},\bar{\mathbf{Y}},\bar{\mathbf{Z}},\bar{\mathbf{T}})=\boldsymbol{\mathcal{R}}(\mathbf{X},\mathbf{Y},\mathbf{Z},\mathbf{T})-\mathbf{B}(\mathbf{X},\mathbf{Z})\mathbf{B}(\mathbf{Y},\mathbf{T})+\mathbf{B}(\mathbf{X},\mathbf{T})\mathbf{B}(\mathbf{Y},\mathbf{Z})\,.
\end{equation}
The Codazzi-Mainardi equation is written as
\begin{equation}\label{co-f}
\bar{\boldsymbol{\mathcal{R}}}(\bar{\mathbf{X}},\bar{\mathbf{Y}},\bar{\mathbf{Z}},\mathbf{N})=({\nabla}_{\mathbf{Y}}\mathbf{B})(\mathbf{X},\mathbf{Z})-({\nabla}_{\mathbf{X}}\mathbf{B})(\mathbf{Y},\mathbf{Z})\,.
\end{equation}
Consider a local coordinate chart $\{X^1,X^2,X^3\}$ for $(\mathcal{B},\bar{\mathbf{G}})$ such that $\{X^1,X^2\}$ is a local chart for $(\mathcal{H},\mathbf{G})$ and $\mathbf{N}$, the unit normal vector to $\mathcal{H}$, is in the direction $\partial/\partial X^3$ at any point of the hypersurface. The metric of $\mathcal{B}$ in this coordinate chart has the following representation
\begin{equation}\label{key-me}
\bar{\mathbf{G}}(X)=\begin{bmatrix}
\bar{G}_{11}(X) & \bar{G}_{12}(X)  & \bar{G}_{13}(X) \\
\bar{G}_{12}(X) & \bar{G}_{22}(X)  & \bar{G}_{23}(X)  \\
\bar{G}_{13}(X) & \bar{G}_{23}(X)  & \bar{G}_{33}(X)  
\end{bmatrix}\,.
\end{equation}
The first fundamental form of the hypersurface is given by
\begin{equation}
\mathbf{G}(X)=\bar{\mathbf{G}}(X)|_{\mathcal{H}}=\begin{bmatrix}
\bar{G}_{11}(X) & \bar{G}_{12}(X) \\
\bar{G}_{12}(X) & \bar{G}_{22}(X)  
\end{bmatrix}\,, \quad \forall ~ X\in\mathcal{H}\,.
\end{equation}
Therefore, the Christoffel symbols associated with the induced connection $\nabla$ read
\begin{equation}\label{key-me1}
\Gamma^A{}_{BC}=\frac{1}{2}G^{AK}\left(G_{KB,C}+G_{KC,B}-G_{BC,K}\right)\,,\quad A,B,C,K=1,2.
\end{equation}
The second fundamental form of the hypersurface is obtained as
\begin{equation}\label{key-me2}
B_{AB}(X)=\bar{\Gamma}^3{}_{AB}(X)\,,\quad A,B=1,2\,,~~\forall \,X\in\mathcal{H}\,,
\end{equation}
where $\bar{\Gamma}^A{}_{BC}$ are the Christoffel symbols of the Levi-Civita connection $\bar{\nabla}$. Thus\footnote{Note that for $X\in\mathcal{H}$, the metric \eqref{key-me} has the following representation
\begin{equation*}
\bar{\mathbf{G}}(X)=\begin{bmatrix}
\bar{G}_{11}(X) & \bar{G}_{12}(X)  & 0 \\
\bar{G}_{12}(X) & \bar{G}_{22}(X)  & 0  \\
0 & 0  & 1  
\end{bmatrix}\,,
\end{equation*}
which using \eqref{key-me4} and \eqref{key-me2}, implies \eqref{key-me3}.
}
\begin{equation}\label{key-me3}
B_{AB}(X)=-\frac{1}{2}\frac{\partial\bar{G}_{AB}}{\partial X^3}\bigg|_{\mathcal{H}}(X), \quad A,B=1,2\,,\quad \forall X\in \mathcal{H}\,. 
\end{equation}
The fundamental theorem of surface theory implies that the geometry of the hypersurface $\mathcal{H}$ is fully determined by its first and second fundamental forms $\mathbf{G}$ and $\mathbf{B}$, respectively.\footnote{Note that the first and the second fundamental forms of $\mathcal H$ can be expressed in terms of the metric of the embedding space $\mathcal B$ given by $\left[
\begin{array}{cc}\bar{G}_{11}&\bar{G}_{12}\\\bar{G}_{12}&\bar{G}_{22}\end{array}
\right]$$(X)$, $X\in\mathcal{B}$, which in turn, fully characterizes the geometry of $\mathcal{H}$.} The Gauss and Codazzi-Mainardi equations given in \eqref{ga-f} and \eqref{co-f} in the local coordinate chart $\{X^1,X^2,X^3\}$ reduce in components to
\begin{equation}
\begin{split}
\bar{\mathcal{R}}_{1212}-\mathcal{R}_{1212}&=B_{12}B_{12}-B_{11}B_{22},\\
\bar{\mathcal{R}}_{1213}&=B_{11|2}-B_{12|1},\\
\bar{\mathcal{R}}_{2123}&=B_{22|1}-B_{12|2},
\end{split}
\end{equation}
where the covariant derivatives correspond to the Levi-Civita connection $\nabla$ of $(\mathcal{H},\mathbf{G})$ with Christoffel symbols $\Gamma^C{}_{AB}$. Note that in components: $B_{AB|C}=B_{AB,C}-\Gamma^K{}_{AC}B_{KB}-\Gamma^K{}_{BC}B_{AK}$.

\subsection{Kinematics of shells}

A shell is a $3$D body whose thickness compared to its other dimensions is very small. Thus, it can be idealized as a two-dimensional Riemannian submanifold $(\mathcal{H},\mathbf{G},\mathbf{B})$ of the Riemannian manifold $(\mathcal{B},\bar{\mathbf{G}})$.\footnote{See \citep{Simo1988} for another equivalent way of characterizing the configuration space of a plate.}  
Let us denote the ambient space by the Riemannian manifold $(\mathcal{S},\bar{\mathbf{g}})$, where $\bar{\mathbf{g}}$ is the standard Euclidean metric. The shell is specified by $(\mathcal{H},\mathbf{G},\mathbf{B})$ and $(\varphi_t(\mathcal{H}),\mathbf{g},\boldsymbol{\theta})$ in the reference and current configurations, respectively, where $\varphi_t:\mathcal{H}\to\mathcal{S}$ is a motion (or a deformation) of $\mathcal{H}$ in $\mathcal{S}$, and the first and the second fundamental forms of the deformed shell $\varphi_t(\mathcal{H})$ are denoted by $\mathbf{g}=\bar{\mathbf{g}}|_{\varphi_t(\mathcal{H})}$ and $\boldsymbol{\theta}\in\Gamma(S^2T^*\varphi(\mathcal{H}))$, respectively.\footnote{Note that in coordinates $(x^1,x^2,x^3)$, for which $x^3$ is the outward normal direction, the second fundamental form of the deformed shell is expressed as
\begin{equation*}
	\theta_{ab}=-\frac{1}{2}\frac{\partial\bar{g}_{ab}}{\partial x^3}\Big|_{\varphi(\mathcal{H})}(x),
	\quad a,b=1,2,\quad \forall x\in \varphi(\mathcal{H})\,.
\end{equation*}
} Let us denote the Levi-Civita connections of $\mathbf{g}$ and $\bar{\mathbf{g}}$ by $\nabla^{\mathbf{g}}$ and $\bar{\nabla}^{\bar{\mathbf{g}}}$, respectively. The smooth unit normal vector field of $\varphi(\mathcal H)$ is denoted by $\mathbf{n}\in \mathcal{X}(\varphi(\mathcal H))^\perp$.
As the ambient space $\mathcal{S}$ is flat, the Gauss and Codazzi-Mainradi equations for the Riemannian hypersuface in its current configuration $(\varphi(\mathcal{H}),\mathbf{g})$ are written as
\begin{equation}\label{am-ga-cod}
\begin{aligned}	
\boldsymbol{\mathsf{R}}(\boldsymbol{x},\boldsymbol{y},\boldsymbol{z},\boldsymbol{w})&=\boldsymbol{\theta}(\boldsymbol{x},\boldsymbol{z})\boldsymbol{\theta}(\boldsymbol{y},\boldsymbol{w})-\boldsymbol{\theta}(\boldsymbol{x},\boldsymbol{w})\boldsymbol{\theta}(\boldsymbol{y},\boldsymbol{z})\,,\\
(\nabla^{\mathbf g}_{\boldsymbol{x}}\boldsymbol{\theta})(\boldsymbol{y},\boldsymbol{z})&=(\nabla^{\mathbf g}_{\boldsymbol{y}}\boldsymbol{\theta})(\boldsymbol{x},\boldsymbol{z})\,,
\end{aligned}
\end{equation}
for any smooth vector fields $\boldsymbol{x}, \boldsymbol{y}, \boldsymbol{z}, \boldsymbol{w}\in \mathcal{X}(\varphi(\mathcal{H}))$, and $\boldsymbol{\mathsf{R}}$ is the Riemannian curvature of the deformed surface $\varphi(\mathcal{H})$. 

The deformation gradient is defined as the tangent map of $\varphi_t:\mathcal{H}\to\varphi_t(\mathcal{H})$, i.e., $\mathbf{F}(X)=T\varphi_t(X):T_X\mathcal{H}\to T_{\varphi_t(X)}\varphi_t(\mathcal{H})$. The right Cauchy-Green deformation tensor is defined as the pull-back of the induced metric on the deformed hypersurface $\varphi_t(\mathcal{H})$ by $\varphi_t$, i.e., $\mathbf{C}^\flat=\varphi_t^*\mathbf{g}$ (in components, $C_{AB}=F^a{}_AF^b{}_Bg_{ab}$, $A,B=1,2$). The Jacobian of the deformation $J$ relates the deformed and undeformed Riemannian surface elements as $ds(\varphi_t(X),\mathbf{g})=JdS(X,\mathbf{G})$, where
\begin{equation}
	J=\sqrt{\frac{\det\mathbf{g}}{\det\mathbf{G}}}\det\mathbf{F}.
\end{equation}
The material (or Lagrangian) strain tensor $\mathbf{E}\in \Gamma(S^2T^*\mathcal{H})$ is defined as $\mathbf{E}=\frac{1}{2}(\mathbf{C}-\mathbf{G})$. Alternatively, one can define the spatial strain tensor as $\mathbf{e}=\frac{1}{2}(\mathbf{g}-\mathbf{c})$, where $\mathbf{c}={\varphi_{t*}}\mathbf{G}$. Note that the material and spatial strain measures are intrinsic in the sense that they only capture changes in the first fundamental form of the surface. Therefore, one needs to define the extrinsic strain measures in order to take into account variations in the second fundamental form of the surface as well. The extrinsic material strain tensor is given by
\begin{equation}
\mathbf{H}=\frac{1}{2}\left(\mathbf{\Theta}-\mathbf{B}\right),
\end{equation}
where $\mathbf{\Theta}=\varphi_t^*\boldsymbol{\theta}$. Similarly, the spatial extrinsic strain tensor can be defined as $\mathbf{h}=\frac{1}{2}\left(\boldsymbol{\theta}-\boldsymbol{\beta}\right)$, where $\boldsymbol{\beta}={\varphi_t}_*\mathbf{B}$. It is straightforward to see that $\mathbf{h}={\varphi_t}_*\mathbf{H}$.


One can pull back the spatial Gauss and Codazzi equations \eqref{am-ga-cod} by $\varphi$ and obtain the following shell compatibility equations (see \citep{angoshtari2015diff,sadik2016shell}) 
\begin{equation}\label{comp-tra}
\begin{aligned}	
\boldsymbol{\mathsf{R}}^{\mathbf C}(\mathbf{X},\mathbf{Y},\mathbf{Z},\mathbf{W})&=\mathbf{\Theta}(\mathbf{X},\mathbf{Z})\mathbf{\Theta}(\mathbf{Y},\mathbf{W})-\mathbf{\Theta}(\mathbf{X},\mathbf{W})\mathbf{\Theta}(\mathbf{Y},\mathbf{Z})\,,\\
(\nabla^{\mathbf C}_{\mathbf{X}}\mathbf{\Theta})(\mathbf{Y},\mathbf{Z})&=(\nabla^{\mathbf C}_{\mathbf{Y}}\mathbf{\Theta})(\mathbf{X},\mathbf{Z})\,,
\end{aligned}
\end{equation}
where $\nabla^{\mathbf C}$ and $\boldsymbol{\mathsf{R}}^{\mathbf C}$ are the Levi-Civita connection and the Riemannian curvature of the Riemannian (convected) manifold $(\mathcal{H},\mathbf{C})$. Note that \eqref{comp-tra} gives the necessary and locally sufficient conditions for the existence (and uniqueness up to isometries of $\mathcal S=\mathbb{R}^3$ when $\mathcal H$ is simply-connected) of a deformation mapping (configuration) of $\mathcal H$ with given deformation tensors $\mathbf C$ and $\mathbf \Theta$. In the coordinate chart $\{X^1,X^2,X^3\}$,  \eqref{comp-tra} is written as
\begin{equation}
\begin{split}
	\mathsf{R}^{\mathbf C}_{1212}&=\Theta_{11}\Theta_{22}-\Theta_{12}\Theta_{12},\\
	\Theta_{11||2}&=\Theta_{12||1},\\
	\Theta_{22||1}&=\Theta_{12|| 2}, 
\end{split}
\end{equation}
where $||$ denotes the covariant derivative with respect to the Levi-Civita connection $\nabla^{\mathbf{C}}$.\footnote{See \citep[P.313]{Ericksen1957} for a discussion on the compatibility equations of a Cosserat shell with deformable directors.} Recall that one may write the components of $\mathbf{C}^\flat$ and $\mathbf{\Theta}^\flat$ in terms of the deformation mapping $\varphi:\mathcal{H}\to\mathbb{R}^3$ in a local chart $\{X,Y\}$ on $\mathcal H$ as follows
\begin{equation}
	C_{AB}=\varphi_{,A}\cdot\varphi_{,B}\,,~~~~~
	\Theta_{AB}=\varphi_{,AB}\cdot\frac{\varphi_{,X}\times\varphi_{,Y}}{\|\varphi_{,X}\times\varphi_{,Y}\|}\,,
\end{equation}
where the dot product, the cross product, and the standard norm in $\mathbb{R}^3$ are denoted by ``$\cdot$'', ``$\times$'', and ``$\|\cdot\|$'', respectively. Note that expressing the first and the second fundamental forms in terms of the motion $\varphi$, the shell compatibility equations \eqref{comp-tra} are trivially satisfied.

\subsection{Velocity and acceleration vector fields}

The material velocity is defined as the mapping $\mathbf{V}:\mathcal{H}\times\mathbb{R}\to T\mathcal{S}$ such that $\mathbf{V}(X,t)=\frac{\partial\varphi_X(t)}{\partial t}\in T_{\varphi_X(t)}\mathcal{S},\, \forall X\in\mathcal{H}$. The material velocity can be decomposed at any point $X\in \mathcal{H}$ as $\mathbf{V}_X(t)=\mathbf{V}^\top_X(t)+\mathbf{V}^\perp_X(t)$, where $\mathbf{V}^\top_X(t)\in T_{\varphi_t(X)}\varphi_t(\mathcal{H})$ is parallel and $\mathbf{V}^\perp_X(t)\in \left(T_{\varphi_t(X)}\varphi_t(\mathcal{H})\right)^\perp$ is normal to the shell in its current configuration.
The spatial velocity is defined as $\mathbf{v}(x,t)=\mathbf{V}(\varphi^{-1}_t(x),t)$, which is a vector field on $\varphi_t(\mathcal H)$ at a fixed time $t$. Note that at any time $t$ the deformation mapping $\varphi_t: \mathcal{H}\to\mathcal{S}$ is a smooth embedding of the shell into the ambient space. The mapping, $\varphi:\mathcal{H}\times\mathbb{R}\to\mathcal{S}$, on the other hand, is not, in general, injective (see \citep{sadik2016shell} for a detailed discussion). The tangent of $\varphi$ is written as
\begin{equation} \label{Tangent-Map}
T_{(X,t)}\varphi=\begin{bmatrix}\frac{\partial\varphi^1}{\partial X^1}& \frac{\partial\varphi^1}{\partial X^2}& \frac{\partial\varphi^1}{\partial t}\\ \frac{\partial\varphi^2}{\partial X^1} &\frac{\partial\varphi^2}{\partial X^2} &\frac{\partial\varphi^2}{\partial t}\\ \frac{\partial\varphi^3}{\partial X^1} &\frac{\partial\varphi^3}{\partial X^2} &\frac{\partial\varphi^3}{\partial t}
\end{bmatrix}\,.
\end{equation}
Notice that the first two columns in \eqref{Tangent-Map} represent the tangent map of the smooth embedding $\varphi_t$, and thus, the tangent map $T_X\varphi_t: T_X\mathcal{H}\to T_{\varphi_t(X)}\mathcal{S}$ is injective. This, in turn, implies that the first two columns are linearly independent. Also, note that the third column represents the material velocity, i.e., $\mathbf{V}(X,t)=\frac{\partial\varphi_X(t)}{\partial t}$. Therefore, $T_{(X,t)}\varphi$ is full rank if and only if $\mathbf{V}(X,t)$ is not purely tangential.
The material acceleration is defined as
\begin{equation}
\mathbf{A}(X,t)=D^{\bar{\mathbf g}}_{\varphi_t}\mathbf{V}(X,t)=\bar{\nabla}^{\bar{\mathbf g}}_{\mathbf V}\mathbf{V}\,,
\end{equation}
where $D^{\bar{\mathbf{g}}}_{\varphi_t}$ denotes the covariant time derivative along the curve $\varphi_X(t)$. 
Note that the material velocity is only defined on the surface $\varphi_t(\mathcal H)$, but one needs to compute the covariant derivative of the velocity along the motion in the ambient space to find the normal and tangential components of the acceleration. Thus, one cannot, in general, compute $\bar{\nabla}^{\bar{\mathbf g}}_{\mathbf V}\mathbf{V}$ to find the acceleration components unless it is possible to define a local extension of $\mathbf V$ to an open neighborhood on $\mathcal S$ (see \citep{Yavari2016} for a detailed discussion).
Provided that $\varphi$ has a nonsingular tangent $T_{(X,t)}\varphi$ at some $(X_o,t_o)\in\mathcal{H}\times\mathbb{R}$, then by the inverse function theorem $\varphi$ is a local diffeomorphism at $(X_o,t_o)$. Therefore, one may construct a local extension vector field $\boldsymbol{\mathcal{V}}$ on $\mathcal{S}$ such that $\boldsymbol{\mathcal{V}}(\varphi(X,t))=\mathbf{V}(X,t)=\mathbf{v}(\varphi(X,t),t)$ in some open neighborhood of $(X_o,t_o)$.\footnote{Note that $T_{(X,t)}\varphi$ (cf. \eqref{Tangent-Map}) is injective, and hence, the local extension vector field always exists, unless $\mathbf V$ is purely tangential, i.e., ${\mathbf V}^\perp=\mathbf{0}$. In this case, however, one does not need a local extension to compute the acceleration unambiguously as $\mathbf{V}=\mathbf{V}^\top$, and hence, $\mathbf{A}(X,t)=\bar{\nabla}^{\bar{\mathbf g}}_{\mathbf V}\mathbf{V}=\bar{\nabla}^{\bar{\mathbf g}}_{{\mathbf V}^\top}\mathbf{V}^\top$.} 
Thus, we may proceed with computing the acceleration as follows
\begin{equation}
	\mathbf{A}(X,t)=D^{\bar{\mathbf g}}_{\varphi_t}\mathbf{V}(X,t)
	:=\bar{\nabla}^{\bar{\mathbf g}}_{\boldsymbol{\mathcal V}}\boldsymbol{\mathcal V}(\varphi(X,t))\,.
\end{equation}
Decomposing the velocity into the normal and parallel components, one obtains
\begin{equation}
\mathbf{A}(X,t)=\bar{\nabla}^{\bar{\mathbf g}}_{\boldsymbol{\mathcal{V}}}{\boldsymbol{\mathcal{V}}}^\top+\bar{\nabla}^{\bar{\mathbf g}}_{\boldsymbol{\mathcal{V}}}{\boldsymbol{\mathcal{V}}}^\perp\,.
\end{equation}
Note that
\begin{equation}
\left[{\boldsymbol{\mathcal{V}}},{\boldsymbol{\mathcal{V}}}^\top \right]=\bar{\nabla}^{\bar{\mathbf{g}}}_{\boldsymbol{\mathcal{V}}}{\boldsymbol{\mathcal{V}}}^\top-\bar{\nabla}^{\bar{\mathbf{g}}}_{\boldsymbol{\mathcal{V}}^\top}{\boldsymbol{\mathcal{V}}}\,.
\end{equation}
Thus
\begin{equation}
\bar{\nabla}^{\bar{\mathbf{g}}}_{\boldsymbol{\mathcal{V}}}{\boldsymbol{\mathcal{V}}}^\top=\left[\boldsymbol{\mathcal{V}},{\boldsymbol{\mathcal{V}}}^\top\right]+\bar{\nabla}^{\bar{\mathbf{g}}}_{{\boldsymbol{\mathcal{V}}}^\top}{\boldsymbol{\mathcal{V}}}^\perp+\bar{\nabla}^{\bar{\mathbf{g}}}_{{\boldsymbol{\mathcal{V}}}^\top}{\boldsymbol{\mathcal{V}}}^\top\,.
\end{equation}
Using the relation \eqref{sec-fun-form} in the ambient space, one has
\begin{equation}
\bar{\nabla}^{\bar{\mathbf g}}_{{\boldsymbol{\mathcal{V}}}^\top}{\boldsymbol{\mathcal{V}}}^\top={\nabla}^{\mathbf g}_{{\boldsymbol{\mathcal{V}}}^\top}{\boldsymbol{\mathcal{V}}}^\top+\boldsymbol{\theta}({\boldsymbol{\mathcal{V}}}^\top,{\boldsymbol{\mathcal{V}}}^\top)\mathbf{n}\,.
\end{equation}
Also, letting ${\boldsymbol{\mathcal{V}}}^\perp={\mathcal{V}}^\perp\mathbf{n}$, one may write 
\begin{equation}
\bar{\nabla}^{\bar{\mathbf g}}_{{\boldsymbol{\mathcal{V}}}^\top}{\boldsymbol{\mathcal{V}}}^\perp=\left({{\boldsymbol{\mathcal{V}}}^\top}\left[{\mathcal{V}}^\perp\right]\right)\mathbf{n}+{\mathcal{V}}^\perp\bar{\nabla}^{\bar{\mathbf g}}_{{\boldsymbol{\mathcal{V}}}^\top}\mathbf{n}=\left(\mathbf{d}{\mathcal V}^\perp\cdot\boldsymbol{\mathcal{V}}^\top\right)\mathbf{n}-{\mathcal{V}}^\perp\mathbf{g}^\sharp\cdot\boldsymbol{\theta}\cdot{\boldsymbol{\mathcal{V}}}^\top,
\end{equation}
where use was made of the relation \eqref{2nd-fun} in the ambient space. Note that 
\begin{equation}
\bar{\nabla}^{\bar{\mathbf g}}_{\boldsymbol{\mathcal V}}{\boldsymbol{\mathcal V}}^\perp=\bar{\nabla}^{\bar{\mathbf g}}_{\boldsymbol{\mathcal V}}({\mathcal V}^\perp\mathbf{n})=\frac{d{\mathcal{V}}^\perp}{dt}\mathbf{n}+\mathcal{V}^\perp\bar{\nabla}^{\bar{\mathbf g}}_{\boldsymbol{\mathcal{V}}}\,\mathbf{n}\,.
\end{equation}
Using the metric compatibility of $\bar{\mathbf g}$ and \eqref{2nd-fun}, one can show that\footnote{Let $\mathbf{W}\in\mathcal{X}(\varphi_t(\mathcal{H}))$ be an arbitrary vector field defined in a neighborhood containing $(X_o,t_o)$. Thus, $\bar{\mathbf g}(\boldsymbol{\mathcal{V}}^\perp,\mathbf{W})=0$, and from \eqref{me-comp}, $\bar{\mathbf g}(\bar{\nabla}^{\bar{\mathbf g}}_{\boldsymbol{\mathcal V}}{\boldsymbol{\mathcal V}}^\perp,\mathbf{W})=-\bar{\mathbf g}({\boldsymbol{\mathcal V}}^\perp,\bar{\nabla}^{\bar{\mathbf g}}_{\boldsymbol{\mathcal V}}\mathbf{W})$. Note that 
\begin{equation*}
\begin{split}
\bar{\nabla}^{\bar{\mathbf g}}_{\boldsymbol{\mathcal V}}\mathbf{W}&=[\boldsymbol{\mathcal V},\mathbf{W}]+\bar{\nabla}^{\bar{\mathbf g}}_{\mathbf W}\boldsymbol{\mathcal V}=[\boldsymbol{\mathcal V},\mathbf{W}]+\bar{\nabla}^{\bar{\mathbf g}}_{\mathbf W}{\boldsymbol{\mathcal V}}^\top+\bar{\nabla}^{\bar{\mathbf g}}_{\mathbf W}{\boldsymbol{\mathcal V}}^\perp\\&=[\boldsymbol{\mathcal V},\mathbf{W}]+{\nabla}^{\mathbf g}_{\mathbf W}{{\boldsymbol{\mathcal V}}}^\top+\boldsymbol{\theta}({{\boldsymbol{\mathcal V}}}^\top,\mathbf{W})\mathbf{n}+(d\mathcal{V}^\perp\cdot\mathbf{W})\mathbf{n}+\mathcal{V}^\perp{\bar{\nabla}}^{\bar{\mathbf g}}_{\mathbf W}\mathbf{n}\,.
\end{split}
\end{equation*}
Thus, noting that $[\boldsymbol{\mathcal V},\mathbf{W}]$, ${\nabla}^{\mathbf g}_{\mathbf W}{{\boldsymbol{\mathcal V}}}^\top$, $\mathcal{V}^\perp{\bar{\nabla}}^{\bar{\mathbf g}}_{\mathbf W}\mathbf{n}\in\mathcal{X}(\varphi_t(\mathcal{H}))$ one concludes that $\bar{\mathbf g}(\boldsymbol{\mathcal V}^\perp,\bar{\nabla}^{\bar{\mathbf g}}_{\boldsymbol{\mathcal V}}\mathbf{W})=\mathcal{V}^\perp\boldsymbol{\theta}({{\boldsymbol{\mathcal V}}}^\top,\mathbf{W})+\mathcal{V}^\perp(d\mathcal{V}^\perp\cdot\mathbf{W})$, which by arbitrariness of $\mathbf W$ together with $\bar{\mathbf g}(\bar{\nabla}^{\bar{\mathbf g}}_{\boldsymbol{\mathcal V}}{\boldsymbol{\mathcal V}}^\perp,\mathbf{W})=-\bar{\mathbf g}({\boldsymbol{\mathcal V}}^\perp,\bar{\nabla}^{\bar{\mathbf g}}_{\boldsymbol{\mathcal V}}\mathbf{W})$ implies \eqref{proof-2}.
}
\begin{equation}\label{proof-2}
\mathcal{V}^\perp\bar{\nabla}^{\bar{\mathbf g}}_{\boldsymbol{\mathcal{V}}}\,\mathbf{n}=-\mathcal{V}^\perp\mathbf{g}^\sharp\cdot\boldsymbol{\theta}\cdot\boldsymbol{\mathcal{V}}^\top-\mathcal{V}^\perp(d\mathcal{V}^\perp)^\sharp\,.
\end{equation}
Hence, replacing $\boldsymbol{\mathcal V}$ by $\mathbf{V}(X,t)=\boldsymbol{\mathcal{V}}(\varphi(X,t))$ the parallel and normal components of the material acceleration are written as
\begin{equation}\label{acc}
\begin{split}
	\mathbf{A}^\top&=\nabla^{\mathbf g}_{{\mathbf V}^\top}{{\mathbf V}^\top}
	+\left[\mathbf{V},\mathbf{V}^\top\right]-2V^\perp\mathbf{g}^\sharp
	\cdot\boldsymbol{\theta}\cdot{{\mathbf V}}^\top-V^\perp(dV^\perp)^\sharp\,,\\
	\mathbf{A}^\perp&=\left(\frac{dV^\perp}{dt}
	+\boldsymbol{\theta}({{\mathbf V}}^\top,{{\mathbf V}}^\top)+dV^\perp\cdot\mathbf{V}^\top\right)\mathbf{n}\,.
\end{split}
\end{equation}


\section{The governing equations of motion of shells}\label{three3}

In this section we use Hamilton's principle of least action to derive the governing equations of a nonlinear elastic shell. We then linearize the governing equations and obtain the equations of motion of a linear elastic shell.

The kinetic energy density per unit surface area is written as
\begin{equation}\label{Kinetic-Energy-Cosserat}
T=\frac{1}{2}\rho~ \bar{\mathbf{g}}(\dot{\varphi},\dot{\varphi})
\,,
\end{equation}
where $\rho$ is the material surface mass density.
The elastic energy density (per unit surface area) of the shell is written as\footnote{
Consider a surface embedded in the ambient space such that the embedding is given as $\varphi: \mathcal{H}\to \mathcal{S}$, where for the sake of simplicity one can assume that $\mathcal{S}=\mathbb{R}^3$. Note that $x^a=\partial \varphi^a/\partial X^A $, where $a=1,2,3$ and $A=1,2$. The fundamental theorem of surface theory proved by \citet{bonnet1867memoire} implies that the surface geometry (up to rigid body motions) is completely characterized by the induced first and second fundamental forms $\mathbf{C}$ and $\mathbf{\Theta}$. Therefore, the surface energy density must depend on $\mathbf C$ and $\mathbf\Theta$.
}
\begin{equation}
 W=W\left(X,\mathbf{C},\mathbf{\Theta},\mathbf{G},\mathbf{B}\right)\,.
\end{equation} 
Let $\{x^1,x^2,x^3\}$ be a local coordinate chart for the ambient space such that at any point of the deformed hypersurface, $\{x^1,x^2\}$ is a local chart for $(\varphi(\mathcal{H}),\mathbf{g})$, and the vector field $\mathbf{n}$ normal to $\varphi(\mathcal{H})$ is tangent to the coordinate curve $x^3$. Thus, the Lagrangian density (per unit surface area) is defined in this coordinate chart as
\begin{equation}
	\mathcal{L}(X,\dot{\varphi},\mathbf{C},\mathbf{\Theta},\mathbf{G},\mathbf{B})
	=\frac{1}{2}\rho\, g_{ab}(\dot{\varphi}^\top)^a(\dot{\varphi}^\top)^b
	+\frac{1}{2}\rho(\dot{\varphi}^\perp)^2-W\left(X,\mathbf{C},\mathbf{\Theta},\mathbf{G},\mathbf{B}\right)\,.
\end{equation}
The action functional is defined as
\begin{equation}
S(\varphi)=\int_{t_0}^{t_1}\int_{\mathcal{H}}\mathcal{L}\Big(X,\dot{\varphi}(X,t),\mathbf{C}(X,t),\mathbf{\Theta}(X,t),\mathbf{G}(X),\mathbf{B}(X)\Big)dA(X)dt\,,
\end{equation}
where $dA(X)=\sqrt{\det \mathbf{G}(X)}~dX^1\wedge dX^2$ is the Riemannian area element. We use the Lagrange--d'Alembert principle to take into account the contribution of non-conservative body forces and body moments associated with the variations of the position $\delta\varphi$, and orientation $\delta\boldsymbol{\mathcal N}$ (where $\boldsymbol{\mathcal{N}}=\mathbf{n}\circ\varphi$ is the normal vector field characterizing the orientation of the deformed surface element). 
The Lagrange-d'Alembert principle \citep{marsden2013introduction} states that the physical motion $\varphi$ of $\mathcal{H}$ satisfies 
\begin{equation}\label{eu-lagrange1}
	\delta S(\varphi)+\int_{t_0}^{t_1}\int_{\mathcal{H}}\bigg(\boldsymbol{\mathfrak{B}}\cdot\delta\varphi
	+\boldsymbol{\mathfrak{L}}\cdot\delta\boldsymbol{\mathcal{N}}\bigg)\rho\, dAdt=0\,,
\end{equation}
where $\boldsymbol{\mathfrak{B}}$ and $\boldsymbol{\mathfrak{L}}$, respectively, denote the external body forces and body moments. Note that
\begin{equation}\label{eu-lagrange}
\delta S(\varphi)=\int_{t_0}^{t_1}\int_{\mathcal{H}}\bigg(\frac{\partial\mathcal{L}}{\partial\dot{\varphi}}\cdot\delta\dot{\varphi}+\frac{\partial\mathcal{L}}{\partial\mathbf{C}}:\delta\mathbf{C}+\frac{\partial\mathcal{L}}{\partial\mathbf{\Theta}}:\delta\mathbf{\Theta}\bigg)dAdt\,.
\end{equation}
Let $\varphi_{\epsilon}$ be a one-parameter family of motions such that $\varphi_{0,t}=\varphi_t$, where, for fixed $X$ and $t$, we denote $\varphi_{\epsilon,t}(X):=\varphi_{\epsilon}(X,t)$. The variation of motion is defined as
\begin{equation}
\delta \varphi(X,t)=\frac{d}{d\epsilon}\Big|_{\epsilon=0}\varphi_{\epsilon,t}(X)\in T_{\varphi_t(X)}\mathcal{S}\,.
\end{equation}
The material velocity is given by $\dot{\varphi}_{\epsilon}=\frac{\partial\varphi_{\epsilon,t}(X)}{\partial t}$. Note that $\dot{\varphi}_{\epsilon}\in T_{\varphi_{\epsilon,t}(X)}\mathcal{S}$, i.e., for fixed time $t$ and $X\in\mathcal{H}$, if $\epsilon$ varies, the velocity lies in different tangent spaces, and thus, a covariant derivative along the curve $\epsilon\to\varphi_{\epsilon,t}(X)$ should be used to find the variation of the material velocity. Thus
\begin{equation}\label{vel-var}
\delta\dot{\varphi}=\bar{\nabla}_{\frac{\partial}{\partial\epsilon}}^{\bar{\mathbf g}}\frac{\partial\varphi_{\epsilon,t}(X)}{\partial t}\Big|_{\epsilon=0}=\bar{\nabla}_{\frac{\partial}{\partial t}}^{\bar{\mathbf g}}\frac{\partial\varphi_{\epsilon,t}(X)}{\partial \epsilon}\Big|_{\epsilon=0}=\bar{\nabla}_t^{\bar{\mathbf g}}\,\delta\varphi_t(X)=D_{\varphi_X(t)}\delta\varphi\,,
\end{equation}
where the symmetry lemma of Riemannian geometry was used to obtain the second equality (see \citep{lee1997riemannian,Nishikawa2002}). The variation of the right Cauchy-Green deformation tensor $\mathbf{C}^{\flat}_{\epsilon}=\varphi^*_{\epsilon,t}\mathbf{g}_{\epsilon}\circ\varphi_{\epsilon,t}$, where $\mathbf{C}^{\flat}_{\epsilon}\in \Gamma(S^2T^*\mathcal{H})$, is obtained as
\begin{equation}
\delta\mathbf{C}^{\flat}=\frac{d}{d\epsilon}\mathbf{C}^{\flat}_{\epsilon}\Big|_{\epsilon=0}=\frac{d}{d\epsilon}(\varphi^*_{\epsilon,t}\mathbf{g}_{\epsilon})\Big|_{\epsilon=0}=\varphi^*_t(\mathbf{L}_{\delta\varphi}\mathbf{g})\,.
\end{equation}
Hence, knowing that (see, e.g., \citep{verpoort2008geometry,kadianakis2013kinematics})
\begin{equation}\label{one-1-key}
 \mathbf{L}_{\delta\varphi}\mathbf{g}=\mathbf{L}_{\delta\varphi^\top}\mathbf{g}-2\,\delta \varphi^\perp\,\boldsymbol{\theta} \,, 
\end{equation}
one obtains (see Appendix \ref{Geometry1} for the details of this derivation)
\begin{equation}\label{cauchy-var}
\delta \mathbf{C}^\flat=\varphi^*_t\,\mathbf{L}_{\delta\varphi^\top}\mathbf{g}-2\,\delta\varphi^\perp\,\mathbf{\Theta}\,.
\end{equation}
In components
\begin{equation}\label{cau-gr-var}
\delta C_{AB}=F^a{}_A\,\delta{\varphi}^{\top}_{a|B}+F^b{}_B\,\delta\varphi^{\top}_{b|A}-2\,\delta\varphi^\perp\,\Theta_{AB}\,.
\end{equation}
Noting that ${\mathbf{\Theta}}^\flat_{\epsilon}\in \Gamma(S^2T^*\mathcal{H})$, the variation of ${\mathbf{\Theta}}^\flat_{\epsilon}=\varphi^*_{\epsilon,t}\boldsymbol{\theta}_{\epsilon}\circ\varphi_{\epsilon,t}$ is calculated as 
\begin{equation}\label{2nd-1}
\delta\mathbf{\Theta}^\flat=\frac{d}{d\epsilon}\mathbf{\Theta}_{\epsilon}^\flat\Big|_{\epsilon=0}=\frac{d}{d\epsilon}(\varphi^*_{\epsilon,t}\boldsymbol{\theta}_{\epsilon})\Big|_{\epsilon=0}=\varphi^*_t(\mathbf{L}_{\delta\varphi}\boldsymbol{\theta})\,.
\end{equation}
For a flat ambient space, the Lie derivative of the second fundamental form is expressed as (see \citep{kadianakis2013kinematics})
\begin{equation}\label{2nd-2}
\mathbf{L}_{\delta\varphi}\boldsymbol{\theta}=\mathbf{L}_{\delta\varphi^\top}\boldsymbol{\theta}-\delta\varphi^\perp\mathbf{III}+\mathrm{Hess}_{\delta\varphi^\perp},
\end{equation}
where $\mathbf{III}$ is the third fundamental form of the deformed hyepersurface, and $\mathrm{Hess}_{\delta\varphi^\perp}$ denotes the Hessian of $\delta\varphi^\perp$ (when viewed as a scalar-valued function on $\varphi_t(\mathcal H)$). The third fundamental form and $\mathrm{Hess}_{\delta\varphi^\perp}$ are given for $\boldsymbol{x},\boldsymbol{y}\in\mathcal{X}(\varphi(\mathcal{H}))$ as 
\begin{align}
	\label{3rd-fundamental-form} \mathbf{III}(\boldsymbol{x},\boldsymbol{y})
	&=\mathbf{g}\left(\bar{\nabla}^{\bar{\mathbf{g}}}_{\boldsymbol{x}}\,
	\mathbf{n},\bar{\nabla}^{\bar{\mathbf{g}}}_{\boldsymbol{y}}\,\mathbf{n}\right),\\
	\label{Hessian} \mathrm{Hess}_{\delta\varphi^\perp}(\boldsymbol{x},\boldsymbol{y})
	&=\mathbf{g}\left(\bar{\nabla}^{\bar{\mathbf{g}}}_{\boldsymbol{x}}\,(\mathbf{d}\,
	\delta\varphi^\perp)^\sharp,\boldsymbol{y}\right),
\end{align}
where $\mathbf{d}$ and $\sharp$, respectively, denote the exterior derivative and the sharp operator for raising indices. Thus, from \eqref{2nd-1} and \eqref{2nd-2}, one obtains
\begin{equation}\label{sec-fun-form-var}
\delta\mathbf{\Theta}^\flat=\varphi^*_t\mathbf{L}_{\delta\varphi^\top}\boldsymbol{\theta}-\delta\varphi^\perp\,\varphi_t^*\mathbf{III}\,+\varphi_t^*\mathrm{Hess}_{\delta\varphi^\perp},
\end{equation}
and in components
\begin{equation}\label{theta-var}
\begin{split}
\delta\Theta_{AB}=&F^a{}_AF^b{}_B\,\theta_{ab|c}\,(\delta\varphi^\top)^c+F^a{}_A\theta_{ac}(\delta\varphi^\top)^c{}_{|B}+F^b{}_B\theta_{bc}(\delta\varphi^\top)^c{}_{|A}\\&-\delta\varphi^\perp F^a{}_AF^b{}_B\theta_{ac}\theta_{bd}\,g^{cd}+F^b{}_A\left(\frac{\partial\,\delta\varphi^\perp}{\partial\, x^b}\right)_{|B}\,.
\end{split}
\end{equation}
The unit normal vector field, $\boldsymbol{\mathcal{N}}_{\epsilon}=\mathbf{n}_{\epsilon}\circ\varphi_{\epsilon,t}$, $\mathbf{n}_{\epsilon}\in\mathcal{X}(\varphi_{\epsilon,t}(\mathcal H))^\perp$, lies in $T_{\varphi_{\epsilon,t}(X)}\mathcal{S}$, for fixed time $t$ and $X\in\mathcal{H}$, and thus, its variation is given by its covariant derivative along the curve $\varphi_{\epsilon}(X,t)$ evaluated at $\epsilon=0$, and therefore (see Appendix \ref{Geometry1})\footnote{$D_{\varphi_{\epsilon}(X,t)}$ denotes the covariant derivative along the curve $\varphi_{\epsilon}(X,t)$.}
\begin{equation}\label{normal-var}
	\delta\boldsymbol{\mathcal{N}}=\frac{d}{d\epsilon}\boldsymbol{\mathcal{N}}_{\epsilon}\Big|_{\epsilon=0}
	=\bar{\nabla}_{\frac{\partial}{\partial\epsilon}}^{\bar{\mathbf g}}
	\boldsymbol{\mathcal{N}}_{\epsilon}\Big|_{\epsilon=0}
	=D_{\varphi_{\epsilon}(X,t)}\boldsymbol{\mathcal{N}}_{\epsilon}\Big|_{\epsilon=0}
	={\bar{\nabla}}^{\bar{\mathbf g} }_{\delta\varphi}\boldsymbol{\mathcal{N}}
	=\bar{\nabla}_{\delta\varphi^\top}^{\bar{\mathbf g}}\boldsymbol{\mathcal{N}}-(\mathbf{d}\,\delta\varphi^\perp)^\sharp\,.
\end{equation}
Using \eqref{2nd-fun}, in components one has
\begin{equation}\label{norm-var}
\delta\mathcal{N}^a=-(\delta\varphi^\top)^c\,\theta^a{}_c-\left(\frac{\partial\,\delta\varphi^\perp}{\partial\,x^b}\right)g^{ab}\,.
\end{equation}
It is straightforward to see that the variation of the ambient space metric vanishes as $\bar{\mathbf g}$ is compatible with the connection, i.e.,
$
\delta\bar{\mathbf{g}}\circ\varphi=D_{\varphi_X(t)}\bar{\mathbf g}\circ\varphi=\bar{\nabla}^{\bar{\mathbf g}}_{\delta\varphi}\,\bar{\mathbf{g}}=\mathbf{0}\,,
$
where $\varphi_X(t)=\varphi(X,t)$ for fix $X$.

Using Hamilton's principle (cf. \eqref{eu-lagrange1}), one obtains the following Euler-Lagrange equations (see Appendix \ref{EU-Lag} for the details of the derivation)
\begin{subequations}\label{eus-lag}
\begin{align}\label{eu-lag-1}
	& \rho(\mathfrak{B}^\top)_a -\frac{d}{dt}\left(\frac{\partial\mathcal{L}}{\partial(\dot{\varphi}^\top)^a}\right)
	-\rho\mathfrak{L}_b\theta^b{}_a-2\left(\frac{\partial\mathcal{L}}{\partial C_{AB}}F^b{}_Bg_{ab}\right)_{|A} 
	\nonumber\\
	&-\left(\frac{\partial\mathcal{L}}{\partial\Theta_{AB}}F^b{}_A\right)_{|B}\theta_{ba}
	-\left(\frac{\partial\mathcal{L}}{\partial\Theta_{AB}}F^b{}_A\theta_{ba}\right)_{|B}=0\,, \\
	& \rho\mathfrak{B}^\perp-\frac{d}{dt}\left(\frac{\partial\mathcal{L}}{\partial\dot{\varphi}^\perp}\right)
	+\left[\rho\mathfrak{L}_b g^{ab}(F^{-1})^A{}_a\right]_{|A}
	-2\frac{\partial\mathcal{L}}{\partial C_{AB}} F^a{}_AF^b{}_B\theta_{ab} \nonumber\\
	&-\frac{\partial\mathcal{L}}{\partial\Theta_{AB}} F^a{}_AF^b{}_B\theta_{ac}\theta_{bd}\,g^{cd}
	+\left[\left(\frac{\partial\mathcal{L}}{\partial\Theta_{AB}}F^b{}_A\right)_{|B}(F^{-1})^D{}_b\right]_{|D}=0\,,
\end{align}
\end{subequations}
where $\boldsymbol{\mathfrak{B}}^\top$ and $\boldsymbol{\mathfrak{B}}^\perp$ are the tangential and normal external body forces, respectively, and $\boldsymbol{\mathfrak{L}}$ is the external body moment. Note that \eqref{am-ga-cod}$_2$ was used in deriving \eqref{eu-lag-1}. The boundary conditions read
\begin{align} \label{eu-bound}
	& \Bigg[2F^a{}_B\left(\frac{\partial\mathcal{L}}{\partial C_{AB}}g_{ac}
	+\frac{\partial\mathcal{L}}{\partial\Theta_{AB}}\theta_{ac}\right)\Bigg]\mathsf{T}_{A}=0\,, \\
	& (F^{-1})^A{}_a\Bigg[\rho\mathfrak{L}_b g^{ab}
	+\left(\frac{\partial\mathcal{L}}{\partial\Theta_{CB}}F^a{}_C\right)_{|B}\Bigg]\mathsf{T}_{A}=0\,, \\
	& \frac{\partial\mathcal{L}}{\partial\Theta_{AB}}F^a{}_B\mathsf{T}_{A}=0\,, 
\end{align}
where $\boldsymbol{\mathsf{T}}$ is the outward vector field normal to $\partial\mathcal{H}$. Note that
\begin{equation}
	\frac{d}{dt} \frac{\partial\mathcal{L}}{\partial(\dot{\varphi}^\top)^a}
	=\frac{d}{dt} \frac{\partial T}{\partial(\dot{\varphi}^\top)^a}
	=\rho \,g_{ac} (A^\top)^c\,,\quad \quad 
	\frac{d}{dt} \frac{\partial\mathcal{L}}{\partial\dot{\varphi}^\perp}
	=\frac{d}{dt} \frac{\partial T}{\partial\dot{\varphi}^\perp} =\rho A^\perp\,.
\end{equation}

\begin{remark}\label{key-prescribe}
	In order to prescribe non-vanishing boundary conditions on $\partial\mathcal{H}$, the Lagrange--d'Alembert principle should be modified. Let $\mathbf{\Upsilon}$, $\mathbf{Q}$, and $\pmb{\mathscr{M}}$ be the boundary surface traction, boundary shear force, and boundary moment, respectively. The Lagrange--d'Alembert's principle is modified to read
\begin{equation}
\begin{split}
\delta S(\varphi)&+\int_{t_0}^{t_1}\int_{\mathcal{H}}\bigg(\boldsymbol{\mathfrak{B}}\cdot\delta\varphi+\boldsymbol{\mathfrak{L}}\cdot\delta\boldsymbol{\mathcal{N}}\bigg)\rho\, dAdt\\&+\int_{t_0}^{t_1}\int_{\partial\mathcal{H}}J\bigg(\Upsilon^ag_{ab}(\delta\varphi^\top)^b+Q\delta\varphi^\perp+\mathscr{M}^a\delta\varphi^\perp_{~,A}(F^{-1})^A{}_a\bigg)dL dt=0\,.
\end{split}
\end{equation}
Therefore, in this case the boundary conditions are obtained as
\begin{align}
	& \Bigg[2F^a{}_B\left(\frac{\partial\mathcal{L}}{\partial C_{AB}}g_{ac}
	+\frac{\partial\mathcal{L}}{\partial\Theta_{AB}}\theta_{ac}\right)\Bigg]\mathsf{T}_{A}=Jg_{bc}\Upsilon^b\,, \\
	& (F^{-1})^A{}_a\Bigg[\rho\mathfrak{L}_b g^{ab}
	+\left(\frac{\partial\mathcal{L}}{\partial\Theta_{CB}}F^a{}_C\right)_{|B}\Bigg]\mathsf{T}_{A}=JQ\,, \\
	& \frac{\partial\mathcal{L}}{\partial\Theta_{AB}}F^a{}_B\mathsf{T}_{A}=J\mathscr{M}^a \,.
\end{align}
\end{remark}

Let us introduce the following tensors


\begin{equation}\label{def-str-quan}
	\mathbf{P}=2\mathbf{F}\frac{\partial W}{\partial\mathbf{C}}\,,\qquad
	\boldsymbol{\mathsf{M}}=\mathbf{F}\frac{\partial W}{\partial\mathbf{\Theta}}\,,
\end{equation}
where $\mathbf{P}$ is the first Piola-Kirchhoff stress tensor, and $\boldsymbol{\mathsf{M}}$ is the couple-stress tensor. Therefore, based on the symmetries of the independent objective measures of strain, i.e., the right Cauchy-Green tensor and the extrinsic deformation tensor (or, equivalently, the symmetries of the first and the second fundamental forms of the deformed shell, i.e., $\mathbf g$ and $\boldsymbol{\theta }$), one has the following symmetries 
\begin{equation}\label{ang-us}
	P^{[aA}F^{b]}{}_A=0\,,\qquad \mathsf{M}^{[aA}F^{b]}{}_A=0\,.
\end{equation}



\begin{remark}
We can regard $\mathbf{C}$ as a function of $\mathbf F$ and the spatial metric $\mathbf g$, and similarly, $\mathbf\Theta$ can be regarded as a function of $\mathbf F$ and $\boldsymbol\theta$. Therefore, we set
\begin{equation}\label{equ-ene}
	\hat{W}\left(X,\mathbf{F},\mathbf{g},\boldsymbol{\theta},\mathbf{G},\mathbf{B}\right)=W\left(X,\mathbf{C},\mathbf{\Theta},\mathbf{G},\mathbf{B}\right).
\end{equation}
Note that
\begin{equation}
\begin{split}
	\frac{\partial\hat{W}}{\partial g_{ab}}&=\frac{\partial W}{\partial C_{AB}}\frac{\partial C_{AB}}{\partial g_{ab}}=\frac{\partial W}{\partial C_{AB}}F^a{}_AF^b{}_B\,,\\
	\frac{\partial \hat{W}}{\partial\theta_{ab}}&=\frac{\partial W}{\partial\Theta_{AB}}\frac{\partial\Theta_{AB}}{\partial\theta_{ab}}=\frac{\partial W}{\partial\Theta_{AB}}F^a{}_AF^b{}_B\,.
\end{split}
\end{equation}
The Cauchy stress tensor and the spatial couple-stress tensor are accordingly defined as\footnote{Note that \eqref{ang-us} is equivalent to $\sigma^{ab}$ and $\mathcal{M}^{ab}$ being symmetric.}
\begin{equation}
\begin{split}
	\sigma^{ab}&=\frac{2}{J}\frac{\partial W}{\partial C_{AB}}F^a{}_AF^b{}_B
	=\frac{2}{J}\frac{\partial\hat{W}}{\partial g_{ab}}\,.\\
	\mathcal{M}^{ab}&=\frac{1}{J}\frac{\partial W}{\partial\Theta_{AB}}F^a{}_AF^b{}_B
	=\frac{1}{J}\frac{\partial\hat{W}}{\partial\theta_{ab}}\,.
	\end{split}
	\end{equation}
Note that $P^{aA}=J (F^{-1})^A{}_b\sigma^{ab}$ and $\mathsf{M}^{aA}=J(F^{-1})^A{}_b\mathcal{M}^{ab}$.
Note that from \eqref{equ-ene}
\begin{equation}
\frac{\partial\hat{W}}{\partial F^a{}_C}=\frac{\partial W}{\partial C_{AB}}\frac{\partial C_{AB}}{\partial F^a{}_C}+\frac{\partial W}{\partial\Theta_{AB}}\frac{\partial\Theta_{AB}}{\partial F^a{}_C}.
\end{equation} 
It is straightforward to show that
\begin{equation}
\begin{split}
\frac{\partial C_{AB}}{\partial F^a{}_C}=g_{ab}\delta^C{}_AF^b{}_B+g_{ac}F^c{}_A\delta^C{}_B,\\
\frac{\partial \Theta_{AB}}{\partial F^a{}_C}={\theta}_{ab}\delta^C{}_AF^b{}_B+{\theta}_{ac}F^c{}_A\delta^C{}_B.
\end{split}
\end{equation}
Thus
\begin{equation}
\frac{\partial\hat{W}}{\partial F^c{}_A}=P^{aA}g_{ac}+2\mathsf{M}^{aA}\theta_{ac}\,.
\end{equation}
Note also that
\begin{equation}\label{key-ss}
	F^c{}_A\frac{\partial\hat{W}}{\partial F^a{}_A}=2\frac{\partial\hat{W}}{\partial g_{bc}}g_{ab}
	+2\frac{\partial\hat{W}}{\partial\theta_{bc}}\theta_{ab}\,.
\end{equation}


\end{remark}

In terms of these tensors, the Euler-Lagrange equations \eqref{eus-lag} are rewritten as
\begin{align}\label{lin-mom-non}
	& \left(P^{aA}+\theta^a{}_b\mathsf{M}^{bA}\right)_{|A}+\mathsf{M}^{bA}{}_{|A}\theta^a{}_b
	+\rho (\mathfrak{B}^\top)^a-\rho\,\theta^a{}_b\mathfrak{L}^b=\rho (A^\top)^a\,, \\
	& \left(P^{aA}+\theta^a{}_b\mathsf{M}^{bA}\right)F^c{}_A\theta_{ac}-\left[\mathsf{M}^{aA}{}_{|A}(F^{-1})^B{}_a\right]_{|B}+\rho \mathfrak{B}^\perp+\left[\rho\mathfrak{L}^a(F^{-1})^A{}_a\right]_{|A}=\rho A^\perp\,, \label{lin-mom-non1}
\end{align}
and the boundary conditions \eqref{eu-bound} are expressed as
\begin{align}
	& \left[g_{ab}P^{aA}+2\theta_{ab}\mathsf{M}^{aA}\right]\mathsf{T}_A=0\,, \\
	& (F^{-1})^A{}_a\bigg[\rho\mathfrak{L}^a-\mathsf{M}^{aB}{}_{|B}\bigg]\mathsf{T}_A=0\,, \\
	& \mathsf{M}^{aA}\mathsf{T}_A=0\,.
\end{align}
The tangential and normal (shear) tractions are given by 
\begin{align}
	(T^\top)^a &= \left[P^{aA}+2\theta_{bc}\mathsf{M}^{bA}g^{ac}\right]\mathsf{T}_A\,,\\ 
	T^\perp &=-(F^{-1})^A{}_a\bigg[\rho\mathfrak{L}^a-\mathsf{M}^{aB}{}_{|B}\bigg]\mathsf{T}_{A}\,.
\end{align}

\begin{remark}[Spatial covariance of the energy density and Noether's theorem]
Let us define the following surface Lagrangian density
\begin{equation}
	\mathcal{L}=\hat{\mathcal{L}}\left(X,\varphi,\dot{\varphi},\mathbf{F},\mathbf{g},\boldsymbol{\theta},
	\mathbf{G},\mathbf{B}\right).
\end{equation}
Let us consider a flow $\psi_s$ generated by a vector field $\mathbf{w}$, i.e.,
\begin{equation}
	\frac{d}{ds}\bigg|_{s=0}\psi_s\circ\varphi=\mathbf{w}\circ\varphi.
\end{equation}
Note that $\psi_s$ is a local diffeomorphism. We assume that $\mathbf{w}$ is tangential, i.e., $\mathbf{w}\in T_{\varphi(X)}\varphi(\mathcal{H})$. 
Let us assume that $\mathcal{L}$ is tangentially covariant, i.e., it is invariant for local diffeomorphisms $\psi_s$ generated by arbitrary $\mathbf{w}\in T_{\varphi(X)}\varphi(\mathcal{H})$. Thus
\begin{equation}
	\hat{\mathcal{L}}\left(X,\psi_s\circ\varphi,\psi_{s*}\dot{\varphi},\psi_{s*}\mathbf{F},\psi_{s*}\mathbf{g},
	\psi_{s*}\boldsymbol{\theta},\mathbf{G},\mathbf{B}\right)
	=\hat{\mathcal{L}}\left(X,\varphi,\dot{\varphi},\mathbf{F},\mathbf{g},\boldsymbol{\theta},
	\mathbf{G},\mathbf{B}\right).
\end{equation}
Taking derivative with respect of $s$ from both sides and evaluating at $s=0$ one obtains\footnote{Note that in the local coordinate chart $\{x^1,x^2,x^3\}$, for which $x^3$ is the outward normal direction
\begin{equation*}	
	 \theta_{ab}=-\frac{1}{2}\frac{\partial\bar{g}_{ab}}{\partial x^3}\Big|_{\varphi(\mathcal H)}\,,\quad a,b=1,2\,.
\end{equation*}	 
Thus
\begin{equation*}
(\psi_{s*}\theta)_{ a' b'}=-\frac{1}{2}\frac{\partial\left((T\psi_s^{-1})^a{}_{ a'}(T\psi_s^{-1})^b{}_{ b'}\bar{g}_{ab}\right)}{\partial x^3}=(T\psi_s^{-1})^a{}_{ a'}(T\psi_s^{-1})^b{}_{ b'}\left(-\frac{1}{2}\frac{\partial\bar{g}_{ab}}{\partial x^3}\right)=(T\psi_s^{-1})^a{}_{ a'}(T\psi_s^{-1})^b{}_{ b'}\theta_{ab}\,,\quad a',b'=1,2\,.
\end{equation*}
Note also that $(T\psi^{-1}_{s})^{a}{}_{a'}=-\delta^b{}_{a'}\delta^a{}_{b'}(T\psi_{s})^{b'}{}_b$.
}
\begin{equation}
	\frac{\partial \hat{\mathcal{L}}}{\partial {\varphi}^a}w^a+\frac{\partial \hat{\mathcal{L}}}
	{\partial \dot{\varphi}^a}w^a{}_{|b}\dot{\varphi}^b
	+F^b{}_A\frac{\partial\hat{\mathcal{L}}}{\partial F^a{}_A}w^a{}_{|b}
	-2\frac{\partial\hat{\mathcal{L}}}{\partial g_{cb}}g_{ca} w^a{}_{|b}
	-2\frac{\partial\hat{\mathcal{L}}}{\partial \theta_{cb}}\theta_{ca}w^a{}_{|b}=0\,.
\end{equation}
Note that
\begin{equation}
	\frac{\partial \hat{\mathcal{L}}}{\partial \dot{\varphi}^a}\dot{\varphi}^b
	-2\frac{\partial\hat{\mathcal{L}}}{\partial g_{cb}}g_{ca}=2\frac{\partial\hat{W}}{\partial g_{cb}}g_{ca}\,.
\end{equation}
Therefore
\begin{equation}
	\frac{\partial \hat{\mathcal{L}}}{\partial {\varphi}^a}w^a
	+\left[-F^b{}_A\frac{\partial\hat{W}}{\partial F^a{}_A}
	+2\frac{\partial\hat{W}}{\partial g_{cb}}g_{ca}
	+2\frac{\partial\hat{W}}{\partial \theta_{cb}}\theta_{ca}\right]w^a{}_{|b}=0\,.
\end{equation}
Knowing that $\mathbf{w}$ is arbitrary one concludes that 
\begin{equation} \label{tangential-Noether}
	\frac{\partial \hat{\mathcal{L}}}{\partial {\varphi}^a}=0,~~~
	F^b{}_A\frac{\partial\hat{W}}{\partial F^a{}_A}=
	2\frac{\partial\hat{W}}{\partial g_{bc}}g_{ac}
	+2\frac{\partial\hat{W}}{\partial \theta_{bc}}\theta_{ac}\,.
\end{equation}
Note that \eqref{tangential-Noether}$_2$ is identical to \eqref{key-ss}, which can be written as
\begin{equation} 
	F^b{}_A\frac{\partial\hat{W}}{\partial F^a{}_A}=
	J\sigma^{bc}g_{ac}
	+2J\mathcal{M}^{bc}\theta_{ac}\,.
\end{equation}
\end{remark}

Noting that $\rho=J{\varrho}$, where $\varrho$ is the spatial mass density, conservation of mass for shells is given by
\begin{equation}\label{key-mass-1}
	\dot{\varrho}+\varrho\frac{\dot{J}}{J}=0\,.
\end{equation}
Using the identity $\frac{d}{dt}\left[\det\mathbf{K}(t)\right]=\det\mathbf{K}(t)\mathrm{tr}\left[\mathbf{K}^{-1}(t)\frac{d}{dt}\mathbf{K}(t)\right]$, one has
\begin{equation}\label{key-mass-2}
	\frac{\dot{J}}{J}=\frac{1}{2}\mathrm{tr}_{{\mathbf C}^\flat}\left(\frac{d}{dt}\varphi^*\mathbf{g}\right)\,,
\end{equation}
where the trace is calculated using the metric $\mathbf{C}^\flat$. From \eqref{cauchy-var}, one obtains
\begin{equation}
\varphi_*\frac{d\varphi^*\mathbf{g}}{dt}=\mathbf{L}_{\mathbf{v}^\top}\mathbf{g}-2v^\perp\boldsymbol{\theta}\,,
\end{equation}
and thus
\begin{equation}\label{key-mass-3}
	\mathrm{tr}_{{\mathbf C}^\flat}\left(\frac{d}{dt}\varphi^*\mathbf{g}\right)
	=2\,\mathrm{div}\,\mathbf{v}^\top-2v^\perp\mathrm{tr}\boldsymbol{\theta}\,.
\end{equation}
Therefore, using \eqref{key-mass-1}, \eqref{key-mass-2}, and \eqref{key-mass-3}, one finds the spatial local form of the conservation of mass as\footnote{Note that $\dot{\varrho}=\frac{\partial\varrho}{\partial t}+\nabla\varrho\cdot\mathbf{v}$.}
\begin{equation}
\dot{\varrho}+\varrho\,\mathrm{div}\,\mathbf{v}^\top-\varrho v^\perp\mathrm{tr}\boldsymbol{\theta}=0\,.
\end{equation}

\subsection{The linearized governing equations}

Next, we linearize the balance of linear momentum and the symmetries \eqref{ang-us} about a motion $\mathring{\varphi}$. We assume that the reference motion is an isometric embedding of an initially stress-free body into the Euclidean space, and thus, $\mathring{F}^{a}{}_A=\delta^a_A$, $\mathring{P}^{aA}=0$, and $\mathring{\mathsf{M}}^{aA}=0$. 
The tangential and normal displacement fields are defined in terms of the variation of the deformation map as  
\begin{equation}
\mathbf{U}^\top(X,t)=\delta\varphi^\top_t,\qquad \mathbf{U}^\perp(X,t)=\delta\varphi^\perp_t.
\end{equation}
Note that the deformation gradient is linearized as 
\begin{equation}\label{key-ff}
\delta F^a{}_A=(\delta\varphi^\top)^a{}_{|A}-\mathring{\theta}^a{}_b\mathring{F}^b{}_A\delta\varphi^\perp=(U^\top)^a{}_{|A}-\mathring{\theta}^a{}_b\mathring{F}^b{}_AU^\perp=\mathring{F}^b{}_A(U^\top)^a{}_{|b}-\mathring{\theta}^a{}_b\mathring{F}^b{}_AU^\perp\,.
\end{equation}
Using \eqref{cau-gr-var}, we have
\begin{equation}\label{C-label}
\delta C_{AB}=\mathring{F}^a{}_A\,{U}^{\top}_{a|B}+\mathring{F}^b{}_B\, U^{\top}_{b|A}-2\, U^\perp\,\mathring{\Theta}_{AB}\,,
\end{equation} 
where $\mathring{\Theta}_{AB}=\mathring{F}^a{}_A\mathring{F}^b{}_B\mathring{\theta}_{ab}=B_{AB}$. From \eqref{theta-var}, one writes
\begin{equation}\label{theta-label}
\begin{split}
\delta\Theta_{AB}=&\mathring{F}^a{}_A\mathring{F}^b{}_B\,\mathring{\theta}_{ab|c}\,( U^\top)^c+\mathring{F}^a{}_A\mathring{\theta}_{ac}( U^\top)^c{}_{|B}+\mathring{F}^b{}_B\mathring{\theta}_{bc}( U^\top)^c{}_{|A}\\&- U^\perp \mathring{F}^a{}_A\mathring{F}^b{}_B\mathring{\theta}_{ac}\mathring{\theta}_{bd}\,g^{cd}+\mathring{F}^b{}_A\left(\frac{\partial\, U^\perp}{\partial\, x^b}\right)_{|B}\,.
\end{split}
\end{equation}
Using \eqref{def-str-quan}, one obtains
\begin{equation}\label{one-1}
\begin{split}
\delta P^{aA}=2 \mathring{F}^a{}_B&\bigg[\frac{\partial^2 W}{\partial C_{AB}\partial C_{CD}}\delta C_{CD}+\frac{\partial^2 W}{\partial C_{AB}\partial\Theta_{CD}}\delta\Theta_{CD}\bigg]\,,
\end{split}
\end{equation}
\begin{equation}\label{four-4}
\begin{split}
\delta\mathsf{M}^{aA}=\mathring{F}^a{}_B&\bigg[\frac{\partial^2W}{\partial\Theta_{AB}\partial\Theta_{CD}}\delta\Theta_{CD}+\frac{\partial^2W}{\partial\Theta_{AB}\partial C_{CD}}\delta C_{CD}\bigg].\,\,
\end{split}
\end{equation}
Substituting \eqref{C-label} and \eqref{theta-label} into \eqref{one-1} and \eqref{four-4} one obtains 
\begin{align}
	 & \delta P^{aA}=\mathbb{A}^{aAbB}
	\left({U}^{\top}_{b|B}- U^\perp\,\mathring{\gamma}_{bB}\right) \label{delta-p} \\ 
	&~~~~~~~~~~+\mathbb{B}^{aAbB}\bigg[\mathring{\theta}_{bc}( U^\top)^c{}_{|B}
	+\frac{1}{2}(\, U^\perp_{~,b})_{|B} -\frac{1}{2}U^\perp\mathring{\theta}_{bc}\mathring{\gamma}_{dB}g^{cd}
	+\frac{1}{2}\mathring{F}^d{}_B\mathring{\theta}_{bd|c}( U^\top)^c\bigg]\,,   \nonumber \\
	& \delta\mathsf{M}^{aA}=\mathbb{C}^{aAbB}\bigg[\mathring{\theta}_{bc}( U^\top)^c{}_{|B}
	+\frac{1}{2}(\, U^\perp_{~,b})_{|B}-\frac{1}{2}U^\perp\mathring{\theta}_{bc}\mathring{\gamma}_{dB}g^{cd}
	+\frac{1}{2}\mathring{F}^d{}_B\mathring{\theta}_{bd|c}( U^\top)^c\bigg]  \label{delta-h} \\ 
	 &~~~~~~~~~~
	+\frac{1}{2}\mathbb{B}^{bBaA}\left({U}^{\top}_{b|B}- U^\perp\,\mathring{\gamma}_{bB}\right)\,,   \nonumber
\end{align}
where $\boldsymbol\gamma$ is a two-point tensor that in components is defined as $\gamma_{aB}=F^b{}_B\theta_{ab}$, and the shell elastic constants are defined as
\begin{equation}
\begin{split}
\mathbb{A}^{aAbB}=4\mathring{F}^a{}_M\mathring{F}^b{}_N\frac{\partial^2 W}{\partial C_{AM}\partial C_{BN}}\,,~~\mathbb{B}^{aAbB}=4\mathring{F}^a{}_M\mathring{F}^b{}_N\frac{\partial^2 W}{\partial C_{AM}\partial \Theta_{BN}}\,,\\ \mathbb{C}^{aAbB}=2\mathring{F}^a{}_M\mathring{F}^b{}_N\frac{\partial^2 W}{\partial \Theta_{AM}\partial \Theta_{BN}}\,.\qquad\qquad\qquad\qquad\qquad\qquad\qquad\quad~~
\end{split}
\end{equation}
Therefore, the linearized governing equations of the shell are expressed in terms of three elasticity tensors. Note that the elastic constants satisfy the following symmetries:  $\mathbb{A}^{aAbB}=\mathbb{A}^{bBaA}$ and $\mathbb{C}^{aAbB}=\mathbb{C}^{bBaA}$.   
The linearized normal and parallel components of the material acceleration are obtained using \eqref{acc} as
$
\delta\mathbf{A}^\top=\ddot{\mathbf{U}}^\top,$ and $ \delta\mathbf{A}^\perp=\ddot{\mathbf{U}}^\perp.
$
Ignoring the body forces and body moments, from \eqref{lin-mom-non} the linearized balance of linear momentum is given by
\begin{subequations}\label{lin-mom}
	\begin{align}
	\label{lin-1} &
	\left(\delta P^{aA}+\mathring{\theta}^a{}_b\,\delta\mathsf{M}^{bA}\right)_{|A}
	+\delta\mathsf{M}^{bA}{}_{|A}\mathring{\theta}^a{}_b=\rho (\delta A^\top)^a, \\
	&\label{lin-2}	
	\left(\delta P^{aA}+\mathring{\theta}^a{}_b\,\delta\mathsf{M}^{bA}\right)\mathring{\gamma}_{aA}
	-\left(\delta\mathsf{M}^{aA}{}_{|A}(\mathring{F}^{-1})^B{}_a\right)_{|B}=\rho \,\delta A^\perp\,.
	\end{align}
\end{subequations}
Similarly, the symmetries \eqref{ang-us} are linearized to read
\begin{equation}\label{angg-1}
\delta P^{[aA}\mathring{F}^{b]}{}_A=0\,,\qquad \delta\mathsf{M}^{[aA}\mathring{F}^{b]}{}_A=0\,.
\end{equation}
Note that from (3.65) and \eqref{angg-1}$_1$, and recalling that the normal and parallel components of the displacement field and their gradients are independent, one concludes that $\mathbb{B}^{[aAbB}\mathring{F}^{e]}{}_A(U^\perp{}_{,b})_{|B}=0$, and thus, $\mathbb{B}^{[aAbB}\mathring{F}^{e]}{}_A=0$. 
Again using (3.65) and \eqref{angg-1}$_1$, and looking at the terms with $U^\top_{b|B}$, one obtains $\mathbb{A}^{[aAbB}\mathring{F}^{e]}{}_AU^\top_{b|B}+\mathbb{B}^{[aAcB}\mathring{F}^{e]}{}_A\mathring{\theta}^b{}_c(U^\top)_{b|B}=0$. Recalling that $\mathbb{B}^{[aAbB}\mathring{F}^{e]}{}_A=0$, one concludes that $\mathbb{A}^{[aAbB}\mathring{F}^{e]}{}_A=0$. Following the same procedure starting from (3.66) and \eqref{angg-1}$_2$, one obtains $\mathbb{B}^{cB[aA} \mathring{F}^{e]}{}_A=0$, and $\mathbb{C}^{[aAcB}\mathring{F}^{e]}{}_A=0$. Therefore, the elastic constants have the following symmetries
\begin{subequations}\label{key-ang1}
\begin{equation}
\mathbb{A}^{[aAcB}\mathring{F}^{b]}{}_A=0\,,\qquad\mathbb{B}^{cB[aA}\mathring{F}^{b]}{}_A=0\,,
\end{equation}
\begin{equation}
\mathbb{B}^{[aAcB}\mathring{F}^{b]}{}_A=0\,,\qquad\mathbb{C}^{[aAcB}\mathring{F}^{b]}{}_A=0\,.
\end{equation}
\end{subequations}
\subsection{The linearized governing equations of pre-stressed shells}\label{prestress-shell}

In this section, we derive the linearized governing equations of a pre-stressed elastic shell. Assume that the shell is initially stressed\footnote{Note that we do not explicitly specify the source of the initial stress or couple-stress. If the initial stress and couple-stress are due to elastic deformations, and the body has an energy function $W$ with respect to its stress-free configuration, then one may express $\mathring{\mathbf P}$ and $\mathring{\boldsymbol{\mathsf{M}}}$ as
\begin{equation*}
\mathring{\mathbf{P}}=2\mathring{\mathbf{F}}\frac{\partial W}{\partial\mathbf{C}}\Big|_{\mathring{\mathbf{F}}}\,,\qquad\mathring{\boldsymbol{\mathsf{M}}}=\mathring{\mathbf{F}}\frac{\partial W}{\partial\mathbf{\Theta}}\Big|_{\mathring{\mathbf{F}}}\,.
\end{equation*}
} such that the initial stress and couple-stress are, respectively, given by $\mathring{\mathbf P}$ and $\mathring{\boldsymbol{\mathsf{M}}}$. Let the initial (normal and parallel) body forces and body moments be given by $\mathring{\boldsymbol{\mathfrak B}}^\top$, $\mathring{\boldsymbol{\mathfrak B}}^\perp$, and $\mathring{\boldsymbol{\mathfrak L}}$, respectively. The shell must be in equilibrium in its initial configuration, i.e., the balance of linear and angular momenta must be satisfied, which read 
\begin{subequations}\label{shell-initial-linear}
\begin{align}
	& \left(\mathring{P}^{aA}+\mathring{\theta}^a{}_b\mathring{\mathsf{M}}^{bA}\right)_{|A}
	+\mathring{\mathsf{M}}^{bA}{}_{|A}\mathring{\theta}^a{}_b+\rho (\mathring{\mathfrak{B}}^\top)^a
	-\rho\,\mathring{\theta}^a{}_b \mathring{\mathfrak{L}}^b=0\,, \\
	& \left(\mathring{P}^{aA}+\mathring{\theta}^a{}_b\mathring{\mathsf{M}}^{bA}\right)
	\mathring{F}^c{}_A\mathring{\theta}_{ac}-\left(\mathring{\mathsf{M}}^{aA}{}_{|A}
	(\mathring{F}^{-1})^B{}_a\right)_{|B}+\rho \mathring{\mathfrak{B}}^\perp
	+\left(\rho\mathring{\mathfrak {L}}^a(\mathring{F}^{-1})^A{}_a\right)_{|A}=0\,,
\end{align}
\end{subequations}
and  
	\begin{equation}\label{shell-initial-ang}
	\mathring{P}^{[aA}\mathring{F}^{b]}{}_A=0\,,\qquad\mathring{\mathsf{M}}^{[aA}\mathring{F}^{b]}{}_A=0\,.
	\end{equation}
We next linearize the governing equations about the motion $\mathring{\varphi}$. The linearized balance of linear momentum reads (cf. \eqref{lin-mom-non} and \eqref{lin-mom-non1})
\begin{subequations}\label{pre-shell}
\begin{align}
	&\left(\delta P^{aA}+\delta\theta^a{}_b\mathring{\mathsf{M}}^{bA}+\mathring{\theta}^a{}_b
	\delta\mathsf{M}^{bA}\right)_{|A}+\delta\mathsf{M}^{bA}{}_{|A}
	\mathring{\theta}^a{}_b+\mathring{\mathsf{M}}^{bA}{}_{|A}\delta\theta^a{}_b  \nonumber \\
	& ~~~+\rho (\delta\mathfrak{B}^\top)^a-\rho\,\delta\theta^a{}_b\mathring{\mathfrak{L}}^b
	-\rho\,\mathring{\theta}^a{}_b\delta\mathfrak{L}^b=\rho (\ddot{U}^\top)^a\,, \\
	&\left(\mathring{P}^{aA}+\mathring{\theta}^a{}_b\mathring{\mathsf{M}}^{bA}\right)
	\left(\delta F^c{}_A\mathring{\theta}_{ac}+\mathring{F}^c{}_A\delta\theta_{ac}\right)
	+\left(\delta P^{aA}+\delta\theta^a{}_b\mathring{\mathsf{M}}^{bA}
	+\mathring{\theta}^a{}_b\delta\mathsf{M}^{bA}\right)\mathring{F}^c{}_A\mathring{\theta}_{ac} \nonumber \\
	&~~~-\left(\delta\mathsf{M}^{aA}{}_{|A}(\mathring{F}^{-1})^B{}_a\right)_{|B}
	-\left(\mathring{\mathsf{M}}^{aA}{}_{|A}(\delta F^{-1})^B{}_a\right)_{|B}
	+\rho \,\delta\mathfrak{B}^\perp+\left(\rho\,\delta\mathfrak{L}^a(\mathring{F}^{-1})^A{}_a\right)_{|A} \nonumber \\
	&~~~+\left(\rho\,\mathring{\mathfrak{L}}^a(\delta F^{-1})^A{}_a\right)_{|A}=\rho\, \ddot{U}^\perp\,.
\end{align}
\end{subequations}
The symmetries \eqref{ang-us} are linearized to read
\begin{equation}\label{angshell}
\delta P^{[aA}\mathring{F}^{b]}{}_A+\mathring{P}^{[aA}\delta F^{b]}{}_A=0\,,\qquad \delta\mathsf{M}^{[aA}\mathring{F}^{b]}{}_A+\mathring{\mathsf{M}}^{[aA}\delta F^{b]}{}_A=0\,.
\end{equation}
Note that $ (\delta F^{-1})^A{}_a=-(\mathring{F}^{-1})^B{}_a(\mathring{F}^{-1})^A{}_b\,\delta F^b{}_B=-(\mathring{F}^{-1})^B{}_a(\mathring{F}^{-1})^A{}_b\, \left[(U^\top)^b{}_{|B}-\mathring{\theta}^b{}_c\mathring{F}^c{}_BU^\perp\right]$, and using the relation $\Theta_{AB}=F^a{}_AF^b{}_B\theta_{ab}$, along with \eqref{theta-label}, one obtains
\begin{equation}\label{delta-teta-pla}
\delta\theta_{ab}
=\mathring{\theta}_{ab|c}\,( U^\top)^c + U^\perp \mathring{\theta}_{ac}\mathring{\theta}_{bd}\,g^{cd}+(\mathring{F}^{-1})^B{}_b\left(\frac{\partial\, U^\perp}{\partial\, x^a}\right)_{|B}\,.
\end{equation}
Knowing that $\boldsymbol{\mathfrak B}^\perp=\bar{\mathbf g}(\boldsymbol{\mathfrak B},\boldsymbol{\mathcal{N}})\boldsymbol{\mathcal{N}}$, the normal component of the body force is linearized as (see also \citep{yavari2008covariance})
\begin{equation}\label{body-var}
\delta\boldsymbol{\mathfrak B}^\perp={\bar{\mathbf g}}(\delta \boldsymbol{\mathfrak B},\boldsymbol{\mathcal{N}})\boldsymbol{\mathcal{N}}+{\bar{\mathbf g}}( \mathring{\boldsymbol{\mathfrak B}},\delta\boldsymbol{\mathcal{N}})\boldsymbol{\mathcal{N}}+ {\bar{\mathbf g}}(\mathring{\boldsymbol{\mathfrak B}},\boldsymbol{\mathcal{N}})\delta\boldsymbol{\mathcal{N}}.
\end{equation}
Assuming that the body force vector $\boldsymbol{\mathfrak B}$ is fixed (dead load), i.e., $\delta \boldsymbol{\mathfrak B}=\mathbf{0}$, we use \eqref{app-deltan} to obtain\footnote{Note that the variation of the normal vector $\delta\boldsymbol{\mathcal N}$ is purely tangential, and hence, so is the term ${\bar{\mathbf g}}(\mathring{\boldsymbol{\mathfrak B}},\boldsymbol{\mathcal{N}})\delta\boldsymbol{\mathcal{N}}$ in \eqref{body-var}. In \eqref{var-body-main}, with an abuse of notation we only consider the term in the normal direction.}
\begin{equation}\label{var-body-main}
\delta\boldsymbol{\mathfrak B}^\perp={\bar{\mathbf g}}( \mathring{\boldsymbol{\mathfrak B}},\delta\boldsymbol{\mathcal{N}})\boldsymbol{\mathcal{N}}=-\mathbf{g}\cdot \mathring{\boldsymbol{\mathfrak B}}^\top\cdot\mathring{\boldsymbol{\theta}}\cdot\mathbf{U}^\top-\mathbf{d}\,\mathbf{U}^\perp\cdot \mathring{\boldsymbol{\mathfrak B}}^\top\,.
\end{equation}
In components
\begin{equation}\label{del-b-perp}
\delta{\mathfrak B}^\perp=-g_{ab}(\mathring{{\mathfrak B}}^\top)^b( U^\top)^c\,\mathring{\theta}^a{}_c-(\mathring{{\mathfrak B}}^\top)^b \frac{\partial\, U^\perp}{\partial\,x^b}\,.
\end{equation}
One obtains the linearized tangent component of the body force as (see also \citep[P.457]{Naghdi1973})  
\begin{equation}
\delta\boldsymbol{\mathfrak B}^\top=-{\bar{\mathbf g}}(\mathring{\boldsymbol{\mathfrak B}},\boldsymbol{\mathcal{N}})\delta\boldsymbol{\mathcal{N}}+\mathring{\boldsymbol{\mathfrak B}}^\top\cdot\left(\nabla^{\mathbf g}\mathbf{U}^\top-U^\perp\mathring{\boldsymbol{\theta}}\right)\cdot\mathbf{g}\,,
\end{equation}
which in components reads
\begin{equation}\label{del-b-top}
	(\delta{\mathfrak B}^\top)^a=\mathring{{\mathfrak B}}^\perp\left[( U^\top)^c\,\mathring{\theta}^a{}_c
	+ \frac{\partial\, U^\perp}{\partial\,x^b} g^{ab}\right]
	+(\mathring{{\mathfrak B}}^\top)^c\left[( U^\top)_{c|b}-\mathring{\theta}_{cb} U^\perp\right]g^{ab}\,.
\end{equation}
As $\mathring{\boldsymbol{\mathfrak L}}$ is purely tangential, i.e., $\mathring{\boldsymbol{\mathfrak L}}^\perp=\mathbf{0}$, the variation of the body moment is given in components by   
\begin{equation}\label{key-ll}
\delta{\mathfrak L}^a=\mathring{{\mathfrak L}}^c\left[( U^\top)_{c|b}-\mathring{\theta}_{cb} U^\perp\right]g^{ab}\,.
\end{equation}

\paragraph{The governing equations for pre-stressed plates.} Let us reduce the governing equations of a pre-stressed shell to that of a pre-stressed plate by setting $\mathring{\theta}_{ab}=0$. Thus, using \eqref{delta-teta-pla}, \eqref{del-b-perp},  \eqref{del-b-top}, and \eqref{key-ll}, Eqs. \eqref{pre-shell} and \eqref{angshell} are simplified to read
\begin{subequations}\label{lin-plate-cosserat}
\begin{align}
	& \left[\delta P^{aA}+g^{ac}(\mathring{F}^{-1})^B{}_b\left(\frac{\partial\, U^\perp}{\partial\, x^c}\right)_{|B}
	\mathring{\mathsf{M}}^{bA}\right]_{|A}+g^{ac}\mathring{\mathsf{M}}^{bA}{}_{|A}
	(\mathring{F}^{-1})^B{}_b\left(\frac{\partial\, U^\perp}{\partial\, x^c}\right)_{|B} \nonumber\\
	&~~~+\rho\mathring{{\mathfrak B}}^\perp g^{ab}\frac{\partial\, U^\perp}{\partial\,x^b}
	+\rho g^{ab}(\mathring{{\mathfrak B}}^\top)^c( U^\top)_{c|b}-\rho\,g^{ac}
	(\mathring{F}^{-1})^B{}_b\left(\frac{\partial\, U^\perp}{\partial\, x^c}\right)_{|B}
	\mathring{\mathfrak{L}}^b=\rho (\ddot{U}^\top)^a\,, \\
	& \mathring{P}^{aA}\left(\frac{\partial\, U^\perp}{\partial\, x^a}\right)_{|A}
	-\left[(\mathring{F}^{-1})^B{}_a\delta\mathsf{M}^{aA}{}_{|A}\right]_{|B}
	+\left[\mathring{\mathsf{M}}^{aA}{}_{|A}\, (\mathring{F}^{-1})^C{}_a
	(\mathring{F}^{-1})^B{}_b(U^\top)^b{}_{|C}\right]_{|B} \nonumber\\
	&~~~-\rho(\mathring{{\mathfrak B}}^\top)^b\frac{\partial\, U^\perp}{\partial\,x^b}
	+\left(\rho\,\mathring{\mathfrak{L}}^c(U^\top)_{c|b}g^{ab}(\mathring{F}^{-1})^A{}_a\right)_{|A}
	-\left[\rho\,\mathring{\mathfrak{L}}^a\, (\mathring{F}^{-1})^B{}_a
	(\mathring{F}^{-1})^A{}_b(U^\top)^b{}_{|B}\right]_{|A}=\rho\, \ddot{U}^\perp\,,
\end{align}
\end{subequations}
and
\begin{equation}\label{plate-ang-sim}
\delta P^{[aA}\mathring{F}^{b]}{}_A+\mathring{P}^{[aA} (U^\top)^{b]}{}_{|A}=0\,,\qquad\delta\mathsf{M}^{[aA}\mathring{F}^{b]}{}_A+\mathring{\mathsf{M}}^{[aA} (U^\top)^{b]}{}_{|A}=0\,.
\end{equation}
From \eqref{def-str-quan}, \eqref{delta-p}, and \eqref{delta-h}, one has
\begin{subequations}\label{sim-delta}
\begin{equation}
\delta P^{aA}=\left[\mathbb{A}^{aAbB}+\mathring{P}^{cA}(\mathring{F}^{-1})^B{}_cg^{ab}\right]{U}^{\top}_{b|B}+\frac{1}{2}\mathbb{B}^{aAbB}( U^\perp_{~,b})_{|B}\,,\quad\quad~
\end{equation}
\begin{equation}
2\,\delta\mathsf{M}^{aA}=\mathbb{C}^{aAbB}( U^\perp_{~,b})_{|B}+\left[\mathbb{B}^{bBaA}+2\mathring{\mathsf{M}}^{cA}(\mathring{F}^{-1})^B{}_cg^{ab}\right]{U}^{\top}_{b|B}\,.\qquad\qquad
\end{equation}
\end{subequations}
Recalling that the normal and parallel displacement gradients are independent from \eqref{plate-ang-sim} one obtains
\begin{equation} \label{key-bal}
\begin{aligned}
	& \left(\mathbb{A}^{[aAcB}+\mathring{P}^{dA}(\mathring{F}^{-1})^B{}_dg^{[ac}\right)\mathring{F}^{b]}{}_A
	+\mathring{P}^{[aB}g^{cb]}=0\,,\\
	& \frac{1}{2}\left(\mathbb{B}^{cB[aA}+2\mathring{\mathsf{M}}^{dA}(\mathring{F}^{-1})^B{}_dg^{[ac}\right)
	\mathring{F}^{b]}{}_A+\mathring{\mathsf M}^{[aB}g^{cb]}=0\,,\\
	& \mathbb{B}^{[aAcB}\mathring{F}^{b]}{}_A=0\,,\qquad\mathbb{C}^{[aAcB}\mathring{F}^{b]}{}_A=0\,.
\end{aligned}
\end{equation}
Note that if one uses the stress and couple-stress symmetries for the finitely-deformed shell \eqref{shell-initial-ang}, \eqref{key-bal} will be simplified to \eqref{key-ang1}.

\begin{remark}
Note that when $\mathbf{U}^\top=\mathbf{0}$, and in the absence of the initial body moments ($\mathring{\boldsymbol{\mathfrak L}}=\mathbf{0}$), the normal body forces ($\mathring{\boldsymbol{\mathfrak B}}^\perp=\mathbf{0}$), and the initial couple stress ($\mathring{\boldsymbol{\mathsf{M}}}=\mathbf{0}$), we recover the governing equations of a classical plate discussed in \citep[P.~379]{timoshenko1959theory}, \citep[P.~289]{lekhnitskii1968anisotropic}, and \citep{colquitt2014tra}. To see this, from \eqref{shell-initial-linear} and \eqref{shell-initial-ang}, the nontrivial equilibrium equations for a classical plate in its initial configuration read 
\begin{equation}\label{initial-cons}
\mathring{P}^{aA}{}_{|A}+\rho (\mathring{\mathfrak{B}}^\top)^a=0\,,\quad \mathrm{and}\quad \mathring{P}^{[aA}\delta^{b]}{}_A=0\,.
\end{equation}
Also, the linearized governing equations \eqref{lin-plate-cosserat} for a classical plate read
\begin{equation}
\delta P^{aA}{}_{|A}=0\,,\quad\mathrm{and}\quad\mathring{P}^{aA}\left(\frac{\partial\, U^\perp}{\partial\, x^a}\right)_{|A}-(\mathring{F}^{-1})^B{}_a\delta\mathsf{M}^{aA}{}_{|A|B}-\rho(\mathring{{\mathfrak B}}^\top)^b\frac{\partial\, U^\perp}{\partial\,x^b}=\rho\, \ddot{U}^\perp\,,
\end{equation}
where from \eqref{sim-delta}, $\delta P^{aA}=\frac{1}{2}\mathbb{B}^{aAbB}( U^\perp_{~,b})_{|B}$, and $\delta\mathsf{M}^{aA}=\frac{1}{2}\mathbb{C}^{aAbB}( U^\perp_{~,b})_{|B}$. Therefore, one obtains
\begin{equation}\label{key-lin-pre}
\begin{split}
-\frac{1}{2}(\mathring{F}^{-1})^B{}_a&\Big[\mathbb{C}^{aAbC}{}_{|A|B}( U^\perp_{~,b})_{|C}+\mathbb{C}^{aAbC}{}_{|A}( U^\perp_{~,b})_{|C|B}+\mathbb{C}^{aAbC}{}_{|B}( U^\perp_{~,b})_{|C|A}\\&+\mathbb{C}^{aAbC}( U^\perp_{~,b})_{|C|A|B}\Big]+\mathring{P}^{aA}\left(\frac{\partial\, U^\perp}{\partial\, x^a}\right)_{|A}-\rho(\mathring{{\mathfrak B}}^\top)^b\frac{\partial\, U^\perp}{\partial\,x^b}=\rho\, \ddot{U}^\perp\,,
\end{split}
\end{equation}
and
\begin{equation}\label{sim-1-plate}
\mathbb{B}^{aAbB}{}_{|A}( U^\perp_{~,b})_{|B}+\mathbb{B}^{aAbB}( U^\perp_{~,b})_{|B|A}=0\,.
\end{equation}
Similarly, after some simplifications \eqref{plate-ang-sim} implies that
\begin{equation}\label{sim-2-plate}
\mathbb{B}^{[aAcB}\mathring{F}^{b]}{}_A=0\,,\qquad\mathbb{C}^{[aAcB}\mathring{F}^{b]}{}_A=0\,.
\end{equation}
Note that in the case of pure bending deformations considered in the above-mentioned references, no dependence of the strain energy function on the Cauchy-Green deformation tensor is assumed, and thus, $\pmb{\mathbb{B}}$ vanishes. Therefore, \eqref{sim-1-plate} is trivially satisfied and \eqref{sim-2-plate} gives $\mathbb{C}^{[aAcB}\mathring{F}^{b]}{}_A=0$, i.e., $\pmb{\mathbb{C}}$ must possess the minor symmetries.
\end{remark}


\section{Transformation cloaking in elastic plates}\label{four4}

\begin{figure}[h]
	\centering
	\includegraphics[width=6.5in]{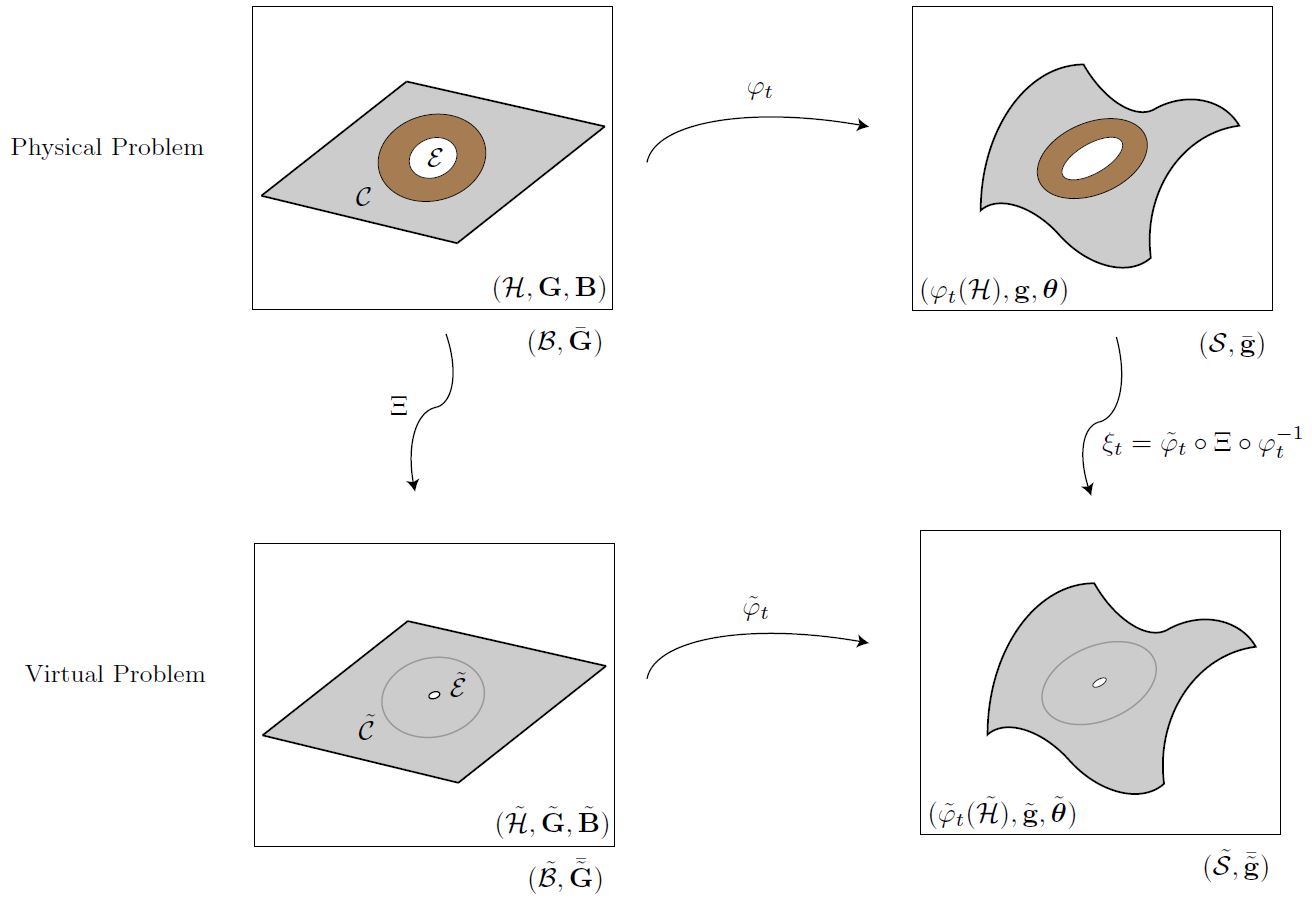}
	\vspace*{-0.1in}
	\caption{\footnotesize A cloaking transformation $\Xi$ (or $\xi$ when the physical plate is pre-stressed) transforms a plate with a finite hole $\mathcal{E}$ to another plate with an infinitesimal hole ($\tilde{\mathcal E}$) that is homogeneous and isotropic. The cloaking transformation is defined to be the identity map outside the cloak $\mathcal{C}$. Note that $\Xi$ is not a referential change of coordinates and $\xi_t$ is not a spatial change of coordinates.} 
	\label{ElasticityshellCloak}
\end{figure}

Let us consider an elastic plate $\mathcal H$ with a finite hole $\mathcal E$ (see Fig.\ref{ElasticityshellCloak}). In transformation cloaking one surrounds the hole with a cloaking device $\mathcal C$, which is an annular elastic plate such that the finite hole has negligible disturbance effects on the upcoming waves, i.e., as if the hole does not exist. The mass density and the elastic properties of the cloak are inhomogeneous and anisotropic, in general. Without loss of generality, we assume that in $\mathcal{H}\setminus\mathcal{C}$ the plate is homogeneous and isotropic. Motion of $\mathcal H$ is represented by a smooth map $\varphi_t:\mathcal{H}\to\mathcal{S}$. The cloaking transformation is a time-independent map $\xi:\mathring{\varphi}(\mathcal{H})\to\mathring{\tilde{\varphi}}(\tilde{\mathcal H})$, which transforms the pre-stressed plate $\mathcal H$ (physical plate) with the initial stress $\mathring{\mathbf P}$ and the initial couple-stress $\mathring{\boldsymbol{\mathsf{M}}}$ in its current configuration to a corresponding homogeneous and isotropic stress-free plate $\tilde{\mathcal H}$ (virtual plate) in its current configuration (see also \citep{yavari2018nonlinear}). 
The physical and virtual plates are endowed with their respective induced Euclidean metrics $\mathbf{G}$ and $\tilde{\mathbf G}$.
The mapping $\xi$ transforms the finite hole $\mathcal E$ to a very small hole $\tilde{\mathcal{E}}$ of radius $\epsilon$ and is assumed to be the identity in $\mathcal{H}\setminus\mathcal{C}$. The corresponding cloaking transformation in the reference configuration is denoted by $\Xi$. We also assume that the virtual plate has the same uniform and isotropic elastic properties as those of the physical plate outside the cloak ($\mathcal H\setminus\mathcal C$). Motion of the virtual plate is represented by $\tilde{\varphi}_t:\tilde{\mathcal H}\to \tilde{\mathcal{S}}$. The initial-boundary value problems corresponding to the motions $\varphi_t:\mathcal{H}\to\mathcal{S}$ and $\tilde{\varphi}_t:\tilde{\mathcal H}\to\tilde{\mathcal S}$ are called the physical and virtual problems, respectively. The deformation gradients corresponding to the physical and virtual problems are, respectively, denoted by $\mathbf{F}=T\varphi_t$ and $\tilde{\mathbf{F}}=T\tilde{\varphi}_t$. The tangent map of the referential and spatial cloaking transformations are, respectively, denoted by $T\Xi=\accentset{\Xi}{\mathbf{F}}$ and $T\xi=\accentset{\xi}{\mathbf{F}}$.
The current configurations of the physical and virtual problems are required to be identical outside the cloaking region, i.e., in $\mathcal{H}\setminus\mathcal{C}$. This implies, in particular, that any elastic measurements performed in the spatial configurations of the virtual and physical plates are identical, and thus, they are indistinguishable by an observer positioned anywhere in $\mathcal H\setminus\mathcal C$.

Note that due to the structure of the governing equations of an elastic plate \eqref{lin-mom-non} and \eqref{lin-mom-non1},  under a cloaking transformation, the (two-point) stress and couple-stress are not necessarily transformed using a Piola transformation\footnote{The Piola transformation of a vector (field) $\mathbf w\in T_{\varphi(X)}\mathcal{S}$ is a vector $\mathbf{W}\in T_X\mathcal{B}$ given by $\mathbf{W}=J\varphi^*\mathbf{w}=J\mathbf{F}^{-1}\mathbf{w}$. 
	In coordinates, one has $W^A=J(F^{-1})^A{}_bw^b$, where $J=\sqrt{\frac{\det\mathbf{g}}{\det\mathbf{G}}}\det \mathbf{F}$ is the Jacobian of $\varphi$ with $\mathbf{G}$ and $\mathbf{g}$ the Riemannian metrics of $\mathcal{B}$ and $\mathcal{S}$, respectively. Note that a Piola transformation can be performed on any index of a given tensor. One can show that $\operatorname{Div}\mathbf{W}=J(\operatorname{div}\mathbf{w})\circ\varphi$, which in coordinates is written as $W^A{}_{|A}=Jw^a{}_{|a}$. This is also known as the Piola identity. Another way of writing the Piola identity is in terms of the unit normal vectors of a surface in $\mathcal{B}$ and its corresponding surface in $\mathcal{S}$, along with the area elements. It is written as $\hat{\mathbf{n}}da=J\mathbf{F}^{-\star}\hat{\mathbf{N}}dA$, or in components, one writes $n_ada=J(F^{-1})^A{}_aN_AdA$. In the literature of continuum mechanics, this is known as Nanson's formula.} (unlike transformation cloaking in $3$D elasticity \citep{yavari2018nonlinear}). This is something that we carefully discuss in \S\ref{four4.1} and \S\ref{linear-elas} for Kirchhoff-Love plates and for elastic plates with both the in-plane and out-of-plane deformations. 
The Jacobian of the referential and spatial \emph{cloaking transformations}, $\Xi$ and $\xi$ are given by
\begin{equation}
J_{\Xi}=\sqrt{\frac{\det\tilde{\mathbf{G}}\circ\Xi}{\det\mathbf{G}}}\det\accentset{\Xi}{\mathbf{F}}\,, \quad J_{\xi}=\sqrt{\frac{\det\tilde{\mathbf{g}}\circ\xi}{\det\mathbf{g}}}\det\accentset{\xi}{\mathbf{F}}\,.
\end{equation}
\paragraph{Shifters in Euclidean ambient space.}We assume that the reference configurations of both the physical and virtual plates are embedded in the Euclidean space. To relate vector fields in the physical problem to those in the virtual problem properly one would need to use shifters. 
The mapping $\boldsymbol{\mathsf{s}}:T\mathcal{S}\rightarrow T\tilde{\mathcal{S}}$, defined as $\boldsymbol{\mathsf{s}}(x,\mathbf{w})=(\tilde{x},\mathbf{w})$ is called the shifter map. The restriction of $\boldsymbol{\mathsf{s}}$ to $x\in\mathcal{S}$ is denoted by $\boldsymbol{\mathsf{s}}_x=\boldsymbol{\mathsf{s}}(x):T_x\mathcal{S}\rightarrow T_{\tilde{x}}\tilde{\mathcal{S}}$, and shifts $\mathbf{w}$ based at $x\in\mathcal{S}$ to $\mathbf{w}$ based at $\tilde{x}\in\tilde{\mathcal{S}}$. Notice that $\boldsymbol{\mathsf{s}}$ simply parallel transports\footnote{The notion of shifter maps in a general Riemannian manifold can be defined similarly if one considers parallel translations along the curves in the two manifolds.} vectors emanating from $x$ to those emanating from $\tilde x$ utilizing the linear structure of the Euclidean ambient space.
For $\mathcal{S}$ and $\tilde{\mathcal S}$ we choose two global collinear Cartesian coordinates $\{\tilde{z}^{\tilde{i}}\}$ and $\{z^i\}$ for the virtual and physical deformed configurations, respectively. Let us also use curvilinear coordinates $\{\tilde{x}^{\tilde{a}}\}$ and $\{x^a\}$ for these configurations.
Note that $\mathsf{s}^{\tilde{i}}{}_i=\delta^{\tilde{i}}_i$. One can show that \citep{MaHu1983}
\begin{equation}
\mathsf{s}^{\tilde{a}}{}_a(x)=\frac{\partial \tilde{x}^{\tilde{a}}}{\partial \tilde{z}^{\tilde{i}}}(\tilde{x})
\frac{\partial z^i}{\partial x^a}(x)	\delta^{\tilde{i}}_i,~~~a,\tilde{a}=1,2,3 \,.
\end{equation}
Note that $\boldsymbol{\mathsf{s}}$ preserves inner products, and thus, $\boldsymbol{\mathsf{s}}^{\mathsf{T}}=\boldsymbol{\mathsf{s}}^{-1}$, where in components, $(\mathsf{s}^{\mathsf{T}})^a{}_{\tilde{a}}=g^{ab}\mathsf{s}^{\tilde{b}}{}_b\tilde{g}_{\tilde{a}\tilde{b}}$. Note also that
\begin{equation}
\mathsf{s}^{\tilde{a}}{}_{a|\tilde{b}}= \frac{\partial \mathsf{s}^{\tilde{a}}{}_a}{\partial \tilde{x}^{\tilde{b}}}
+\tilde{\gamma}^{\tilde{a}}{}_{\tilde{b}\tilde{c}}\mathsf{s}^{\tilde{c}}{}_a
-\frac{\partial x^b}{\partial \tilde{x}^{\tilde{b}}}\gamma^c{}_{ab} \mathsf{s}^{\tilde{a}}{}_c\,,
\end{equation}
where $\gamma^a{}_{bc}=\frac{\partial x^a}{\partial z^k}\frac{\partial^2 z^k}{\partial x^b\partial x^c}$ and $\tilde{\gamma}^{\tilde{a}}{}_{\tilde{b}\tilde{c}}=\frac{\partial \tilde{x}^{\tilde{a}}}{\partial \tilde{z}^{\tilde{k}}}\frac{\partial^2 \tilde{z}^{\tilde{k}}}{\partial \tilde{x}^{\tilde{b}}\partial \tilde{x}^{\tilde{c}}}$ are the Christoffel symbols associated with $\mathcal S$ and $\tilde{\mathcal S}$ (with their induced Euclidean metrics), respectively. It is straightforward to verify that $\mathsf{s}^{\tilde{a}}{}_{a|\tilde{b}}=0$, i.e., the shifter is covariantly constant. As the reference configurations of both the physical and virtual plates are embedded in the Euclidean space, referential shifters may also be defined similarly.
As an example consider polar coordinates $(r,\theta)$ and $(\tilde{r},\tilde{\theta})$ at $x\in\mathbb{R}^2$ and $\tilde{x}\in\mathbb{R}^2$, respectively. The shifter map has the following matrix representation with respect to these coordinates
\begin{equation}
\boldsymbol{\mathsf{s}}=
\begin{bmatrix} 
\cos(\tilde{\theta}-\theta) &  r\sin(\tilde{\theta}-\theta)   \\
-\sin(\tilde{\theta}-\theta)/\tilde{r} & r\cos(\tilde{\theta}-\theta)/\tilde{r}    
\end{bmatrix} \,.
\end{equation}


In order to ensure that the (tangential and normal) components of the acceleration term remain form invariant under the cloaking map,
the tangential and normal components of the displacement field in the physical and virtual plates are related as
\begin{equation}
({\tilde{U}^\top})^{\tilde a}=\mathsf{s}^{\tilde a}{}_a(U^\top)^a\,,\quad 
{\tilde{U}^\perp}=U^\perp\,, \quad a,\tilde{a}=1,2,
\end{equation}
where $\boldsymbol{\mathsf s}$ is the shifter in the local tangent plane to plates in $\mathbb{R}^2$.\footnote{Note that for both plates we may use two global collinear Cartesian coordinates $\{z^1,z^2,z^3\}$ and $\{\tilde{z}^1,\tilde{z}^2,\tilde{z}^3\}$ such that $z^3$ and $\tilde{z}^3$ are the outward normal directions to the physical and virtual plates, respectively. Therefore, $\{z^1,z^2\}$ and $\{\tilde{z}^1,\tilde{z}^2\}$ are two global collinear Cartesian coordinates for $\varphi(\mathcal H)$ and $\tilde{\varphi}(\tilde{\mathcal H})$, respectively, where $\Xi:\mathcal{H}\to\tilde{\mathcal H}$, $\varphi:\mathcal{H}\to\varphi(\mathcal H)$,  $\tilde{\varphi}:\tilde{\mathcal H}\to\tilde{\varphi}(\tilde{\mathcal{H}})$, and $\xi:\varphi(\mathcal H)\to\tilde{\varphi}(\tilde{\mathcal H})$, and $\boldsymbol{\mathsf{s}}$ is defined as
\begin{equation*}
\mathsf{s}^{\tilde{a}}{}_a(x)=\frac{\partial \tilde{x}^{\tilde{a}}}{\partial \tilde{z}^{\tilde{i}}}(\tilde{x})
\frac{\partial z^i}{\partial x^a}(x)	\delta^{\tilde{i}}_i,~~~a,\tilde{a},i,\tilde{i}=1,2 \,,
\end{equation*}
where $\{x^a\}$ and $\{\tilde{x}^{\tilde a}\}$ are local coordinate charts for $\varphi(\mathcal H)$ and $\tilde{\varphi}(\tilde{\mathcal H})$, respectively.
} 
Let $\{X^1,X^2,X^3\}$ and $\{\tilde{X}^1,\tilde{X}^2,\tilde{X}^3\}$ be local coordinate charts for $\mathcal{B}$ and $\tilde{\mathcal{B}}$ such that $\{X^1,X^2\}$ and $\{\tilde{X}^1,\tilde{X}^2\}$ are local charts for $\mathcal{H}$ and $\tilde{\mathcal{H}}$, respectively (with $\partial/\partial X^3$ and $\partial/\partial\tilde{X}^3$ being, respectively, normal to $\mathcal{H}$ and $\tilde{\mathcal{H}}$). We assume that $\mathcal{B}$ and $\tilde{\mathcal{B}}$ are both embedded in the Euclidean space using two global collinear Cartesian coordinates $\{\tilde{Z}^{\tilde{I}}\}$ and $\{Z^I\}$, respectively.  
The referential shifter is similarly defined as
\begin{equation}
\mathsf{S}^{\tilde{A}}{}_A(X)=\frac{\partial \tilde{X}^{\tilde{A}}}{\partial \tilde{Z}^{\tilde{I}}}(\tilde{X})
\frac{\partial Z^I}{\partial X^A}(X)	\delta^{\tilde{I}}{}_I \,,\quad A,\tilde{A},I,\tilde{I}=1,2,3.
\end{equation}
Thus, one obtains $\bar{G}_{AB}=\mathsf{S}^{\tilde{A}}{}_A\mathsf{S}^{\tilde{B}}{}_B\bar{\tilde{{G}}}_{\tilde A\tilde B}$, where $A,B,\tilde{A},\tilde{B}=1,2,3$. 

\paragraph{Boundary conditions in the physical and virtual problems.} Let $\partial \mathcal{H}=\partial\mathcal{E}\cup\partial_o\mathcal{H}$, where $\partial\mathcal{E}$ is the boundary of the physical hole and $\partial_o\mathcal{H}$ is the outer boundary of $\mathcal H$. Let us assume that $\partial_o\mathcal{H}$ is the disjoint union of $\partial_o\mathcal{H}_d$ and $\partial_o\mathcal{H}_t$, i.e., $\partial_o\mathcal{H}=\partial_o\mathcal{H}_t\cup\partial_o\mathcal{H}_d$, such that the Neumann and Dirichlet boundary conditions read\footnote{See Remark. \ref{key-prescribe} for a discussion on how the boundary surface traction, boundary shear force, and boundary moment as well as their corresponding Dirichlet boundary conditions are prescribed in the boundary-value problem.}  
\begin{equation}
\begin{split}
\begin{cases}
~\qquad\left[P^{aA}+2\theta_{bc}\mathsf{M}^{bA}g^{ac}\right]\mathsf{T}_A=(\bar{T}^\top)^a\,,\\ -(F^{-1})^A{}_a\left[\rho\mathfrak{L}^a-\mathsf{M}^{aB}{}_{|B}\right]\mathsf{T}_{A}=\bar{T}^\perp\,, \\
~\qquad\qquad\qquad\qquad\quad\,\mathsf{M}^{aA}\mathsf{T}_A=\bar{\mathsf{m}}^a\,,
\end{cases}\quad \mathrm{on}\quad \partial_o\mathcal{H}_t\\
\begin{cases}
\varphi^\top(X,t)=\bar{\varphi}^\top(X,t)\,,\\
\varphi^\perp(X,t)=\bar{\varphi}^\perp(X,t)\,,
\\
\,\frac{\partial \varphi^\perp(X,t)}{\partial X^A}=\frac{\partial \bar{\varphi}^\perp(X,t)}{\partial X^A}\,,
\end{cases}\quad \mathrm{on}\quad \partial_o\mathcal{H}_d
\end{split}
\end{equation}
where $\boldsymbol{\mathsf{T}}$ is the unit normal one-form on $\partial_o\mathcal{H}$. Similarly, for the virtual problem, one has
\begin{equation}
\begin{split}
\begin{cases}
\,\,\quad\qquad\left[\tilde{P}^{\tilde{a}\tilde{A}}+2\tilde{\theta}_{\tilde{b}\tilde{c}}\tilde{\mathsf{M}}^{\tilde{b}\tilde{A}}\tilde{g}^{\tilde{a}\tilde{c}}\right]\tilde{\mathsf T}_{\tilde{A}}=(\bar{\tilde T}^\top)^{\tilde a}\,,\\~ -(\tilde{F}^{-1})^{\tilde A}{}_{\tilde a}\left[\tilde{\rho}\tilde{\mathfrak{L}}^{\tilde a}-\tilde{\mathsf{M}}^{\tilde a\tilde B}{}_{|\tilde B}\right]{\tilde {\mathsf T}}_{\tilde A}=\bar{\tilde T}^\perp\,, \\
\,\,\,\quad\qquad\qquad\qquad\qquad\quad\,\tilde{\mathsf{M}}^{\tilde a\tilde A}\tilde{\mathsf T}_{\tilde A}=\bar{\tilde{\mathsf{m}}}^{\tilde a}\,,
\end{cases}\quad \mathrm{on}\quad \partial_o\tilde{\mathcal{H}}_t\\
\begin{cases}
\tilde{\varphi}^\top(\tilde X,t)=\bar{\tilde{\varphi}}^\top(\tilde X,t)\,,\\ \tilde{\varphi}^\perp(\tilde X,t)=\bar{\tilde{\varphi}}^\perp(\tilde X,t)\,,
\\
\frac{\partial \tilde{\varphi}^\perp({\tilde X},t)}{\partial \tilde X^{\tilde A}}=\frac{\partial \bar{\tilde \varphi}^\perp(\tilde X,t)}{\partial \tilde{X}^{\tilde A}}\,.
\end{cases}\quad \mathrm{on}\quad \partial_o\tilde{\mathcal{H}}_d
\end{split}
\end{equation}
Note that the cloaking map $\Xi:\mathcal{H}\to\tilde{\mathcal H}$ is set to be the identity outside the cloak, i.e., in $\mathcal H\setminus\mathcal C$ (or $\tilde{\mathcal H}\setminus\tilde{\mathcal C}$), and thus, one is able to impose identical boundary conditions on the outer boundaries $\partial_o \mathcal H$, and $\partial_o\tilde{\mathcal H}$. Thus, noting that $\partial(\mathcal H\setminus\mathcal C)=\partial_o\mathcal H\cup\partial_o\mathcal C$ and $\partial(\tilde{\mathcal H}\setminus\tilde{\mathcal C})=\partial_o\tilde{\mathcal H}\cup\partial_o\tilde{\mathcal C}$, in order for the two problems to have identical current configurations (and thus, elastic measurements) outside the cloak, it remains to make sure that the boundary data of $\partial_o\mathcal{C}$ and $\partial_o\tilde{\mathcal C}$ are identical, i.e., for $X\in\partial_o\mathcal{C}$ and $\tilde{X}\in\partial_o\tilde{\mathcal{C}}$ (note that $\Xi|_{\partial_o\mathcal{C}}=id$), one needs to have


\begin{equation}\label{key-bound-cond}
\begin{split}
\begin{cases}
\begin{split}
(\tilde{T}^\top)^{\tilde a}&=\left[\tilde{P}^{\tilde{a}\tilde{A}}+2\tilde{\theta}_{\tilde b\tilde c}\tilde{\mathsf{M}}^{\tilde b\tilde A}\tilde{g}^{\tilde a\tilde c}\right]\tilde{\mathsf T}_{\tilde A}=\mathsf{s}^{\tilde a}{}_{a}\left[P^{aA}+2\theta_{bc}\mathsf{M}^{bA}g^{ac}\right]\mathsf{T}_A=\mathsf{s}^{\tilde a}{}_{a}({T}^\top)^a\,,\\ \tilde{T}^\perp&=-(\tilde{F}^{-1})^{\tilde A}{}_{\tilde a}\left[\tilde{\rho}\tilde{\mathfrak{L}}^{\tilde a}-\tilde{\mathsf{M}}^{\tilde a\tilde B}{}_{|\tilde B}\right]{\tilde {\mathsf T}}_{\tilde A}=-(F^{-1})^A{}_a\left[\rho\mathfrak{L}^a-\mathsf{M}^{aB}{}_{|B}\right]\mathsf{T}_{A}={T}^\perp\,, \\
\tilde{\mathsf{m}}^{\tilde a}&=\tilde{\mathsf{M}}^{\tilde a\tilde A}{\tilde {\mathsf T}}_{\tilde A}=\mathsf{s}^{\tilde a}{}_a\mathsf{M}^{aA}\mathsf{T}_A=\mathsf{s}^{\tilde a}{}_a{\mathsf{m}}^a\,,
\end{split}
\end{cases}~ \mathrm{on}~~ \partial_o\mathcal{C}\\
\begin{cases}
\begin{split}
\tilde{\varphi}^\top\circ\Xi( X,t)&=\varphi^\top(X,t)\,,\\
\tilde{\varphi}^\perp\circ\Xi( X,t)&=\varphi^\perp(X,t)\,,
\\
\frac{\partial \tilde{\varphi}^\perp}{\partial \tilde{X}^{\tilde{A}}}\circ\Xi( X,t)&=(\mathsf{S}^{-1})^A{}_{\tilde A}\frac{\partial \varphi^\perp}{\partial X^A}(X,t)\,.
\end{split}
\end{cases}~ \mathrm{on}~~ \partial_o\mathcal{C}
\end{split}
\end{equation}
Moreover, the hole in the virtual shell is assumed to be traction-free, i.e.,
\begin{equation}\label{key-hole-virtual}
\begin{split}
\left[\tilde{P}^{\tilde a\tilde A}+2\tilde{\theta}_{\tilde b\tilde c}\tilde{\mathsf{M}}^{\tilde b\tilde A}\tilde{g}^{\tilde a\tilde c}\right]\tilde{\mathsf T}_{\tilde A}=0\,,\quad (\tilde{F}^{-1})^{\tilde A}{}_{\tilde a}\left[\tilde{\rho}\tilde{\mathfrak{L}}^{\tilde a}-\tilde{\mathsf{M}}^{\tilde a\tilde B}{}_{|\tilde B}\right]\tilde{\mathsf T}_{\tilde A}=0\,,\\  
\quad\quad\quad\quad\,\tilde{\mathsf{M}}^{\tilde a\tilde A}\tilde{\mathsf T}_{\tilde A}=0\,.
\end{split}\quad \mathrm{on}\quad \partial\tilde{\mathcal{E}}
\end{equation}
The surface of the hole in the physical shell must be traction-free as well, and hence
\begin{equation}\label{key-shell-hole}
\begin{split}
\left[{P}^{ a A}+2{\theta}_{ b c}{\mathsf{M}}^{ b A}{g}^{ a c}\right]\mathsf{T}_{ A}=0\,,\quad ({F}^{-1})^{ A}{}_{ a}\left[{\rho}{\mathfrak{L}}^{a}-{\mathsf{M}}^{ a B}{}_{| B}\right]\mathsf{T}_{ A}=0\,,\\ 
\quad\quad\quad\quad\,{\mathsf{M}}^{ a A}\mathsf{T}_{ A}=0\,.
\end{split}\quad \mathrm{on}\quad \partial{\mathcal{E}}
\end{equation}

\begin{remark}
	In the linearized setting, the condition \eqref{key-bound-cond} is written as
\begin{equation}\label{key-bound-2}
\begin{split}
\begin{cases}
\begin{split}
(\delta\tilde{T}^\top)^{\tilde a}&=\left[\delta\tilde{P}^{\tilde{a}\tilde{A}}+2\delta\tilde{\theta}_{\tilde b\tilde c}\mathring{\tilde{\mathsf{M}}}^{\tilde b\tilde A}\tilde{g}^{\tilde a\tilde c}+2\mathring{\tilde{\theta}}_{\tilde b\tilde c}\delta\tilde{\mathsf{M}}^{\tilde b\tilde A}\tilde{g}^{\tilde a\tilde c}\right]\tilde{\mathsf T}_{\tilde A}\\&=\mathsf{s}^{\tilde a}{}_{a}\left[\delta P^{aA}+2\delta\theta_{bc}\mathring{\mathsf{M}}^{bA}g^{ac}+2\mathring{\theta}_{bc}\delta\mathsf{M}^{bA}g^{ac}\right]\mathsf{T}_A=\mathsf{s}^{\tilde a}{}_{a}(\delta T^\top)^a\,,\\ \delta\tilde{T}^\perp&=-(\mathring{\tilde{F}}^{-1})^{\tilde A}{}_{\tilde a}\left[\tilde{\rho}\delta{\tilde{\mathfrak{L}}}^{\tilde a}-\delta\tilde{\mathsf{M}}^{\tilde a\tilde B}{}_{|\tilde B}\right]{\tilde {\mathsf T}}_{\tilde A}-(\delta{\tilde{F}}^{-1})^{\tilde A}{}_{\tilde a}\left[\tilde{\rho}\mathring{\tilde{\mathfrak{L}}}^{\tilde a}-\mathring{\tilde{\mathsf{M}}}^{\tilde a\tilde B}{}_{|\tilde B}\right]{\tilde {\mathsf T}}_{\tilde A}\\&=-(\mathring{F}^{-1})^A{}_a\left[\rho\delta{\mathfrak{L}}^a-\delta\mathsf{M}^{aB}{}_{|B}\right]\mathsf{T}_{A}-(\delta{F}^{-1})^A{}_a\left[\rho\mathring{\mathfrak{L}}^a-\mathring{\mathsf{M}}^{aB}{}_{|B}\right]\mathsf{T}_{A}=\delta{T}^\perp\,,\\  
\delta\tilde{\mathsf{m}}^{\tilde a}&=\delta\tilde{\mathsf{M}}^{\tilde a\tilde A}{\tilde {\mathsf T}}_{\tilde A}=\mathsf{s}^{\tilde a}{}_a\delta\mathsf{M}^{aA}\mathsf{T}_A=\mathsf{s}^{\tilde a}{}_a\delta{\mathsf{m}}^a\,,
\end{split}
\end{cases} \mathrm{on}~~ \partial_o\mathcal{C}\\
\begin{cases}
\begin{split}
\tilde{\mathbf U}^\top\circ\Xi( X,t)&=\boldsymbol{\mathsf{s}}{\mathbf U}^\top(X,t)\,,\\\quad\tilde{\mathbf U}^\perp\circ\Xi( X,t)&={\mathbf{U}}^\perp(X,t)\,,
\\
\frac{\partial \tilde{U}^\perp}{\partial \tilde{X}^{\tilde{A}}}\circ\Xi( X,t)&=(\mathsf{S}^{-1})^A{}_{\tilde A}\frac{\partial U^\perp}{\partial X^A}(X,t)\,.
\end{split}
\end{cases}~ \mathrm{on}~~ \partial_o\mathcal{C}
\end{split}
\end{equation}
\end{remark}
Next, we discuss transformation cloaking in Kirchhoff-Love plates, for which only the out-of-plane displacement is allowed. We also examine the possibility of transformation cloaking when the pure bending assumption is relaxed and the plate is allowed to have both in-plane and out-of-plane displacements.

\subsection{Elastodynamic transformation cloaking in Kirchhoff-Love plates}\label{four4.1}
In this section, we discuss transformation cloaking in classical elastic plates in the absence of in-plane deformations (pure bending). 
For the sake of brevity, let us denote the normal displacement of the physical plate $U^\perp$ and the normal displacement of the virtual plate $\tilde{U}^\perp$ by $\mathsf{W}$ and $\widetilde{\mathsf{W}}$, respectively.
For the virtual plate with uniform elastic properties and vanishing pre-stress and initial body forces, \eqref{key-lin-pre} is simplified to read
\begin{equation}\label{virtual-plate}
-\frac{1}{2}(\mathring{\tilde F}^{-1})^{\tilde{B}}{}_{\tilde a}\tilde{\mathbb{C}}^{\tilde{a}\tilde{A}\tilde{b}\tilde{C}}( \widetilde{\mathsf{W}}{}_{,\tilde{b}})_{|\tilde{C}|\tilde{A}|\tilde{B}}=\tilde{\rho}\, \ddot{\widetilde{\mathsf W} }\,.
\end{equation}
Note that in the absence of in-plane deformations ($\mathbf{U}^\top=\mathbf{0}$) and in the case of thin plate bending ($\pmb{\mathbb B}=\mathbf{0}$), $\delta P^{aA}=\delta\tilde{P}^{\tilde a\tilde A}=0$, and \eqref{virtual-plate} is the only non-trivial linearized balance of linear momentum equation.
We assume a Saint Venant-Kirchhoff constitutive model\footnote{See \citep{fox1993justification,le1993modele,lods1998nonlinearly,friesecke2002theorem} for details on the derivation of the Saint Venant-Kirchhoff shell constitutive model.} for the virtual plate, for which the energy density is given by
\begin{equation}\label{saint}
\begin{split}
\tilde{W}=&\frac{Eh}{8(1+\nu)}\left\{\mathrm{tr}\left[\left(\tilde{\mathbf{C}}-\tilde{\mathbf{G}}\right)^2\right]+\frac{\nu}{1-\nu}\left[\mathrm{tr}\left(\tilde{\mathbf{C}}-\tilde{\mathbf{G}}\right)\right]^2\right\}\\&+\frac{Eh^3}{24(1+\nu)}\left\{\mathrm{tr}\left[\left(\tilde{\mathbf{\Theta}}-\tilde{\mathbf{B}}\right)^2\right]+\frac{\nu}{1-\nu}\left[\mathrm{tr}\left(\tilde{\mathbf{\Theta}}-\tilde{\mathbf{B}}\right)\right]^2\right\}\,,
\end{split}
\end{equation}
where $E$ is Young's modulus and $\nu$ is Poisson's ratio. Therefore, in the case of pure bending deformations, the flexural rigidity tensor for the virtual plate is written with a slight abuse of notation as
\begin{equation}\label{key-vir-ct}
	\tilde{\mathbb{C}}^{\tilde{a}\tilde{A}\tilde{b}\tilde{C}}=\frac{Eh^3}{12(1+\nu)}
	\mathring{\tilde F}^{\tilde a}{}_{\tilde M}\mathring{\tilde F}^{\tilde b}{}_{\tilde N}
	\left[\tilde{G}^{\tilde{A}\tilde{N}}\tilde{G}^{\tilde{C}\tilde{M}}+\tilde{G}^{\tilde{A}\tilde{C}}
	\tilde{G}^{\tilde{M}\tilde{N}}+\frac{2\nu}{1-\nu}\tilde{G}^{\tilde{A}\tilde{M}}\tilde{G}^{\tilde{C}\tilde{N}}\right]\,.
\end{equation}
Note that \eqref{key-vir-ct} is the most general isotropic constitutive equation for a Kirchhoff-Love plate.
Assuming that $\mathsf{W}=\widetilde{\mathsf{W}}\circ\Xi$, the derivatives of the normal displacement field are transformed as  
\begin{equation}\label{disp-grad}
\begin{split}
\widetilde{\mathsf{W}}{}_{,\tilde b}=&(\accentset{\Xi}{F}^{-1})^{ b}{}_{\tilde b}\,\mathsf{W}{}_{, b}\,,\\
(\widetilde{\mathsf{W}}{}_{, \tilde b})_{|\tilde C}=&(\accentset{\Xi}{F}^{-1})^{ b}{}_{\tilde b|\tilde C}\,{\mathsf{W}}{}_{, b}+(\accentset{\Xi}{F}^{-1})^{ b}{}_{\tilde b}(\accentset{\Xi}{F}^{-1})^{ C}{}_{\tilde C}\,({\mathsf{W}}{}_{, b})_{|{C}}\,,\\
(\widetilde{\mathsf{W}}{}_{, \tilde{b}})_{|\tilde{C}|\tilde{A}}=&(\accentset{\Xi}{F}^{-1})^{ b}{}_{\tilde{b}|\tilde{C}|\tilde{A}}\,{\mathsf{W}}{}_{, b}+\Big[(\accentset{\Xi}{F}^{-1})^{ b}{}_{\tilde{b}|\tilde{A}}(\accentset{\Xi}{F}^{-1})^{ C}{}_{\tilde C}+(\accentset{\Xi}{F}^{-1})^{ C}{}_{\tilde C|\tilde A}(\accentset{\Xi}{F}^{-1})^{ b}{}_{\tilde b}\\&+(\accentset{\Xi}{F}^{-1})^{ b}{}_{\tilde b|\tilde C}(\accentset{\Xi}{F}^{-1})^{ C}{}_{\tilde A}\Big]({\mathsf{W}}{}_{, b})_{|{C}}+(\accentset{\Xi}{F}^{-1})^{ b}{}_{\tilde b}(\accentset{\Xi}{F}^{-1})^{ C}{}_{\tilde C}(\accentset{\Xi}{F}^{-1})^{ A}{}_{\tilde A}\,({\mathsf{W}}{}_{, b})_{|{C}|{A}}\,,\\
(\widetilde{\mathsf{W}}{}_{, \tilde b})_{|\tilde C|\tilde A|\tilde B}=&(\accentset{\Xi}{F}^{-1})^{ b}{}_{\tilde b|\tilde C|\tilde A|\tilde B}\,{\mathsf{W}}{}_{, b}+\Big[(\accentset{\Xi}{F}^{-1})^{ b}{}_{\tilde b|\tilde A}(\accentset{\Xi}{F}^{-1})^{ A}{}_{\tilde C|\tilde B}+(\accentset{\Xi}{F}^{-1})^{ b}{}_{\tilde b|\tilde C}(\accentset{\Xi}{F}^{-1})^{ A}{}_{\tilde A|\tilde B}\\&+(\accentset{\Xi}{F}^{-1})^{ b}{}_{\tilde b|\tilde B}(\accentset{\Xi}{F}^{-1})^{ A}{}_{\tilde C|\tilde A}+(\accentset{\Xi}{F}^{-1})^{ b}{}_{\tilde b|\tilde C|\tilde B}(\accentset{\Xi}{F}^{-1})^{ A}{}_{\tilde A}+(\accentset{\Xi}{F}^{-1})^{ b}{}_{\tilde b|\tilde A|\tilde B}(\accentset{\Xi}{F}^{-1})^{ A}{}_{\tilde C}\\&+(\accentset{\Xi}{F}^{-1})^{ b}{}_{\tilde b|\tilde C|\tilde A}(\accentset{\Xi}{F}^{-1})^{ A}{}_{\tilde B}+(\accentset{\Xi}{F}^{-1})^{ A}{}_{\tilde C|\tilde A|\tilde B}(\accentset{\Xi}{F}^{-1})^{ b}{}_{\tilde b}\Big]({\mathsf{W}}{}_{, b})_{|{A}}\\&+\Big[(\accentset{\Xi}{F}^{-1})^{ b}{}_{\tilde b|\tilde B}(\accentset{\Xi}{F}^{-1})^{ C}{}_{\tilde C}(\accentset{\Xi}{F}^{-1})^{ A}{}_{\tilde A}+(\accentset{\Xi}{F}^{-1})^{ C}{}_{\tilde C|\tilde B}(\accentset{\Xi}{F}^{-1})^{ b}{}_{\tilde b}(\accentset{\Xi}{F}^{-1})^{ A}{}_{\tilde A}\\&+(\accentset{\Xi}{F}^{-1})^{ A}{}_{\tilde A|\tilde B}(\accentset{\Xi}{F}^{-1})^{ b}{}_{\tilde b}(\accentset{\Xi}{F}^{-1})^{ C}{}_{\tilde C}+(\accentset{\Xi}{F}^{-1})^{ b}{}_{\tilde b|\tilde C}(\accentset{\Xi}{F}^{-1})^{ C}{}_{\tilde A}(\accentset{\Xi}{F}^{-1})^{ A}{}_{\tilde B}\\&+(\accentset{\Xi}{F}^{-1})^{ b}{}_{\tilde b|\tilde A}(\accentset{\Xi}{F}^{-1})^{ A}{}_{\tilde B}(\accentset{\Xi}{F}^{-1})^{ C}{}_{\tilde C}+(\accentset{\Xi}{F}^{-1})^{ C}{}_{\tilde C|\tilde A}(\accentset{\Xi}{F}^{-1})^{ A}{}_{\tilde B}(\accentset{\Xi}{F}^{-1})^{ b}{}_{\tilde b}\Big]({\mathsf{W}}{}_{, b})_{|{C}|{A}}\\&+(\accentset{\Xi}{F}^{-1})^{ b}{}_{\tilde b}(\accentset{\Xi}{F}^{-1})^{ C}{}_{\tilde C}(\accentset{\Xi}{F}^{-1})^{ A}{}_{\tilde A}(\accentset{\Xi}{F}^{-1})^{ B}{}_{\tilde B}\,({\mathsf{W}}{}_{, b})_{|{C}|{A}|{B}}\,,
\end{split}
\end{equation}
The equilibrium equation for the physical plate is given by (cf. \eqref{key-lin-pre})
\begin{equation}\label{virtual-lin-pre}
	\begin{split}
		-\frac{1}{2}(\mathring{F}^{-1})^B{}_a&\Big[\mathbb{C}^{aAbC}{}_{|A|B}( \mathsf{W}{}_{,b})_{|C}+\mathbb{C}^{aAbC}{}_{|A}( \mathsf{W}{}_{,b})_{|C|B}+\mathbb{C}^{aAbC}{}_{|B}( \mathsf{W}{}_{,b})_{|C|A}\\&+\mathbb{C}^{aAbC}( \mathsf{W}{}_{,b})_{|C|A|B}\Big]+\mathring{P}^{aA}\left(\frac{\partial\, \mathsf{W}}{\partial\, x^a}\right)_{|A}-\rho(\mathring{{\mathfrak B}}^\top)^b\frac{\partial\, \mathsf{W}}{\partial\,x^b}=\rho\, \ddot{\mathsf{W}}\,.
	\end{split}
\end{equation}
We next multiply both sides of \eqref{virtual-plate} by some positive function $k=k(X)$ to be determined,\footnote{Note that introducing the scalar field $k=k(X)$  provides an extra degree of freedom in the cloaking problem.} and substitute for derivatives from \eqref{disp-grad}. Then compare the coefficients of different derivatives with those in \eqref{virtual-lin-pre}. Comparing the coefficients of the fourth-order derivatives gives us the elastic constants of the physical plate, comparing the first-order derivatives gives the tangential body force in the finitely-deformed physical plate, and finally comparing the second-order derivatives will give the pre-stress in the physical plate. In addition, comparing the coefficients of the third-order derivatives will result in a set of constraints on the cloaking map that we call \emph{cloaking compatibility equations}.
The flexural rigidity tensor of the physical plate is obtained as
\begin{equation}\label{key-trans-C}
	\mathbb{C}^{aAbC}=k(\accentset{\Xi}{F}^{-1})^{ a}{}_{\tilde a}(\accentset{\Xi}{F}^{-1})^{ A}{}_{\tilde A}
	(\accentset{\Xi}{F}^{-1})^{ b}{}_{\tilde b}(\accentset{\Xi}{F}^{-1})^{ C}{}_{\tilde C}
	\tilde{\mathbb{C}}^{\tilde{a}\tilde{A}\tilde{b}\tilde{C}}\,.
\end{equation}
Notice that $\pmb{\mathbb{C}}$ has the minor symmetries.
The mass density of the physical plate is given by $\rho=k\tilde{\rho}\circ\Xi$. 
The tangential body force and the pre-stress in the physical plate are obtained as
\begin{align}
	\label{key-ini-body}  \rho(\mathring{{\mathfrak B}}^\top)^b
	= & \frac{1}{2}k(\mathring{\tilde F}^{-1})^{\tilde{B}}{}_{\tilde a}
	\tilde{\mathbb{C}}^{\tilde{a}\tilde{A}\tilde{b}\tilde{C}}
	(\accentset{\Xi}{F}^{-1})^{ b}{}_{\tilde{b}|\tilde{C}|\tilde{A}|\tilde{B}}\,, \\
	\label{key-ini-stress}   \mathring{P}^{bA}
	= & \frac{1}{2}(\mathring{F}^{-1})^B{}_a\mathbb{C}^{aCbA}{}_{|C|B}
	-\frac{1}{2}k(\mathring{\tilde F}^{-1})^{\tilde{B}}{}_{\tilde a}\tilde{\mathbb{C}}^{\tilde{a}\tilde{A}\tilde{b}\tilde{C}}
	\Big[(\accentset{\Xi}{F}^{-1})^{ b}{}_{\tilde b|\tilde A}(\accentset{\Xi}{F}^{-1})^{ A}{}_{\tilde C|\tilde B} \nonumber \\
	&+(\accentset{\Xi}{F}^{-1})^{ b}{}_{\tilde b|\tilde C}(\accentset{\Xi}{F}^{-1})^{ A}{}_{\tilde A|\tilde B}
	+(\accentset{\Xi}{F}^{-1})^{ b}{}_{\tilde b|\tilde B}(\accentset{\Xi}{F}^{-1})^{ A}{}_{\tilde C|\tilde A}
	+(\accentset{\Xi}{F}^{-1})^{ b}{}_{\tilde b|\tilde C|\tilde B}(\accentset{\Xi}{F}^{-1})^{ A}{}_{\tilde A} \nonumber\\
	&+(\accentset{\Xi}{F}^{-1})^{ b}{}_{\tilde b|\tilde A|\tilde B}(\accentset{\Xi}{F}^{-1})^{ A}{}_{\tilde C}
	+(\accentset{\Xi}{F}^{-1})^{ b}{}_{\tilde b|\tilde C|\tilde A}(\accentset{\Xi}{F}^{-1})^{ A}{}_{\tilde B}
	+(\accentset{\Xi}{F}^{-1})^{ A}{}_{\tilde C|\tilde A|\tilde B}(\accentset{\Xi}{F}^{-1})^{ b}{}_{\tilde b}\Big]\,,
\end{align}
The cloaking compatibility equations read
\begin{equation}\label{key-constraint}
\begin{split}
(\mathring{F}^{-1})^B{}_a\mathbb{C}^{aAbC}{}_{|B}&+(\mathring{F}^{-1})^A{}_a\mathbb{C}^{aBbC}{}_{|B}=k(\mathring{\tilde F}^{-1})^{\tilde{B}}{}_{\tilde a}\tilde{\mathbb{C}}^{\tilde{a}\tilde{A}\tilde{b}\tilde{C}}\Big[(\accentset{\Xi}{F}^{-1})^{ b}{}_{\tilde b|\tilde B}(\accentset{\Xi}{F}^{-1})^{ C}{}_{\tilde C}(\accentset{\Xi}{F}^{-1})^{ A}{}_{\tilde A}\\&+(\accentset{\Xi}{F}^{-1})^{ C}{}_{\tilde C|\tilde B}(\accentset{\Xi}{F}^{-1})^{ b}{}_{\tilde b}(\accentset{\Xi}{F}^{-1})^{ A}{}_{\tilde A}+(\accentset{\Xi}{F}^{-1})^{ A}{}_{\tilde A|\tilde B}(\accentset{\Xi}{F}^{-1})^{ b}{}_{\tilde b}(\accentset{\Xi}{F}^{-1})^{ C}{}_{\tilde C}\\&+(\accentset{\Xi}{F}^{-1})^{ b}{}_{\tilde b|\tilde C}(\accentset{\Xi}{F}^{-1})^{ C}{}_{\tilde A}(\accentset{\Xi}{F}^{-1})^{ A}{}_{\tilde B}+(\accentset{\Xi}{F}^{-1})^{ b}{}_{\tilde b|\tilde A}(\accentset{\Xi}{F}^{-1})^{ A}{}_{\tilde B}(\accentset{\Xi}{F}^{-1})^{ C}{}_{\tilde C}\\&+(\accentset{\Xi}{F}^{-1})^{ C}{}_{\tilde C|\tilde A}(\accentset{\Xi}{F}^{-1})^{ A}{}_{\tilde B}(\accentset{\Xi}{F}^{-1})^{ b}{}_{\tilde b}\Big]\,,
\end{split}
\end{equation}
The initial body force and the pre-stress need to satisfy \eqref{initial-cons}. Therefore
\begin{equation}\label{key-finite-linear}
\begin{split}
	&k(\mathring{\tilde F}^{-1})^{\tilde{B}}{}_{\tilde a}\tilde{\mathbb{C}}^{\tilde{a}\tilde{A}
	\tilde{b}\tilde{C}}(\accentset{\Xi}{F}^{-1})^{ b}{}_{\tilde{b}|\tilde{C}|\tilde{A}|\tilde{B}}
	+\bigg\{(\mathring{F}^{-1})^B{}_a\mathbb{C}^{aCbA}{}_{|C|B}
	-k(\mathring{\tilde F}^{-1})^{\tilde{B}}{}_{\tilde a}\tilde{\mathbb{C}}^{\tilde{a}\tilde{A}\tilde{b}\tilde{C}}
	\Big[(\accentset{\Xi}{F}^{-1})^{ b}{}_{\tilde b|\tilde A}(\accentset{\Xi}{F}^{-1})^{ A}{}_{\tilde C|\tilde B}\\&
	+(\accentset{\Xi}{F}^{-1})^{ b}{}_{\tilde b|\tilde C}(\accentset{\Xi}{F}^{-1})^{ A}{}_{\tilde A|\tilde B}
	+(\accentset{\Xi}{F}^{-1})^{ b}{}_{\tilde b|\tilde B}(\accentset{\Xi}{F}^{-1})^{ A}{}_{\tilde C|\tilde A}
	+(\accentset{\Xi}{F}^{-1})^{ b}{}_{\tilde b|\tilde C|\tilde B}(\accentset{\Xi}{F}^{-1})^{ A}{}_{\tilde A}\\&
	+(\accentset{\Xi}{F}^{-1})^{ b}{}_{\tilde b|\tilde A|\tilde B}(\accentset{\Xi}{F}^{-1})^{ A}{}_{\tilde C}
	+(\accentset{\Xi}{F}^{-1})^{ b}{}_{\tilde b|\tilde C|\tilde A}(\accentset{\Xi}{F}^{-1})^{ A}{}_{\tilde B}
	+(\accentset{\Xi}{F}^{-1})^{ A}{}_{\tilde C|\tilde A|\tilde B}(\accentset{\Xi}{F}^{-1})^{ b}{}_{\tilde b}
	\Big]\bigg\}_{|A}=0\,,
\end{split}
\end{equation}
and, with an abuse of notation
\begin{equation}\label{key-symm}
\begin{split}
&(\mathring{F}^{-1})^B{}_a\mathbb{C}^{aC[bA]}{}_{|C|B}-k(\mathring{\tilde F}^{-1})^{\tilde{B}}{}_{\tilde a}\tilde{\mathbb{C}}^{\tilde{a}\tilde{A}\tilde{b}\tilde{C}}\Big[(\accentset{\Xi}{F}^{-1})^{ [b}{}_{\tilde b|\tilde A}(\accentset{\Xi}{F}^{-1})^{ A]}{}_{\tilde C|\tilde B}\\&+(\accentset{\Xi}{F}^{-1})^{ [b}{}_{\tilde b|\tilde C}(\accentset{\Xi}{F}^{-1})^{ A]}{}_{\tilde A|\tilde B}+(\accentset{\Xi}{F}^{-1})^{ [b}{}_{\tilde b|\tilde B}(\accentset{\Xi}{F}^{-1})^{ A]}{}_{\tilde C|\tilde A}+(\accentset{\Xi}{F}^{-1})^{ [b}{}_{\tilde b|\tilde C|\tilde B}(\accentset{\Xi}{F}^{-1})^{ A]}{}_{\tilde A}\\&+(\accentset{\Xi}{F}^{-1})^{ [b}{}_{\tilde b|\tilde A|\tilde B}(\accentset{\Xi}{F}^{-1})^{ A]}{}_{\tilde C}+(\accentset{\Xi}{F}^{-1})^{ [b}{}_{\tilde b|\tilde C|\tilde A}(\accentset{\Xi}{F}^{-1})^{ A]}{}_{\tilde B}+(\accentset{\Xi}{F}^{-1})^{ [b}{}_{\tilde b}(\accentset{\Xi}{F}^{-1})^{ A]}{}_{\tilde C|\tilde A|\tilde B}\Big]=0\,.
\end{split}
\end{equation}
Note that
\begin{equation}
\begin{split}
(\accentset{\Xi}{F}^{-1})^{ A}{}_{\tilde A|\tilde{B}}=&\frac{\partial}{\partial\tilde{X}^{\tilde B}}\left[(\accentset{\Xi}{F}^{-1})^{ A}{}_{\tilde A}\right]+\Gamma^A{}_{CB}(\accentset{\Xi}{F}^{-1})^{ B}{}_{\tilde B}(\accentset{\Xi}{F}^{-1})^{ C}{}_{\tilde A}-\tilde{\Gamma}^{\tilde C}{}_{\tilde A\tilde B}(\accentset{\Xi}{F}^{-1})^{ A}{}_{\tilde C}\,,\\
(\accentset{\Xi}{F}^{-1})^{ A}{}_{\tilde A|\tilde{B}|\tilde{C}}=&\frac{\partial}{\partial\tilde{X}^{\tilde C}}\left[(\accentset{\Xi}{F}^{-1})^{ A}{}_{\tilde A|\tilde B}\right]+\Gamma^A{}_{CB}(\accentset{\Xi}{F}^{-1})^{ B}{}_{\tilde C}(\accentset{\Xi}{F}^{-1})^{ C}{}_{\tilde A|\tilde B}\\&-\tilde{\Gamma}^{\tilde D}{}_{\tilde A\tilde C}(\accentset{\Xi}{F}^{-1})^{ A}{}_{\tilde D|\tilde B}-\tilde{\Gamma}^{\tilde D}{}_{\tilde B\tilde C}(\accentset{\Xi}{F}^{-1})^{ A}{}_{\tilde A|\tilde D}\,,\\
(\accentset{\Xi}{F}^{-1})^{ A}{}_{\tilde A|\tilde{B}|\tilde{C}|\tilde{D}}=&\frac{\partial}{\partial\tilde{X}^{\tilde D}}\left[(\accentset{\Xi}{F}^{-1})^{ A}{}_{\tilde A|\tilde B|\tilde C}\right]+\Gamma^A{}_{EB}(\accentset{\Xi}{F}^{-1})^{ B}{}_{\tilde D}(\accentset{\Xi}{F}^{-1})^{ E}{}_{\tilde A|\tilde B|\tilde C}\\&-\tilde{\Gamma}^{\tilde E}{}_{\tilde A\tilde D}(\accentset{\Xi}{F}^{-1})^{ A}{}_{\tilde E|\tilde B|\tilde C}-\tilde{\Gamma}^{\tilde E}{}_{\tilde B\tilde D}(\accentset{\Xi}{F}^{-1})^{ A}{}_{\tilde A|\tilde E|\tilde C}-\tilde{\Gamma}^{\tilde E}{}_{\tilde C\tilde D}(\accentset{\Xi}{F}^{-1})^{ A}{}_{\tilde A|\tilde B|\tilde E}\,,
\end{split}
\end{equation}
where $\Gamma^A{}_{BC}$ and $\tilde{\Gamma}^{\tilde A}{}_{\tilde B \tilde C}$ are, respectively, the Christoffel symbols associated with the induced connections on the physical and virtual plates. Also\footnote{
The covariant derivative of a two-point tensor $\mathbf{T}$ is given by
\begin{equation*}
\begin{aligned}
T^{AB\cdots F}{}_{G\cdots Q}{}^{ab\cdots f}{}_{g\cdots q|K}=&\frac{\partial}{\partial X^K}T^{AB\cdots F}{}_{G\cdots Q}{}^{ab\cdots f}{}_{g\cdots q}\\&+T^{RB\cdots F}{}_{G\cdots Q}{}^{ab\cdots f}{}_{g\cdots q}\Gamma^A{}_{RK}+\mathrm{(all\,\,upper\,\, referential\,\, indices)}\\&-T^{AB\cdots F}{}_{R\cdots Q}{}^{ab\cdots f}{}_{g\cdots q}\Gamma^R{}_{GK}-\mathrm{(all\,\,lower\,\, referential\,\, indices)}\\&+T^{AB\cdots F}{}_{G\cdots Q}{}^{lb\cdots f}{}_{g\cdots q}\gamma^a{}_{lr}F^r{}_K+\mathrm{(all\,\,upper\,\, spatial\,\, indices)}\\&-T^{AB\cdots F}{}_{G\cdots Q}{}^{ab\cdots f}{}_{l\cdots q}\gamma^l{}_{gr}F^r{}_K-\mathrm{(all\,\,lower\,\, spatial\,\, indices)}\,.
\end{aligned}
\end{equation*}
}
\begin{equation}
\begin{split}
\mathbb{C}^{aAbB}{}_{|C}=&\frac{\partial}{\partial X^C}\left[\mathbb{C}^{aAbB}\right]+\mathbb{C}^{kAbB}\gamma^a{}_{kl}\mathring{F}^l{}_C+\mathbb{C}^{aAkB}\gamma^b{}_{kl}\mathring{F}^l{}_C\\&+\mathbb{C}^{aKbB}\Gamma^A{}_{KC}+\mathbb{C}^{aAbK}\Gamma^B{}_{KC}\,,\\
\mathbb{C}^{aAbB}{}_{|C|D}=&\frac{\partial}{\partial X^D}\left[\mathbb{C}^{aAbB}{}_{|C}\right]+\mathbb{C}^{kAbB}{}_{|C}\gamma^a{}_{kl}\mathring{F}^l{}_D+\mathbb{C}^{aAkB}{}_{|C}\gamma^b{}_{kl}\mathring{F}^l{}_D\\&+\mathbb{C}^{aKbB}{}_{|C}\Gamma^A{}_{KD}+\mathbb{C}^{aAbK}{}_{|C}\Gamma^B{}_{KD}-\mathbb{C}^{aAbB}{}_{|K}\Gamma^K{}_{CD}\,.
\end{split}
\end{equation}
Using \eqref{key-trans-C} and \eqref{disp-grad}, the couple-stress is transformed as
\begin{equation}
\begin{split}
	\delta\mathsf{M}^{aA}&=\frac{1}{2}\mathbb{C}^{aAbB}( \mathsf{W}_{,b})_{|B}\\&
	=\frac{1}{2}k\tilde{\mathbb{C}}^{\tilde{a}\tilde{A}\tilde{b}\tilde{B}}(\accentset{\Xi}{F}^{-1})^{ b}{}_{\tilde b}
	(\accentset{\Xi}{F}^{-1})^{ B}{}_{\tilde B}(\accentset{\Xi}{F}^{-1})^{ A}{}_{\tilde A}
	(\accentset{\Xi}{F}^{-1})^{ a}{}_{\tilde a}\bigg[\accentset{\Xi}{F}^{ \tilde{c}}{}_{ b| B}
	\,{\widetilde{\mathsf{W}}}{}_{, \tilde c}+\accentset{\Xi}{F}^{ \tilde c}{}_{ b}\accentset{\Xi}{F}^{ \tilde C}{}_{ B}\,
	({\widetilde{\mathsf{W}}}{}_{, \tilde c})_{|{\tilde C}}\bigg]\\&=\frac{1}{2}k\tilde{\mathbb{C}}^{\tilde{a}\tilde{A}
	\tilde{b}\tilde{B}}(\accentset{\Xi}{F}^{-1})^{ b}{}_{\tilde b}(\accentset{\Xi}{F}^{-1})^{ B}{}_{\tilde B}
	(\accentset{\Xi}{F}^{-1})^{ A}{}_{\tilde A}(\accentset{\Xi}{F}^{-1})^{ a}{}_{\tilde a}
	\accentset{\Xi}{F}^{ \tilde{c}}{}_{ b| B}\,{\widetilde{\mathsf{W}}}{}_{, \tilde c}
	+k(\accentset{\Xi}{F}^{-1})^{ A}{}_{\tilde A}(\accentset{\Xi}{F}^{-1})^{ a}{}_{\tilde a}
	\delta\tilde{\mathsf{M}}^{\tilde{a}\tilde{A}}\,.
\end{split}
\end{equation}
From \eqref{key-bound-2}, $\widetilde{\mathsf{W}}_{,\tilde{A}}=\mathsf{W}_{,A}(\mathsf{S}^{-1})^A{}_{\tilde A}$, on $\partial_o\mathcal{C}$, whence, together with \eqref{disp-grad}, it follows that $T\Xi|_{\partial_o\mathcal{C}}=\accentset{\Xi}{\mathbf F}|_{\partial_o\mathcal{C}}=id$. Moreover, \eqref{key-bound-2} also requires that $(\mathring{\tilde{F}}^{-1})^{\tilde A}{}_{\tilde a}\delta\tilde{\mathsf{M}}^{\tilde a\tilde B}{}_{|\tilde{B}}\tilde{\mathsf T}_{\tilde A}=(\mathring{F}^{-1})^A{}_a\delta\mathsf{M}^{aB}{}_{|B}\mathsf{T}_A$, and $\delta\tilde{\mathsf M}^{\tilde a\tilde A}\tilde{\mathsf T}_{\tilde A}=\mathsf{s}^{\tilde a}{}_a\delta\mathsf{M}^{aA}\mathsf{T}_A$, on $\partial_o\mathcal{C}$, which imply that $\accentset{\Xi}{F}^{\tilde A}{}_{ A|B}\Big|_{\partial_o\mathcal{C}}=0$, $\accentset{\Xi}{F}^{\tilde A}{}_{ A|B|C}\Big|_{\partial_o\mathcal{C}}=0$, $k|_{\partial_o\mathcal{C}}=1$, and $k_{,A}|_{\partial_o\mathcal{C}}=0$. Similarly, given that the virtual plate is not pre-stressed ($\mathring{\tilde{\mathbf{P}}}=\mathbf{0}$), \eqref{key-bound-cond} implies that the initial traction must vanish on the outer boundary of the cloak, i.e., $(\mathring{{T}}^\top)^a|_{\partial_o\mathcal{C}}=\mathring{P}^{aA}\mathsf{T}_A\big|_{\partial_o\mathcal{C}}={0}$.

Knowing that the hole surface in the virtual plate is traction-free (cf. \eqref{key-hole-virtual}), i.e., $(\mathring{\tilde{F}}^{-1})^{\tilde A}{}_{\tilde a}\delta\tilde{\mathsf{M}}^{\tilde a\tilde B}{}_{|\tilde{B}}\tilde{\mathsf T}_{\tilde A}=0$, and $\delta\tilde{\mathsf M}^{\tilde a\tilde A}\tilde{\mathsf T}_{\tilde A}=0$, on $\partial\tilde{\mathcal{E}}$, if $k_{,A}|_{\partial\mathcal{E}}=0$, $\accentset{\Xi}{F}^{\tilde A}{}_{A|B}\Big|_{\partial\mathcal{E}}=0$, and $\accentset{\Xi}{F}^{\tilde A}{}_{A|B|C}\Big|_{\partial\mathcal{E}}=0$, then the hole inner surface $\partial\mathcal{E}$ will be traction-free in the physical plate as well. Note that \eqref{key-shell-hole} requires that the initial traction vanish on the boundary of the physical hole, viz., $(\mathring{{T}}^\top)^a|_{\partial\mathcal{E}}=\mathring{P}^{aA}\mathsf{T}_A\big|_{\partial\mathcal{E}}={0}$. 


\begin{remark}
It is important to note that contrary to  transformation cloaking in $3$D elasticity, where stress and couple-stress are transformed using the Piola transformation under a cloaking map, for Kirchhoff-Love plates, the couple-stress is not 
transformed via the Piola transformation. However, this should not be surprising, as in $3$D ealsticity, divergence of stress (and couple-stress) appear in the balance of linear momentum, and one uses the Piola transformation in order to preserve the divergence terms (and thus, the governing equations) up to the Jacobian of the cloaking map.   
\end{remark}
\begin{remark}
The mass form is not necessarily preserved under the cloaking transformation $\Xi$ in Kirchhoff-Love plates (unlike transformation cloaking in $3$D elasticity, where the mass form is preserved under $\Xi$, see \citep[Remark 6]{yavari2018nonlinear}). To see this let us denote the virtual and physical mass forms by $\tilde{\boldsymbol{\mathsf{m}}}=\tilde{\rho}d\tilde{A}_{\tilde{\mathbf G}}$, and $\boldsymbol{\mathsf{m}}=\rho dA_{\mathbf G}$, respectively. Therefore, one obtains  
\begin{equation}
\Xi^*\tilde{\boldsymbol{\mathsf{m}}}=\Xi^*(\tilde{\rho}d\tilde{A}_{\tilde{\mathbf G}})=(\tilde{\rho}\circ\Xi)\Xi^*d\tilde{A}_{\tilde{\mathbf G}}=(\tilde{\rho}\circ\Xi)J_{\Xi}dA_{\mathbf G}=\frac{J_{\Xi}}{k}\rho dA_{\mathbf G}=\frac{J_{\Xi}}{k}\boldsymbol{\mathsf{m}}\,.
\end{equation}
\end{remark}

\begin{remark}\label{key-pp}
It is straightforward to see that when $(\accentset{\Xi}{F}^{-1})^A{}_{\tilde A|\tilde B}=0$ (i.e., when $\accentset{\Xi}{\mathbf F}^{-1}$ is covariantly constant) and $k_{,A}=0$, one has $\mathring{\mathbf{P}}=\mathbf{0}$ and $\mathring{\boldsymbol{\mathfrak B}}^\top=\mathbf{0}$, and thus, the cloaking compatibility equations  \eqref{key-constraint} and the balance equations in the finitely-deformed configuration \eqref{key-finite-linear} and \eqref{key-symm} are trivially satisfied. However, if one uses Cartesian coordinates $\{Z^I\}$ and $\{\tilde{Z}^{\tilde I}\}$ for $\mathcal{H}$ and $\tilde{\mathcal H}$, respectively, $\accentset{\Xi}{\mathbf F}$ would have constant components if it is covariantly constant. Knowing that on the outer boundary of the cloak one needs to have $T\Xi|_{\partial_o\mathcal{C}}=\accentset{\Xi}{\mathbf F}|_{\partial_o\mathcal{C}}=id$, it follows that $\accentset{\Xi}{\mathbf F}$ is the identity everywhere (so is the cloaking map $\Xi$), and thus, cloaking is not possible if one assumes that the tangent map of the cloaking transformation is covariantly constant.\footnote{\citet{pomot2019form} used a linear cloaking transformation, which has a covarianlty constant tangent map. However, a linear cloaking map does not satisfy the required traction boundary condition on $\partial_o\mathcal{C}$, i.e., $T\Xi|_{\partial_o\mathcal{C}}=\accentset{\Xi}{\mathbf F}|_{\partial_o\mathcal{C}}=id$, and therefore, using a linear cloaking map is not acceptable (see \citep{Brun2009} for another improper use of this type of mapping).}	
\end{remark}

\subsubsection{A circular cloak in a Kirchhoff-Love plate} 

Consider a circular hole $\mathcal E$ in the physical plate in its reference configuration with radius $R_i$ that needs to be cloaked from the out-of-plane excitations using an annular cloak having inner and outer radii $R_i$ and $R_o$, respectively. Let us map the reference configuration to the reference configuration of the virtual plate via a cloaking map, $\Xi:\mathcal{H}\to\tilde{\mathcal H}$, where for $R_i\leq R\leq R_o$, it is defined as, $(\tilde R,\tilde \Theta)=\Xi(R,\Theta)=(f(R),\Theta)$ such that $f(R_o)=R_o$ and $f(R_i)=\epsilon$, and for $R\geq R_o$ is the identity map. The physical and virtual plates are endowed with the Euclidean metrics $\mathbf{G}=\mathrm{diag}(1,R^2)$, and $\tilde{\mathbf{G}}=\mathrm{diag}(1,\tilde{R}^2)$ in the polar coordinates, respectively. Thus
\begin{equation}
\accentset{\Xi}{\mathbf F}=\begin{bmatrix}
f'(R) & 0 \\
0 & 1     
\end{bmatrix}\,.
\end{equation}
From \eqref{key-vir-ct}, the flexural rigidity of the virtual plate is given by\footnote{Note that the physical components of the flexural rigidity tensor are given by $\hat{\tilde{\mathbb{C}}}^{\tilde{a}\tilde{A}\tilde{b}\tilde{B}}=\sqrt{\tilde{g}_{\tilde a\tilde a}}\sqrt{\tilde{G}_{\tilde A\tilde A}}\sqrt{\tilde{g}_{\tilde b\tilde b}}\sqrt{\tilde{G}_{\tilde B\tilde B}}\tilde{\mathbb{C}}^{\tilde a\tilde A\tilde b\tilde B}$ (no summation).}
\begin{equation}\label{key-vir-rig}
\hat{\tilde{\pmb{\mathbb{C}}}}=\left[\hat{\tilde{\mathbb{C}}}^{\tilde{a}\tilde{A}\tilde{b}\tilde{B}}\right]=\frac{Eh^3}{12(1+\nu)}\begin{bmatrix}
\begin{bmatrix}
\frac{2 }{1-\nu} & 0 \\
0 & \frac{2\nu }{1-\nu}  \\
\end{bmatrix} & \begin{bmatrix}
0 & 1 \\
1  & 0 \\
\end{bmatrix} \\
\begin{bmatrix}
0 & 1  \\
1 & 0 \\
\end{bmatrix} & \begin{bmatrix}
\frac{2\nu }{1-\nu}  & 0 \\
0 & \frac{2 }{1-\nu} \\
\end{bmatrix} \\
\end{bmatrix}\,,
\end{equation}
where the first two indices identify the submatrix and the last two specify the components of that submatrix. The (surface) mass density of the physical plate is given by $\rho=k(R)\tilde{\rho}$.
Using \eqref{key-trans-C}, the flexural rigidity of the cloak are determined up to the scalar $k(R)$ as follows
\begin{equation}
\hat{{\pmb{\mathbb{C}}}}=\left[\hat{{\mathbb{C}}}^{{a}{A}{b}{B}}\right]=\frac{Eh^3k(R)}{12(1+\nu)}\begin{bmatrix}
\begin{bmatrix}
\frac{2 }{1-\nu}\frac{1}{{f'}^4(R)} & 0 \\
0 & \frac{2\nu }{1-\nu}\frac{R^2}{f^2(R){f'}^2(R)}  \\
\end{bmatrix} & \begin{bmatrix}
0 & \frac{R^2}{f^2(R){f'}^2(R)} \\
\frac{R^2}{f^2(R){f'}^2(R)}  & 0 \\
\end{bmatrix} \\
\begin{bmatrix}
0 & \frac{R^2}{f^2(R){f'}^2(R)}  \\
\frac{R^2}{f^2(R){f'}^2(R)} & 0 \\
\end{bmatrix} & \begin{bmatrix}
\frac{2\nu }{1-\nu}\frac{R^2}{f^2(R){f'}^2(R)}  & 0 \\
0 & \frac{2 }{1-\nu}\frac{R^4}{f^4(R)} \\
\end{bmatrix} \\
\end{bmatrix}\,.
\end{equation}
From \eqref{key-ini-body}, the circumferential component of the tangential body force vanishes, i.e., $(\mathring{{\mathfrak B}}^\top)^\theta=0$, and its radial component reads  
\begin{equation}
\begin{split}
(\mathring{{\mathfrak B}}^\top)^r=&\frac{E h^3 }{12\tilde{\rho} \left(\nu ^2-1\right) f^4(R) {f'}^7(R)} \Big[3 R {f'}^7(R)-3 f(R) {f'}^6(R)-3 f^2(R) {f'}^4(R) f''(R)\\&+2 f^3(R) {f'}^2(R)
	\left(f^{(3)}(R) f'(R)-3 {f''}^2(R)\right)\\&+f^4(R) \left(15 {f''}^3(R)+f^{(4)}(R) {f'}^2(R)-10 f^{(3)}(R)
	f'(R) f''(R)\right)\Big]\,.
\end{split}	
\end{equation}
The initial stress is given by (cf. \eqref{key-ini-stress})
\begin{equation}
\begin{split}
\hat{\mathring{P}}^{r\Theta}=&\hat{\mathring{P}}^{\theta R}=0\,,\\
\hat{\mathring{P}}^{rR}=&\frac{E h^3}{12 \left(\nu ^2-1\right) R f^4(R) {f'}^6(R)} \bigg[R^3 k(R) {f'}^6(R)+2 (\nu -1) R^2 f(R) k(R) {f'}^5(R)\\&+R
	f^2(R) {f'}^3(R) \left[2 (\nu -1) R k(R) f''(R)+f'(R) \left\{(1-2 \nu ) k(R)-(\nu
	-2) R k'(R)\right\}\right]\\&-f^4(R) \left[5 R k(R) {f''}^2(R)+{f'}^2(R) \left(R
	k''(R)+2 k'(R)\right)-8 f'(R) f''(R) \left(R k'(R)+k(R)\right)\right]\\&-6 R f^3(R)
	k(R) {f'}^2(R) f''(R)\bigg]\,,\\
\hat{\mathring{P}}^{\theta\Theta}=&\frac{E h^3}{12
	\left(1-\nu ^2\right) f^5(R) {f'}^4(R)} \bigg[-R^2 f(R) {f'}^4(R) \left(R k'(R)-6 (\nu -1) k(R)\right)+4 R^3 k(R) {f'}^5(R)\\&-2 \nu 
	R f^2(R) {f'}^2(R) \left[2 f'(R) \left(R k'(R)+2 k(R)\right)-3 R k(R) f''(R)\right]\\&+f^3(R) \Big(R f'(R)
	\left[f'(R) \left(\nu  R k''(R)+(4 \nu +2) k'(R)\right)-4 \nu  R f''(R) k'(R)\right]\\&+2 k(R) \left[3 \nu
	R^2 {f''}^2(R)+(\nu +1) {f'}^2(R)-\nu  R f'(R) \left\{R f^{(3)}(R)+4 f''(R)\right\}\right]\Big)\bigg]\,.\\
\end{split}
\end{equation}
Note that $\mathring{\mathbf{P}}$ is diagonal, and thus, \eqref{key-symm} is already satisfied. The constraint \eqref{key-constraint} gives the following two ODEs:
\begin{subequations}
\begin{equation}\label{key-1}
f''(R)=f'(R)\left[\frac{1}{R}-\frac{f'(R)}{f(R)}+\frac{k'(R)}{k(R)}\right]\,,
\end{equation}
\begin{equation}\label{key-2}
\begin{split}
&(1-2 \nu ) R f^2(R) k(R) f''(R)\\&+f'(R) \left[k(R)
\left(f(R)-R f'(R)\right) \left(R f'(R)+2 (\nu -1)
f(R)\right)+(\nu -1) R f^2(R) k'(R)\right]=0\,.
\end{split}
\end{equation}
\end{subequations}
Using \eqref{key-1} and \eqref{key-2}, one obtains the following second-order nonlinear ODE for $f(R)$:
\begin{equation}\label{key-3}
f'(R) \left[f(R)-R f'(R)\right] \left[R f'(R)+(\nu -1)
f(R)\right]-\nu  R f^2(R) f''(R)=0\,.
\end{equation}
It is interesting to observe that the differential equation governing the gradient of the cloaking map involves the Poisson's ratio of the virtual plate. Note that for $\nu=0$, from \eqref{key-3}, the cloaking map is forced to be the identity. If $\nu\neq0$, then \eqref{key-1} and \eqref{key-2} imply that
\begin{equation}\label{key-1-key}
\frac{k'(R)}{k(R)}=-\frac{ \left(f(R)-R f'(R)\right)^2}{\nu  R f^2(R)}\,,
\end{equation}
and \eqref{key-3} can be rewritten as
\begin{equation}\label{key-2-key}
f''(R)=\frac{f'(R)}{\nu  R f^2(R)} \left[f(R)-R f'(R)\right] \left[R f'(R)+(\nu -1) f(R)\right]\,.
\end{equation}
Note, however, that \eqref{key-3} is a second-order ODE, and hence, one cannot enforce the required boundary conditions $f(R_o)=R_o$, $f(R_i)=\epsilon$, and $f'(R_o)=1$ (i.e., $\accentset{\Xi}{\mathbf F}|_{\partial_o\mathcal{C}}=id$) simultaneously. Therefore, cloaking is not possible.
The finite (in-plane) balance of linear momentum \eqref{key-finite-linear}, i.e., $\mathring{P}^{aA}{}_{|A}+\rho (\mathring{\mathfrak{B}}^\top)^a=0$, is simplified to read
\begin{equation}\label{key-satis}
\begin{split}
	&6 R^2 k(R) f'(R)^7-3 R f(R) f'(R)^5 \left[f'(R) \left(2 R k'(R)+3 k(R)\right)-2 R k(R) f''(R)\right]\\
	&-2f(R)^3 f'(R)^2 \left[2 R f^{(3)}(R) k(R) f'(R)+3 f''(R) \left(f'(R) 
	\left(R k'(R)+k(R)\right)-3 R k(R)f''(R)\right)\right]\\
	&+f(R)^2 f'(R)^3 \Big\{R f'(R) \left[f'(R) \left(2 R k''(R)+7 k'(R)\right)-6 Rf''(R) k'(R)\right]\\
	&+k(R) \left[6 R^2 f''(R)^2+3 f'(R)^2-R f'(R) \left(2 R f^{(3)}(R)+3f''(R)\right)\right]\Big\}\\
	&+f(R)^4 \bigg\{45 R k(R) f''(R)^3+f'(R)^3 
	\left[-\left(R k^{(3)}(R)+3k''(R)\right)\right]\\
	&-5 f'(R) f''(R) \left[4 R f^{(3)}(R) k(R)+9 f''(R) \left(Rk'(R)+k(R)\right)\right]\\
	&+f'(R)^2 \Big[R f^{(4)}(R) k(R)+8 f^{(3)}(R) k(R)+12 R f''(R) k''(R)\\
	&+8 \left(Rf^{(3)}(R)+3 f''(R)\right) k'(R)\Big]\bigg\}=0\,.
\end{split}
\end{equation}
One can recursively use \eqref{key-1-key} and \eqref{key-2-key} to express $k''(R)$, $k^{(3)}(R)$, $f^{(3)}(R)$, and $f^{(4)}(R)$ in terms of $f'(R)$, $f(R)$, $k'(R)$, and $k(R)$. Plugging these expressions into \eqref{key-satis}, one can verify that \eqref{key-satis} holds. Therefore, the satisfaction of the balance of linear and angular momenta in the physical plate in its finitely deformed configuration (i.e., \eqref{key-finite-linear} and \eqref{key-symm}) does not impose any additional restriction on the cloaking map; the cloaking compatibility equations \eqref{key-constraint} is the only constraint on $\Xi$.
\begin{remark}\label{key-biham}
For the isotropic and homogeneous flexural rigidity \eqref{key-vir-rig} and a flat ambient space, the governing equation of the virtual plate \eqref{virtual-plate} is expanded and is written in the form of the following biharmonic equation
\begin{equation}\label{key-wave2}
	-D^{(0)}\widetilde{\nabla}^4\widetilde{\mathsf{W}}=\tilde{\rho}\,\ddot{\widetilde{\mathsf{W}}}\,.
\end{equation}
We first show that the simplified governing equation \eqref{key-wave2} does not correspond to a unique flexural rigidity tensor for an isotropic and homogeneous plate, and thus, that of the cloak when the cloaking map is the identity. Without loss of generality, we use Cartesian coordinates, for which \eqref{virtual-plate} is simplified to read\footnote{Note that
\begin{equation*}
\widetilde{\mathsf W}_{\tilde A\tilde B\tilde C\tilde D}=\frac{\partial^4\widetilde{\mathsf W}}{\partial\tilde{X}^{\tilde A}\partial\tilde{X}^{\tilde B}\partial\tilde{X}^{\tilde C}\partial\tilde{X}^{\tilde D}}\,.
\end{equation*}
}
\begin{equation}
\begin{split}
	-\frac{1}{2}&\bigg[\tilde{\mathbb C}^{\tilde X\tilde X\tilde X\tilde X}\widetilde{\mathsf W}_{\tilde X\tilde X\tilde X\tilde X}+\tilde{\mathbb C}^{\tilde Y\tilde Y\tilde Y\tilde Y}\widetilde{\mathsf W}_{\tilde Y\tilde Y\tilde Y\tilde Y}+4\tilde{\mathbb C}^{\tilde X\tilde X\tilde X\tilde Y}\widetilde{\mathsf W}_{\tilde X\tilde X\tilde X\tilde Y}\\&+4\tilde{\mathbb C}^{\tilde X\tilde Y\tilde X\tilde Y}\widetilde{\mathsf W}_{\tilde X\tilde X\tilde Y\tilde Y}+4\tilde{\mathbb C}^{\tilde X\tilde Y\tilde Y\tilde Y}\widetilde{\mathsf W}_{\tilde X\tilde Y\tilde Y\tilde Y}+2\tilde{\mathbb C}^{\tilde X\tilde X\tilde Y\tilde Y}\widetilde{\mathsf W}_{\tilde X\tilde X\tilde Y\tilde Y}\bigg]=\tilde{\rho}\, \ddot{\widetilde{\mathsf W} }\,,
\end{split}
\end{equation}
where the minor and major symmetries of the flexural rigidity tensor were used. From \eqref{key-wave2} we have
\begin{equation}
	-D^{(0)}\bigg[\widetilde{\mathsf W}_{\tilde X\tilde X\tilde X\tilde X}+\widetilde{\mathsf W}_{\tilde Y\tilde Y\tilde Y\tilde Y}+2\widetilde{\mathsf W}_{\tilde X\tilde X\tilde Y\tilde Y}\bigg]=\tilde{\rho}\, \ddot{\widetilde{\mathsf W}}\,.
\end{equation}
Comparing the coefficients of different derivatives, one obtains
\begin{equation}\label{key-non-uni}
\begin{aligned}
	& \tilde{\mathbb C}^{\tilde X\tilde X\tilde X\tilde X}=2D^{(0)}\,,\quad
	\tilde{\mathbb C}^{\tilde Y\tilde Y\tilde Y\tilde Y}=2D^{(0)}\,, \\
	& \tilde{\mathbb C}^{\tilde X\tilde X\tilde X\tilde Y}=0\,,\quad
	\tilde{\mathbb C}^{\tilde X\tilde Y\tilde Y\tilde Y}=0\,,\quad 
	2\tilde{\mathbb C}^{\tilde X\tilde Y\tilde X\tilde Y}+\tilde{\mathbb C}^{\tilde X\tilde X\tilde Y\tilde Y}=2D^{(0)}\,.
\end{aligned}
\end{equation}
Therefore, there are infinitely many choices for $\tilde{\pmb{\mathbb C}}$; one cannot uniquely determine the flexural rigidity tensor starting from the simplified governing equation \eqref{key-wave2} and comparing it with the initial governing equation in its tensorial form \eqref{virtual-plate}.
We next transform the biharmonic equation \eqref{key-wave2} under a cloaking map. In Cartesian coordinates 
\begin{equation}
	\accentset{\Xi}{\mathbf F}^{-1}(X,Y)=\begin{bmatrix}
	\mathsf{F}^{X}{}_{\tilde{X}} & \mathsf{F}^{X}{}_{\tilde{Y}} \\
	\mathsf{F}^{Y}{}_{\tilde{X}} & \mathsf{F}^{Y}{}_{\tilde{Y}}     
	\end{bmatrix}\,.
\end{equation}
Knowing that $\mathsf{W}=\widetilde{\mathsf W}\circ\Xi^{-1}$, using the chain rule, one finds
\begin{equation}\label{key-3tr}
\begin{split}
	\frac{\partial\widetilde{\mathsf W}}{\partial \tilde{X}}=\frac{\partial \mathsf{W}}{\partial X}\mathsf{F}^{X}{}_{\tilde{X}}+\frac{\partial\mathsf{W}}{\partial Y}\mathsf{F}^{Y}{}_{\tilde{X}}\,,\\
	\frac{\partial\widetilde{\mathsf W}}{\partial\tilde{Y}}=\frac{\partial \mathsf{W}}{\partial X}\mathsf{F}^{X}{}_{\tilde{Y}}+\frac{\partial\mathsf{W}}{\partial Y}\mathsf{F}^{Y}{}_{\tilde{Y}}\,.
\end{split}
\end{equation}
One may recursively use	\eqref{key-3tr} to find the transformed higher order derivatives of $\widetilde{\mathsf W}$, and eventually, obtain the transformation of the biharmonic term $\widetilde{\nabla}^4\widetilde{\mathsf W}=\frac{\partial^4\widetilde{\mathsf W}}{\partial\tilde{X}^4}+\frac{\partial^4\widetilde{\mathsf W}}{\partial\tilde{Y}^4}+2\frac{\partial^4\widetilde{\mathsf W}}{\partial\tilde{X}^2\partial\tilde{Y}^2}$.\footnote{In a general curvilinear coordinate system, the biharmonic term is given by
\begin{equation*}
\widetilde{\nabla}^4\widetilde{\mathsf W}=\frac{1}{\sqrt{\det\tilde{\mathbf{G}}}}\frac{\partial}{\partial \tilde{X}^{\tilde A}}\left[\sqrt{\det\tilde{\mathbf{G}}}\frac{\partial}{\partial\tilde{X}^{\tilde B}}\left(\frac{1}{\sqrt{\det\tilde{\mathbf{G}}}}\frac{\partial}{\partial \tilde{X}^{\tilde C}}\left[\sqrt{\det\tilde{\mathbf{G}}}\frac{\partial\widetilde{\mathsf W}}{\partial\tilde{X}^{\tilde D}}\tilde{G}^{\tilde C\tilde D}\right]\right)\tilde{G}^{\tilde A\tilde B}\right]\,.
\end{equation*}
} Comparing the coefficients of the different derivatives in the transformed biharmonic equation with those in the governing equation of the physical plate \eqref{virtual-lin-pre}, one determines the unknown fields. Comparing the fourth-order derivatives one finds
\begin{equation}\label{key-consistent}
\begin{split}
	\mathbb{C}^{XXXX}=&2D^{(0)}\left[\left(\mathsf{F}^{X}{}_{\tilde{X}}\right)^2
	+\left(\mathsf{F}^{X}{}_{\tilde{Y}}\right)^2\right]^2\,,\quad\mathbb{C}^{YYYY}
	=2D^{(0)}\left[\left(\mathsf{F}^{Y}{}_{\tilde{X}}\right)^2+\left(\mathsf{F}^{Y}{}_{\tilde{Y}}\right)^2\right]^2\,,\\
	\mathbb{C}^{XYYY}=&2D^{(0)}\left[\mathsf{F}^{X}{}_{\tilde{X}}\mathsf{F}^{Y}{}_{\tilde{X}}
	+\mathsf{F}^{X}{}_{\tilde{Y}}\mathsf{F}^{Y}{}_{\tilde{Y}}\right]\left[\left(\mathsf{F}^{Y}{}_{\tilde{X}}\right)^2
	+\left(\mathsf{F}^{Y}{}_{\tilde{Y}}\right)^2\right]\,,\\
	\mathbb{C}^{YXXX}=&2D^{(0)}\left[\mathsf{F}^{X}{}_{\tilde{X}}\mathsf{F}^{Y}{}_{\tilde{X}}
	+\mathsf{F}^{X}{}_{\tilde{Y}}\mathsf{F}^{Y}{}_{\tilde{Y}}\right]\left[\left(\mathsf{F}^{X}{}_{\tilde{X}}\right)^2
	+\left(\mathsf{F}^{X}{}_{\tilde{Y}}\right)^2\right]\,,\\
	2\mathbb{C}^{XYXY}+\mathbb{C}^{XXYY}=&2D^{(0)}
	\bigg\{4\mathsf{F}^{X}{}_{\tilde{X}}\mathsf{F}^{X}{}_{\tilde{Y}}\mathsf{F}^{Y}{}_{\tilde{X}}\mathsf{F}^{Y}{}_{\tilde{Y}}
	+\left(\mathsf{F}^{X}{}_{\tilde{X}}\right)^2\left[3\left(\mathsf{F}^{Y}{}_{\tilde{X}}\right)^2
	+\left(\mathsf{F}^{Y}{}_{\tilde{Y}}\right)^2\right]\\
	&+\left(\mathsf{F}^{X}{}_{\tilde{Y}}\right)^2
	\left[3\left(\mathsf{F}^{Y}{}_{\tilde{Y}}\right)^2+\left(\mathsf{F}^{Y}{}_{\tilde{X}}\right)^2\right]\bigg\}\,.
\end{split}
\end{equation}
In particular, we note that transforming the biharmonic equation under the cloaking map does not fully determine the flexural rigidity tensor of the cloak. This should not be surprising as the biharmonic equation \eqref{key-wave2} does not correspond to a unique flexural rigidity tensor for the virtual plate (cf. \eqref{key-non-uni}). Also, note that \eqref{key-consistent} is consistent with \eqref{key-trans-C} in the sense that, if one only knows the flexural rigidity of the virtual plate $\tilde{\pmb{\mathbb C}}$ up to \eqref{key-non-uni}, then \eqref{key-trans-C} gives us \eqref{key-consistent}.
\end{remark}

\subsubsection{The work of \citet{colquitt2014tra} on flexural cloaking} 

Next we show that the transformation cloaking formulation of Kirchhoff-Love plates given in \citep{colquitt2014tra} is, unfortunately, incorrect. They start from the biharmonic governing equation of an isotropic and homogeneous elastic plate and apply \citep[Lemma 2.1]{Norris2008} twice to transform the governing equation under the cloaking map. Using this lemma, one imposes certain constraints on the gradients of the displacements and the gradients of the Laplacian of the displacements in the virtual and physical plates. These constraints are incompatible with the way the displacement field is transformed under a cloaking map. In particular, these constraints will force the cloaking transformation to isometrically map the governing equations of the virtual plate to those of the physical plate. Therefore, the virtual and physical plates are essentially the same elastic plate and this formulation of transformation cloaking does not result in any new information. Ignoring these constraints and what restrictions they impose on the cloaking map have resulted in deriving incorrect transformed fields for the physical plate, and in particular, we show that \citet{colquitt2014tra}'s transformed flexural rigidity is incorrect.
 The governing equation of flexural waves in an isotropic and homogeneous thin plate with the flexural rigidity $D^{(0)}=Eh^3/12(1-\nu^2)$, mass density $P$, thickness $h$, and in the presence of time-harmonic anti-plane excitations with frequency $\omega$ reads  
\begin{equation}\label{key-wave}
D^{(0)}\nabla_{\mathbf X}^4W{(\mathbf X)}-Ph\omega^2W{(\mathbf X)}=0\,,\quad\mathbf{X}\in\chi\subseteq\mathbb{R}^2\,.
\end{equation}
\citet{colquitt2014tra} rewrite the governing equation as
\begin{equation}\label{key-wrong}
\left(\nabla_{\mathbf X}^4-\frac{Ph}{D^{(0)}}\omega^2\right)W{(\mathbf X)}=0\,,\quad\mathbf{X}\in\chi\subseteq\mathbb{R}^2\,.
\end{equation}
They transform \eqref{key-wrong} by applying \citep[Lemma 2.1]{Norris2008} twice via an invertible map $\mathcal{F}:\chi\to\Omega$, where $\mathbf{x}=\mathcal{F}(\mathbf X)$,  $\mathbf{F}=\nabla_{\mathbf X}\mathbf{x}$, and $J=\det\mathbf{F}$, and obtain 
\begin{equation}\label{key-fl-tr}
\left(\nabla\cdot J^{-1}\mathbf{F}\mathbf{F}^{\mathsf T}\nabla J\nabla\cdot J^{-1}\mathbf{F}\mathbf{F}^{\mathsf T}\nabla -\frac{Ph}{JD^{(0)}}\omega^2\right)W(\mathbf x)=0\,,\quad\mathbf{x}\in{\Omega}\,.
\end{equation}
In particular, their transformed rigidity tensor is given by
\begin{equation}\label{key-stu}
D_{ijkl}=D^{(0)}JG_{ij}G_{kl}\,,
\end{equation}
where $G_{ij}=J^{-1}F_{ip}F_{jp}$.

Let us discuss the implication of applying \citep[Lemma 2.1]{Norris2008} to the out-of-plane displacement field. In particular, we show that the way the fields are transformed in this lemma is incompatible with the underlying assumption of the transformation of the displacement fields under a cloaking map, i.e., $\widetilde{\mathsf W}\circ\Xi=\mathsf W$. We work in general curvilinear coordinates and distinguish between the out-of-plane displacement fields in the virtual and physical plates, i.e., $\widetilde{\mathsf W}$ and $\mathsf W$. The gradients of $\widetilde{\mathsf W}$ and $\mathsf W$ are written in components as
\begin{equation}
(\widetilde{\nabla}\widetilde{\mathsf W})^{\tilde A}=\tilde{G}^{\tilde A\tilde B}\frac{\partial\widetilde{\mathsf W}}{\partial\tilde{X}^{\tilde B}}\,,\quad (\nabla\mathsf{W})^A=G^{AB}\frac{\partial\mathsf{W}}{\partial X^B}\,.
\end{equation}
Note that	
\begin{equation}
\begin{split}
\widetilde{\nabla}^2\widetilde{\mathsf W}&=\widetilde{\mathrm{DIV}}(\widetilde{\nabla}\widetilde{\mathsf W})=(\widetilde{\nabla}\widetilde{\mathsf W})^{\tilde A}{}_{|\tilde{A}}=\frac{1}{\sqrt{\det\tilde{\mathbf{G}}}}\frac{\partial}{\partial \tilde{X}^{\tilde A}}\left[\sqrt{\det\tilde{\mathbf{G}}}\frac{\partial\widetilde{\mathsf W}}{\partial\tilde{X}^{\tilde B}}\tilde{G}^{\tilde A\tilde B}\right]\,,\\
{\nabla}^2{\mathsf W}&={\mathrm{DIV}}({\nabla}{\mathsf W})=({\nabla}{\mathsf W})^{ A}{}_{|{A}}=\frac{1}{\sqrt{\det{\mathbf{G}}}}\frac{\partial}{\partial {X}^{ A}}\left[\sqrt{\det{\mathbf{G}}}\frac{\partial{\mathsf W}}{\partial{X}^{ B}}{G}^{ A B}\right]\,,
\end{split}
\end{equation}
where $\nabla^2$ is known as the Laplace-Beltrami operator. Applying \citep[Lemma 2.1]{Norris2008} to the gradient of the displacement field, one assumes that $(\nabla\mathsf{W})^A=J_{\Xi}(\accentset{\Xi}{F}^{-1})^A{}_{\tilde A}(\widetilde{\nabla}\widetilde{\mathsf W})^{\tilde A}$, or in components
\begin{equation}\label{key-under}
\frac{\partial\mathsf{W}}{\partial X^B}G^{AB}=J_{\Xi}(\accentset{\Xi}{F}^{-1})^A{}_{\tilde A}\frac{\partial\widetilde{\mathsf W}}{\partial\tilde{X}^{\tilde{B}}}\tilde{G}^{\tilde A\tilde B}\,,
\end{equation}
i.e., the gradients of the out-of-plane displacements in the virtual and physical plates are related by the Piola transformation. Then, under this assumption, one obtains
\begin{equation}\label{key-st1}
\left[\frac{\partial\mathsf{W}}{\partial X^B}G^{AB}\right]_{|A}=\left[J_{\Xi}(\accentset{\Xi}{F}^{-1})^A{}_{\tilde A}\frac{\partial\widetilde{\mathsf W}}{\partial\tilde{X}^{\tilde B}}\tilde{G}^{\tilde A\tilde B}\right]_{|A}=J_{\Xi}\left[\frac{\partial\widetilde{\mathsf W}}{\partial\tilde{X}^{\tilde B}}\tilde{G}^{\tilde A\tilde B}\right]_{|\tilde A}\,,
\end{equation}
or, equivalently, $\nabla^2\mathsf{W}=\mathrm{DIV}(\nabla\mathsf{W})=J_{\Xi}\widetilde{\mathrm{DIV}}(\widetilde{\nabla}\widetilde{\mathsf{W}})=J_{\Xi}\widetilde{\nabla}^2\widetilde{\mathsf W}$. Applying the lemma to the Laplacian $\nabla^2\mathsf{W}$, one finds
\begin{equation}\label{key-st2}
\begin{split}
\nabla^4\mathsf{W}&=\left[\frac{\partial}{\partial X^B}\left(\left[\frac{\partial\mathsf{W}}{\partial X^D}G^{CD}\right]_{|C}\right)G^{AB}\right]_{|A}=\left[J_{\Xi}(\accentset{\Xi}{F}^{-1})^A{}_{\tilde A}\frac{\partial}{\partial\tilde{X}^{\tilde B}}\left(\left[\frac{\partial\widetilde{\mathsf W}}{\partial\tilde{X}^{\tilde D}}\tilde{G}^{\tilde C\tilde D}\right]_{|\tilde C}\right)\tilde{G}^{\tilde A\tilde B}\right]_{|A}\\&=J_{\Xi}\left[\frac{\partial}{\partial\tilde{X}^{\tilde B}}\left(\left[\frac{\partial\widetilde{\mathsf W}}{\partial\tilde{X}^{\tilde D}}\tilde{G}^{\tilde C\tilde D}\right]_{|\tilde C}\right)\tilde{G}^{\tilde A\tilde B}\right]_{|\tilde A}=J_{\Xi}\widetilde{\nabla}^4\widetilde{\mathsf W}\,.
\end{split}
\end{equation}
Using \eqref{key-st1}, one rewrites \eqref{key-st2} as
\begin{equation}\label{key-st3}
\begin{split}
	\nabla^4\mathsf{W}
	=\left[J_{\Xi}(\accentset{\Xi}{F}^{-1})^A{}_{\tilde A}\frac{\partial}
	{\partial\tilde{X}^{\tilde B}}\left[J_{\Xi}^{-1}\left(J_{\Xi}(\accentset{\Xi}{F}^{-1})^C{}_{\tilde C}
	\frac{\partial\widetilde{\mathsf W}}{\partial\tilde{X}^{\tilde D}}\tilde{G}^{\tilde C\tilde D}
	\right)_{| C}\right]\tilde{G}^{\tilde A\tilde B}\right]_{|A}
	=J_{\Xi}\widetilde{\nabla}^4\widetilde{\mathsf W}\,.
\end{split}
\end{equation}

\citet{colquitt2014tra} and many other researchers start from the biharmonic equation for the virtual plate, i.e., $D^{(0)}\widetilde{\nabla}^4\widetilde{\mathsf W}-Ph\omega^2\widetilde{\mathsf W}=0$. Then, they use \eqref{key-st3} and rewrite the biharmonic equation as
\begin{equation}\label{key-st4}
	\left[J_{\Xi}(\accentset{\Xi}{F}^{-1})^A{}_{\tilde A}\frac{\partial}{\partial\tilde{X}^{\tilde B}}
	\left[J_{\Xi}^{-1}\left(J_{\Xi}(\accentset{\Xi}{F}^{-1})^C{}_{\tilde C}
	\frac{\partial\widetilde{\mathsf W}}{\partial\tilde{X}^{\tilde D}}\tilde{G}^{\tilde C\tilde D}
	\right)_{| C}\right]\tilde{G}^{\tilde A\tilde B}\right]_{|A}
	-J_{\Xi}\frac{Ph}{D^{(0)}}\omega^2\widetilde{\mathsf W}=0\,.
\end{equation}
They implicitly assume that the virtual and physical plates have the same displacement fields, i.e., $\mathsf{W}=\widetilde{\mathsf W}\circ\Xi$,\footnote{This assumption ensures that the second term in \eqref{key-wave} remains form invariant under the cloaking map.} or, $\mathsf{W}=\widetilde{\mathsf{W}}\circ\mathcal{F}^{-1}$ (cf. \eqref{key-fl-tr}), and write \eqref{key-st4} as (note that their mapping $\mathcal F$ corresponds to $\Xi^{-1}$)
\begin{equation}\label{key-5tr}
\begin{split}
	& \left[J_{\Xi}(\accentset{\Xi}{F}^{-1})^A{}_{\tilde A}(\accentset{\Xi}{F}^{-1})^B{}_{\tilde B}
	\frac{\partial}{\partial{X}^{ B}}\left[J_{\Xi}^{-1}
	\left(J_{\Xi}(\accentset{\Xi}{F}^{-1})^C{}_{\tilde C}(\accentset{\Xi}{F}^{-1})^D{}_{\tilde D}
	\frac{\partial{\mathsf W}}{\partial{X}^{ D}}\tilde{G}^{\tilde C\tilde D}
	\right)_{| C}\right]\tilde{G}^{\tilde A\tilde B}\right]_{|A}\\
	& ~~~-J_{\Xi}\frac{Ph}{D^{(0)}}\omega^2{\mathsf W}=0\,.
\end{split}
\end{equation}
However, this formulation of the transformation cloaking problem is problematic for a number of reasons: 

(i) Once one assumes $\mathsf{W}=\widetilde{\mathsf W}\circ\Xi$, the gradients of displacements are related by the chain rule as $\frac{\partial\widetilde{\mathsf W}}{\partial\tilde{X}^{\tilde D}}=(\accentset{\Xi}{F}^{-1})^D{}_{\tilde D}\frac{\partial{\mathsf W}}{\partial{X}^{ D}}$, and not by the Piola transformation \eqref{key-under}, which is the underlying assumption of \citep[Lemma 2.1]{Norris2008}. 
In other words, one cannot assume that $\mathsf{W}=\widetilde{\mathsf W}\circ\Xi$, and at the same time use the Piola transform, which is what \citet{colquitt2014tra} did; in order to derive \eqref{key-5tr} from \eqref{key-st4} they used the chain rule to relate the gradients of displacements as $\frac{\partial\widetilde{\mathsf W}}{\partial\tilde{X}^{\tilde D}}=(\accentset{\Xi}{F}^{-1})^D{}_{\tilde D}\frac{\partial{\mathsf W}}{\partial{X}^{ D}}$. 

(ii) Applying \citep[Lemma 2.1]{Norris2008} twice requires the following extra constraint on the gradients of the Laplacian terms (this is similar to \eqref{key-under}): 
\begin{equation}
	\frac{\partial}{\partial X^B}\left[\left(\frac{\partial\mathsf{W}}{\partial X^D}G^{CD}\right)_{|C}\right]G^{AB}
	=J_{\Xi}(\accentset{\Xi}{F}^{-1})^A{}_{\tilde A}\frac{\partial}{\partial\tilde{X}^{\tilde{B}}}
	\left[\left(\frac{\partial\widetilde{\mathsf W}}{\partial\tilde{X}^{\tilde D}}
	\tilde{G}^{\tilde C\tilde D}\right)_{|\tilde C}\right]\tilde{G}^{\tilde A\tilde B}\,,
\end{equation}
where using \eqref{key-under}, it is simplified to read
\begin{equation}\label{key-assum2}
	\accentset{\Xi}{F}^{\tilde E}{}_{ B}\frac{\partial}{\partial\tilde{X}^{\tilde E}}
	\left[J_{\Xi}\left(\frac{\partial\widetilde{\mathsf W}}{\partial\tilde{X}^{\tilde D}}\tilde{G}^{\tilde C\tilde D}
	\right)_{|\tilde C}\right]G^{AB}
	=J_{\Xi}(\accentset{\Xi}{F}^{-1})^A{}_{\tilde A}\tilde{G}^{\tilde A\tilde B}
	\frac{\partial}{\partial\tilde{X}^{\tilde B}}\left[\left(
	\frac{\partial\widetilde{\mathsf W}}{\partial\tilde{X}^{\tilde D}}\tilde{G}^{\tilde C\tilde D}\right)_{|\tilde C}\right]\,.
\end{equation}
\citet{colquitt2014tra} use the lemma twice and write their Eq. (2) without mentioning the constraint \eqref{key-assum2} or \eqref{key-under} and even checking if these constraints are compatible with the underlying assumption that $\mathsf{W}=\widetilde{\mathsf W}\circ\Xi$. Let us see what restrictions these constraints impose on the cloaking map. From \eqref{key-under} and the chain rule $\frac{\partial\widetilde{\mathsf W}}{\partial\tilde{X}^{\tilde D}}=(\accentset{\Xi}{F}^{-1})^D{}_{\tilde D}\frac{\partial{\mathsf W}}{\partial{X}^{ D}}$, one obtains
\begin{equation}\label{key-mee1}
G^{AB}=J_{\Xi}(\accentset{\Xi}{F}^{-1})^A{}_{\tilde A}(\accentset{\Xi}{F}^{-1})^B{}_{\tilde B}\tilde{G}^{\tilde A\tilde B}\,.
\end{equation}
Substituting \eqref{key-mee1} into \eqref{key-assum2}, one finds
\begin{equation}\label{key-hold}
	\frac{\partial}{\partial\tilde{X}^{\tilde B}}\left[\left(J_{\Xi}-1\right)
	\left(\frac{\partial\widetilde{\mathsf W}}{\partial\tilde{X}^{\tilde D}}
	\tilde{G}^{\tilde C\tilde D}\right)_{|\tilde C}\right]
	=\frac{\partial}{\partial\tilde{X}^{\tilde B}}\left[\left(J_{\Xi}-1\right)
	\widetilde{\nabla}^2\widetilde{\mathsf W}\right]=0\,.
\end{equation}
Knowing that \eqref{key-hold} must hold for an arbitrary displacement field $\widetilde{\mathsf W}$, one concludes that $J_{\Xi}=1$,\footnote{Note that \eqref{key-hold} implies that $\left(J_{\Xi}-1\right)\widetilde{\nabla}^2\widetilde{\mathsf W}=C$, where $C$ is a constant. Recalling that on the outer boundary of the cloak $\accentset{\Xi}{\mathbf{F}}|_{\partial_o\mathcal{C}}=id$, and thus, $J_{\Xi}|_{\partial_o\mathcal{C}}=1$, and given that $\widetilde{\mathsf W}$ is smooth, one concludes that $C=0$. Therefore, $J_{\Xi}=1$.} and thus, $\mathbf{G}=\Xi^*\tilde{\mathbf G}$, meaning that the physical and virtual plates are isometric and are essentially the same elastic plate with the same mechanical response. To see this more clearly, using the fact that $J_{\Xi}=1$, and $\mathbf{G}=\Xi^*\tilde{\mathbf G}$, or in components, $G^{AB}=(\accentset{\Xi}{F}^{-1})^A{}_{\tilde A}(\accentset{\Xi}{F}^{-1})^B{}_{\tilde B}\tilde{G}^{\tilde A\tilde B}$ and the metric compatibility of $\mathbf G$, i.e., $G_{AB|C}=0$, \eqref{key-5tr} is simplified to read
\begin{equation}\label{key-ey}
	D^{(0)}G^{AB}G^{CD}\left(\frac{\partial\mathsf{W}}{\partial X^D}\right)_{|C|B|A}-Ph\omega^2\mathsf{W}=0\,,
\end{equation}
i.e., $D^{(0)}\nabla^4\mathsf{W}-Ph\omega^2\mathsf{W}=0$, which is identical to the biharmonic equation for the virtual plate $D^{(0)}\widetilde{\nabla}^4\widetilde{\mathsf{W}}-Ph\omega^2\widetilde{\mathsf{W}}=0$. Therefore, the physical and the virtual plates are the same elastic plate and the application of \citep[Lemma 2.1]{Norris2008} maps the biharmonic equation to itself; it does not result in any new information. 

(iii) \citet{colquitt2014tra} express \eqref{key-5tr} in Cartesian coordinates and looking at the fourth-order derivatives of the displacement, i.e., $J_{\Xi}(\accentset{\Xi}{F}^{-1})^A{}_{\tilde A}(\accentset{\Xi}{F}^{-1})^B{}_{\tilde A}(\accentset{\Xi}{F}^{-1})^C{}_{\tilde C}(\accentset{\Xi}{F}^{-1})^D{}_{\tilde C}\frac{\partial^4\mathsf{W}}{\partial X^D\partial X^C\partial X^B\partial X^A}$ (summations on $\tilde A$ and $\tilde C$), incorrectly conclude that the flexural rigidity of the cloak is given by \eqref{key-stu}, or in our notation, $D^{ABCD}=D^{(0)}J_{\Xi}(\accentset{\Xi}{F}^{-1})^A{}_{\tilde A}(\accentset{\Xi}{F}^{-1})^B{}_{\tilde A}(\accentset{\Xi}{F}^{-1})^C{}_{\tilde C}(\accentset{\Xi}{F}^{-1})^D{}_{\tilde C}$. 
The reason for this mistake is that \eqref{key-5tr} has some hidden strong assumptions. Incorporating these assumptions one arrives at \eqref{key-ey}. In other words, one needs to look at \eqref{key-ey} in order to calculate the transformed elastic constants, and not \eqref{key-5tr}.
Looking at the term in \eqref{key-ey} with fourth-order derivatives the transformed elasticity tensor is given by $D^{(0)}\frac{\partial^4\mathsf{W}}{\partial X^C\partial X^C\partial X^A\partial X^A}$ in Cartesian coordinates. However, \eqref{key-ey} is nothing but the governing equation of the virtual plate, i.e., the virtual and physical plates are identical. 
Note that when the cloaking transformation is the identity, i.e., $\mathcal{F}=id$, \citet{colquitt2014tra}'s transformed rigidity tensor \eqref{key-stu} is not reduced to that of the homogeneous and isotropic plate \eqref{key-vir-rig} and is not even positive definite (see also \citep{pomot2019form}). To see this note that for the identity cloaking map \eqref{key-stu} is simplified to read
\begin{equation}\label{key-vir-rig11}
{{{\boldsymbol{D}}}}=\left[{{{D}}}^{ijkl}\right]=\frac{Eh^3}{12(1+\nu)}
\begin{bmatrix}
\begin{bmatrix}
1 & 0 \\
0 & 1  \\
\end{bmatrix} & \begin{bmatrix}
0 & 0 \\
0  & 0 \\
\end{bmatrix} \\
\\
\begin{bmatrix}
0 & 0  \\
0 & 0 \\
\end{bmatrix} & \begin{bmatrix}
1  & 0 \\
0 & 1 \\
\end{bmatrix} \\
\end{bmatrix}\,,
\end{equation}
which clearly does not agree with the flexural rigidity of the isotropic and homogeneous elastic plate \eqref{key-vir-rig}. It is also immediate to see that \eqref{key-vir-rig11} has zero eigenvalues, and thus, is not positive definite. 

(iv) The traction due to the so-called membrane forces obtained in \citep{colquitt2014tra} does not vanish on the boundary of the hole, i.e., the hole surface is not traction-free. This means that in the numerical simulations presented in \citep{colquitt2014tra}, one needs to apply some forces on the boundary of the hole even if the transformation cloaking problem had  been properly formulated. Furthermore, the finite traction due to the membrane forces does not vanish on the boundary of the cloak either, and their cloaking transformation does not have the identity tangent map on the outer boundary of the cloak $\partial_o\mathcal{C}$, i.e., $T\mathcal{F}|_{\partial_o\mathcal{C}}=\accentset{\mathcal{F}}{\mathbf F}|_{\partial_o\mathcal{C}}\neq id$. Therefore, the physical and virtual plates cannot have identical current configurations outside the cloaking region.

\begin{remark}\label{key-helm-holtz}
In this remark we show that the work of \citet{Colquitt2013} on cloaking of the out-of-plane shear waves for the Helmholtz equation is also incorrect because similar to their flexural cloaking formulation \citep{colquitt2014tra} their use of the Piola transformation is inconsistent with their displacement transformation. \citet{Colquitt2013} start from the Helmholtz equation for an isotropic and homogeneous medium
\begin{equation}\label{key-hem}
	 \mu\nabla_X\cdot(\nabla_X)u(X)+\varrho\,\omega^2u(X)=0\,,\quad X\in\chi\subset\mathbb{R}^2\,, 
\end{equation}	  
where $\mu$ and $\varrho$ are, respectively, the shear modulus and the mass density of the isotropic and homogeneous medium, and $u$ is the out-of-plane displacement. Applying \citep[Lemma 2.1]{Norris2008} using an invertible map $\mathcal{F}:\chi\to\Omega$ such that  $\mathbf{x}=\mathcal{F}(\mathbf X)$, $\mathbf{F}=\nabla_{\mathbf X}\mathbf{x}$, and $J=\det\mathbf{F}$, they transform \eqref{key-hem} as
\begin{equation}\label{key-tus}
\left[\nabla\cdot\left(\boldsymbol{\mathcal C}(x)\nabla\right)+\rho(x)\omega^2\right]u(x)=0\,,\quad x\in\Omega\subset\mathbb{R}^2\,,
\end{equation}
where $\boldsymbol{\mathcal C}(x)=\mu/J(x)\mathbf{F}(x)\mathbf{F}^{\mathsf T}(x)$ is the transformed stiffness matrix and the transformed mass density is given by $\rho(x)=\varrho/J(x)$. To write this in our notation, one starts with the Helmholtz equation for the virtual medium $\tilde{\mu}\,\widetilde{\nabla}^2\widetilde{\mathsf W}+\tilde{\rho}\,\omega^2\widetilde{\mathsf W}=0$, where $\tilde{\mu}$, $\tilde{\rho}$, and $\widetilde{\mathsf W}$ are, respectively, the shear modulus, the mass density, and the displacement in the virtual medium. Then, provided that \eqref{key-under} holds, one may use \citep[Lemma 2.1]{Norris2008} (see \eqref{key-st1}), to obtain the transformed equation as
\begin{equation}
\tilde{\mu}\left[J_{\Xi}(\accentset{\Xi}{F}^{-1})^A{}_{\tilde A}\frac{\partial\widetilde{\mathsf W}}{\partial\tilde{X}^{\tilde B}}\tilde{G}^{\tilde A\tilde B}\right]_{|A}+\tilde{\rho}\,J_{\Xi}\omega^2\widetilde{\mathsf W}=0\,.
\end{equation}
Next, assuming that the virtual and the physical media have identical displacements, i.e., $\mathsf{W}=\widetilde{\mathsf W}\circ\Xi$ (which is what \citet{Colquitt2013} implicitly assume) and using the chain rule, one finds 
\begin{equation}
\left[\tilde{\mu}J_{\Xi}(\accentset{\Xi}{F}^{-1})^A{}_{\tilde A}(\accentset{\Xi}{F}^{-1})^B{}_{\tilde B}\frac{\partial{\mathsf W}}{\partial{X}^{ B}}\tilde{G}^{\tilde A\tilde B}\right]_{|A}+\tilde{\rho}\,J_{\Xi}\omega^2{\mathsf W}=0\,.
\end{equation}
This is identical to what they have in \eqref{key-tus}, recalling that $\mathcal F$ corresponds to $\Xi^{-1}$. As we discussed in the case of flexural transformation cloaking, \eqref{key-under} imposes a strong constraint on the cloaking map, which is given by \eqref{key-mee1}. In fact, in Cartesian coordinates, \eqref{key-mee1} is simplified to read $J_{\Xi}(\accentset{\Xi}{F}^{-1})^A{}_{\tilde A}(\accentset{\Xi}{F}^{-1})^B{}_{\tilde A}=\delta^{AB}$ (summation on $\tilde A$). Solving this relation, one obtains the tangent of the (inverse) cloaking map as\footnote{Notice that \eqref{key-form-cons} can be represented as the rotation matrix multiplied by the scalar $\left(\alpha^2+\beta^2\right)$.}
\begin{equation}\label{key-form-cons}
\accentset{\Xi}{\mathbf F}^{-1}(\tilde X,\tilde Y)=\begin{bmatrix}
\alpha(\tilde X,\tilde Y) & \beta(\tilde X,\tilde Y) \\
-\beta(\tilde X,\tilde Y) & \alpha(\tilde X,\tilde Y)     
\end{bmatrix}\,,
\end{equation}
where $\alpha^2+\beta^2>0$. Note that the (bulk) compatibility of $\accentset{\Xi}{\mathbf{F}}^{-1}$ are written as $(\accentset{\Xi}{F}^{-1})^A{}_{\tilde A|\tilde B}=(\accentset{\Xi}{F}^{-1})^A{}_{\tilde B|\tilde A}$, and thus, one has\footnote{Note that if one defines the complex function $f(\tilde{X}+i\tilde{Y})=\beta(\tilde{X},\tilde{Y})+i\alpha(\tilde{X},\tilde{Y})$, then \eqref{key-holomorphicity} are the Cauchy-Riemann equations, and hence, the complex function $f$ is holomorphic.}
\begin{equation}\label{key-holomorphicity}
\frac{\partial \alpha}{\partial \tilde X}=-\frac{\partial \beta}{\partial\tilde{Y}}\,,\quad \mathrm{and}\quad \frac{\partial\alpha}{\partial\tilde Y}=\frac{\partial \beta}{\partial \tilde X}\,.
\end{equation}
Hence, one concludes that $\alpha$ and $\beta$ are harmonic, i.e., 
\begin{equation}
	\frac{\partial^2\alpha}{\partial\tilde{X}^2}+\frac{\partial^2\alpha}{\partial\tilde{Y}^2}=0\,,
	\quad \mathrm{and}\quad \frac{\partial^2\beta}{\partial\tilde{X}^2}+\frac{\partial^2\beta}{\partial\tilde{Y}^2}=0\,.
\end{equation}
The mapping $\mathcal F$ that \citet{Colquitt2013} introduce to design a square-shaped cloak by shrinking a finite rectangular cavity to a small one is not of the form \eqref{key-form-cons}.
\end{remark}
\subsection{Elastodynamic transformation cloaking in plates in the presence of in-plane and out-of-plane displacements}\label{linear-elas}

In this section, we relax the pure bending assumption and formulate the transformation cloaking problem of a classical elastic plate in the presence of both in-plane and out-of-plane displacements. In doing so, we let the physical plate undergo finite in-plane deformations while staying flat, i.e., ($\mathring{\varphi}^\perp= id$ and $\mathring{\boldsymbol{\theta}}=\mathbf{0}$), whereas the virtual plate remains undeformed ($\mathring{\tilde\varphi}=id$). For simplicity of notation, let us drop the superscripts and denote the normal and tangential displacement fields of the physical (virtual) plate $\mathbf{U}^\perp$ and $\mathbf{U}^\top$ ($\tilde{\mathbf{U}}^\perp$ and $\tilde{\mathbf{U}}^\top$) by $\boldsymbol{\mathsf{W}}$ and $\boldsymbol{\mathsf{U}}$ 
($\widetilde{\boldsymbol{\mathsf{W}}}$ and $\tilde{\boldsymbol{\mathsf{U}}}$), respectively.
The tangential and normal displacement fields are transformed as
\begin{equation}
\tilde{\mathsf U}^{\tilde a}=\mathsf{s}^{\tilde a}{}_a\circ\mathring{\varphi}\,\mathsf{U}^a\circ\Xi\,,\quad\mathrm{and}\quad \widetilde{\mathsf W}={\mathsf{W}}\circ\Xi^{-1}\,.
\end{equation}
The linearized balance of linear momentum for the virtual plate, which has uniform elastic properties, reads
\begin{equation}\label{key-vir-both}
\delta\tilde{P}^{\tilde a\tilde A}{}_{|\tilde A}=\tilde{\rho}_0\,\ddot{\tilde{\mathsf{U}}}^{\tilde a}\,,\quad\mathrm{and}\quad -(\mathring{\tilde{F}}^{-1})^{\tilde B}{}_{\tilde a}\delta\tilde{\mathsf{M}}^{\tilde a\tilde A}{}_{|\tilde A|\tilde B}=\tilde{\rho}_0\, \ddot{\widetilde{\mathsf W} }\,,
\end{equation}
where
\begin{equation}
\delta \tilde{P}^{\tilde a\tilde A}=\tilde{\mathbb{A}}^{\tilde a\tilde A\tilde b\tilde B}{\tilde{\mathsf{U}}}{}_{\tilde b|\tilde B}+\frac{1}{2}\tilde{\mathbb{B}}^{\tilde a\tilde A\tilde b\tilde B}( \widetilde{\mathsf{W}}{}_{,\tilde b})_{|\tilde B}\,,\quad\mathrm{and}\quad \,\delta\tilde{\mathsf{M}}^{\tilde a\tilde A}=\frac{1}{2}\tilde{\mathbb{C}}^{\tilde a\tilde A\tilde b\tilde B}( {\widetilde {\mathsf W}}{}_{,\tilde b})_{|\tilde B}+\frac{1}{2}\tilde{\mathbb{B}}^{\tilde b\tilde B\tilde a\tilde A}{\tilde{\mathsf{U}}}{}_{\tilde b|\tilde B}\,.
\end{equation}
The symmetries of the elastic constants \eqref{key-ang1} for the virtual plate implies that
\begin{subequations}\label{key-ang-vir}
\begin{equation}
\tilde{\mathbb{A}}^{[\tilde{a}\tilde{A}\tilde{c}\tilde{B}}\mathring{\tilde F}^{\tilde b]}{}_{\tilde A}=0\,,\qquad\tilde{\mathbb{B}}^{\tilde{c}\tilde{B}[\tilde{a}\tilde{A}}\mathring{\tilde F}^{\tilde{b}]}{}_{\tilde A}=0\,,
\end{equation}
\begin{equation}
\tilde{\mathbb{B}}^{[\tilde{a}\tilde{A}\tilde{c}\tilde{B}}\mathring{\tilde F}^{\tilde b]}{}_{\tilde A}=0\,,\qquad\tilde{\mathbb{C}}^{[\tilde{a}\tilde{A}\tilde{c}\tilde{B}}\mathring{\tilde F}^{\tilde b]}{}_{\tilde A}=0\,.
\end{equation}
\end{subequations}
Knowing that the virtual plate is uniform, its elastic properties (constants) are (covariantly) constant. Thus, \eqref{key-vir-both} is simplified to read
\begin{equation}\label{key-main}
\begin{split}
\tilde{\mathbb{A}}^{\tilde a\tilde A\tilde b\tilde B}{\tilde {\mathsf{U}}}{}_{\tilde b|\tilde B|\tilde A}+\frac{1}{2}\tilde{\mathbb{B}}^{\tilde a\tilde A\tilde b\tilde B}( \widetilde{\mathsf{W}}{}_{,\tilde b})_{|\tilde B|\tilde A}=\tilde{\rho}_0\,\ddot{\tilde {\mathsf{U}}}^{\tilde a}\,,\\ -\frac{1}{2}(\mathring{\tilde{F}}^{-1})^{\tilde B}{}_{\tilde a}\left(\tilde{\mathbb{C}}^{\tilde a\tilde A\tilde b\tilde C}( {\widetilde {\mathsf W}}{}_{,\tilde b})_{|\tilde C|\tilde A|\tilde B}+\tilde{\mathbb{B}}^{\tilde b\tilde C\tilde a\tilde A}{\tilde {\mathsf{U}}}{}_{\tilde b|\tilde C|\tilde A|\tilde B}\right)=\tilde{\rho}_0\, \ddot{\widetilde{\mathsf W} }\,.
\end{split}
\end{equation}
The balance of linear momentum for the physical plate in the absence of initial couple-stress ($\mathring{\boldsymbol{\mathsf{M}}}=\mathbf{0}$) reads (cf. \eqref{lin-plate-cosserat})
\begin{subequations}
\begin{equation}\label{key-1-one}
\delta P^{aA}{}_{|A}+\rho_0\mathring{{\mathfrak B}}^\perp g^{ab}{ \mathsf{W}}_{,b}+\rho_0 g^{ab}(\mathring{{\mathfrak B}}^\top)^c {\mathsf U}^{}_{c|b}-\rho_0\,g^{ac}(\mathring{F}^{-1})^B{}_b\left({ \mathsf{W}}_{,c}\right)_{|B}\mathring{\mathfrak{L}}^b=\rho_0 \ddot{\mathsf U}^a\,,
\end{equation}
\begin{equation}\label{key-2-two}
\begin{split}
\mathring{P}^{aA}\left( \mathsf{W}_ {,a}\right)_{|A}&-\left[(\mathring{F}^{-1})^B{}_a\delta\mathsf{M}^{aA}{}_{|A}\right]_{|B}-\rho_0(\mathring{{\mathfrak B}}^\top)^b{\mathsf{W} }_{,b}+\left(\rho_0\,\mathring{\mathfrak{L}}^c{\mathsf U}^{}_{c|b}g^{ab}(\mathring{F}^{-1})^A{}_a\right)_{|A}\\&-\left[\rho_0\,\mathring{\mathfrak{L}}^a\, (\mathring{F}^{-1})^B{}_a(\mathring{F}^{-1})^A{}_b{\mathsf U}^b{}_{|B}\right]_{|A}=\rho_0 \ddot{\mathsf{W}}\,,
\end{split}
\end{equation}
\end{subequations}
where using \eqref{sim-delta}
\begin{equation}
\begin{split}
\delta {P}^{ a A}&=\left[{\mathbb{A}}^{ a A b B}+\mathring{P}^{cA}(\mathring{F}^{-1})^B{}_cg^{ab}\right]{ \mathsf U}{}_{ b| B}+\frac{1}{2}{\mathbb{B}}^{ a A b B}( {\mathsf{W}}{}_{, b})_{| B}\,,\\ \delta{\mathsf{M}}^{ a A}&=\frac{1}{2}{\mathbb{C}}^{ a A b B}( { {\mathsf W}}{}_{, b})_{| B}+\frac{1}{2}{\mathbb{B}}^{ b B a A}{ \mathsf U}{}_{ b| B}\,.
\end{split}
\end{equation}
Note that
\begin{equation}\label{key-grad-disp}
\begin{split}
\tilde{\mathsf{U}}_{\tilde b|\tilde C}=&(\mathsf{s}^{-1})^b{}_{\tilde b}(\accentset{\Xi}{F}^{-1})^C{}_{\tilde C}\mathsf{U}_{b|C}\,,\\
\tilde{\mathsf U}_{\tilde b|\tilde C|\tilde A}=&(\mathsf{s}^{-1})^b{}_{\tilde b}\left[(\accentset{\Xi}{F}^{-1})^C{}_{\tilde C|\tilde A}\mathsf{U}_{b|C}+(\accentset{\Xi}{F}^{-1})^C{}_{\tilde C}(\accentset{\Xi}{F}^{-1})^A{}_{\tilde A}\mathsf{U}_{b|C|A}\right]\,,\\
\tilde{\mathsf U}_{\tilde b|\tilde C|\tilde A|\tilde B}=&(\mathsf{s}^{-1})^b{}_{\tilde b}\Big[(\accentset{\Xi}{F}^{-1})^C{}_{\tilde C|\tilde A|\tilde B}\mathsf{U}_{b|C}+\Big((\accentset{\Xi}{F}^{-1})^C{}_{\tilde C|\tilde A}(\accentset{\Xi}{F}^{-1})^A{}_{\tilde B}+(\accentset{\Xi}{F}^{-1})^C{}_{\tilde C|\tilde B}(\accentset{\Xi}{F}^{-1})^A{}_{\tilde A}\\&+(\accentset{\Xi}{F}^{-1})^C{}_{\tilde C}(\accentset{\Xi}{F}^{-1})^A{}_{\tilde A|\tilde B}\Big)\mathsf{U}_{b|C|A}+(\accentset{\Xi}{F}^{-1})^C{}_{\tilde C}(\accentset{\Xi}{F}^{-1})^A{}_{\tilde A}(\accentset{\Xi}{F}^{-1})^B{}_{\tilde B}\mathsf{U}_{b|C|A|B}\Big]\,,
\end{split}
\end{equation}
\begin{equation}\label{disp-grad-both}
\begin{split}
\widetilde{\mathsf{W}}{}_{,\tilde b}=&(\accentset{\xi}{F}^{-1})^{ b}{}_{\tilde b}\,\mathsf{W}{}_{, b}\,,\\
(\widetilde{\mathsf{W}}{}_{, \tilde b})_{|\tilde C}=&(\accentset{\xi}{F}^{-1})^{ b}{}_{\tilde b|\tilde C}\,{\mathsf{W}}{}_{, b}+(\accentset{\xi}{F}^{-1})^{ b}{}_{\tilde b}(\accentset{\Xi}{F}^{-1})^{ C}{}_{\tilde C}\,({\mathsf{W}}{}_{, b})_{|{C}}\,,\\
(\widetilde{\mathsf{W}}{}_{, \tilde{b}})_{|\tilde{C}|\tilde{A}}=&(\accentset{\xi}{F}^{-1})^{ b}{}_{\tilde{b}|\tilde{C}|\tilde{A}}\,{\mathsf{W}}{}_{, b}+\Big[(\accentset{\xi}{F}^{-1})^{ b}{}_{\tilde{b}|\tilde{A}}(\accentset{\Xi}{F}^{-1})^{ C}{}_{\tilde C}+(\accentset{\Xi}{F}^{-1})^{ C}{}_{\tilde C|\tilde A}(\accentset{\xi}{F}^{-1})^{ b}{}_{\tilde b}\\&+(\accentset{\xi}{F}^{-1})^{ b}{}_{\tilde b|\tilde C}(\accentset{\Xi}{F}^{-1})^{ C}{}_{\tilde A}\Big]({\mathsf{W}}{}_{, b})_{|{C}}+(\accentset{\xi}{F}^{-1})^{ b}{}_{\tilde b}(\accentset{\Xi}{F}^{-1})^{ C}{}_{\tilde C}(\accentset{\Xi}{F}^{-1})^{ A}{}_{\tilde A}\,({\mathsf{W}}{}_{, b})_{|{C}|{A}}\,,\\
(\widetilde{\mathsf{W}}{}_{, \tilde b})_{|\tilde C|\tilde A|\tilde B}=&(\accentset{\xi}{F}^{-1})^{ b}{}_{\tilde b|\tilde C|\tilde A|\tilde B}\,{\mathsf{W}}{}_{, b}+\Big[(\accentset{\xi}{F}^{-1})^{ b}{}_{\tilde b|\tilde A}(\accentset{\Xi}{F}^{-1})^{ A}{}_{\tilde C|\tilde B}+(\accentset{\xi}{F}^{-1})^{ b}{}_{\tilde b|\tilde C}(\accentset{\Xi}{F}^{-1})^{ A}{}_{\tilde A|\tilde B}\\&+(\accentset{\xi}{F}^{-1})^{ b}{}_{\tilde b|\tilde B}(\accentset{\Xi}{F}^{-1})^{ A}{}_{\tilde C|\tilde A}+(\accentset{\xi}{F}^{-1})^{ b}{}_{\tilde b|\tilde C|\tilde B}(\accentset{\Xi}{F}^{-1})^{ A}{}_{\tilde A}+(\accentset{\xi}{F}^{-1})^{ b}{}_{\tilde b|\tilde A|\tilde B}(\accentset{\Xi}{F}^{-1})^{ A}{}_{\tilde C}\\&+(\accentset{\xi}{F}^{-1})^{ b}{}_{\tilde b|\tilde C|\tilde A}(\accentset{\Xi}{F}^{-1})^{ A}{}_{\tilde B}+(\accentset{\Xi}{F}^{-1})^{ A}{}_{\tilde C|\tilde A|\tilde B}(\accentset{\xi}{F}^{-1})^{ b}{}_{\tilde b}\Big]({\mathsf{W}}{}_{, b})_{|{A}}\\&+\Big[(\accentset{\xi}{F}^{-1})^{ b}{}_{\tilde b|\tilde B}(\accentset{\Xi}{F}^{-1})^{ C}{}_{\tilde C}(\accentset{\Xi}{F}^{-1})^{ A}{}_{\tilde A}+(\accentset{\Xi}{F}^{-1})^{ C}{}_{\tilde C|\tilde B}(\accentset{\xi}{F}^{-1})^{ b}{}_{\tilde b}(\accentset{\Xi}{F}^{-1})^{ A}{}_{\tilde A}\\&+(\accentset{\Xi}{F}^{-1})^{ A}{}_{\tilde A|\tilde B}(\accentset{\xi}{F}^{-1})^{ b}{}_{\tilde b}(\accentset{\Xi}{F}^{-1})^{ C}{}_{\tilde C}+(\accentset{\xi}{F}^{-1})^{ b}{}_{\tilde b|\tilde C}(\accentset{\Xi}{F}^{-1})^{ C}{}_{\tilde A}(\accentset{\Xi}{F}^{-1})^{ A}{}_{\tilde B}\\&+(\accentset{\xi}{F}^{-1})^{ b}{}_{\tilde b|\tilde A}(\accentset{\Xi}{F}^{-1})^{ A}{}_{\tilde B}(\accentset{\Xi}{F}^{-1})^{ C}{}_{\tilde C}+(\accentset{\Xi}{F}^{-1})^{ C}{}_{\tilde C|\tilde A}(\accentset{\Xi}{F}^{-1})^{ A}{}_{\tilde B}(\accentset{\xi}{F}^{-1})^{ b}{}_{\tilde b}\Big]({\mathsf{W}}{}_{, b})_{|{C}|{A}}\\&+(\accentset{\xi}{F}^{-1})^{ b}{}_{\tilde b}(\accentset{\Xi}{F}^{-1})^{ C}{}_{\tilde C}(\accentset{\Xi}{F}^{-1})^{ A}{}_{\tilde A}(\accentset{\Xi}{F}^{-1})^{ B}{}_{\tilde B}\,({\mathsf{W}}{}_{, b})_{|{C}|{A}|{B}}\,,
\end{split}
\end{equation}
where $\accentset{\xi}{\mathbf F}=\mathring{\tilde{\mathbf{F}}}\accentset{\Xi}{\mathbf F}\mathring{\mathbf F}^{-1}$.\footnote{Note that $\mathring{\tilde{\mathbf F}}=id$, while $\mathring{\mathbf F}$ is not the identity, in general. However, for thin plates (due to the inextensibility constraint) $\mathring{\mathbf{F}}=id$, and thus, the mappings $\xi$ and $\Xi$ are identical, whence \eqref{disp-grad-both} reduces to \eqref{disp-grad} with a slight abuse of notation.} Under the cloaking transformation $\Xi:\mathcal{H}\to\tilde{\mathcal H}$ and using \eqref{key-grad-disp} and \eqref{disp-grad-both}, the divergence term in \eqref{key-vir-both}$_1$ is transformed via the Piola transformation as
\begin{equation}\label{key-1st-tr}
\begin{split}
	&\left[\tilde{\mathbb{A}}^{\tilde a\tilde A\tilde b\tilde B}{\tilde {\mathsf{U}}}{}_{\tilde b|\tilde B}
	+\frac{1}{2}\tilde{\mathbb{B}}^{\tilde a\tilde A\tilde b\tilde B}
	( \widetilde{\mathsf{W}}{}_{,\tilde b})_{|\tilde B}\right]_{|\tilde A}\\&
	=J_{\Xi}^{-1}\bigg[J_{\Xi}(\accentset{\Xi}{F}^{-1})^A{}_{\tilde A}\tilde{\mathbb{A}}^{\tilde a\tilde A\tilde b\tilde B}
	{\tilde {\mathsf{U}}}{}_{\tilde b|\tilde B}+\frac{1}{2}J_{\Xi}(\accentset{\Xi}{F}^{-1})^A{}_{\tilde A}
	\tilde{\mathbb{B}}^{\tilde a\tilde A\tilde b\tilde B}( \widetilde{\mathsf{W}}{}_{,\tilde b})_{|\tilde B}
	\bigg]_{|A}\\&=J_{\Xi}^{-1}\bigg[J_{\Xi}(\accentset{\Xi}{F}^{-1})^A{}_{\tilde A}
	(\accentset{\Xi}{F}^{-1})^B{}_{\tilde B}(\mathsf{s}^{-1})^b{}_{\tilde b}\tilde{\mathbb{A}}^{\tilde a\tilde A
	\tilde b\tilde B}{ \mathsf U}{}_{ b| B}+\frac{1}{2}J_{\Xi}(\accentset{\Xi}{F}^{-1})^A{}_{\tilde A}
	(\accentset{\xi}{F}^{-1})^{ b}{}_{\tilde b}(\accentset{\Xi}{F}^{-1})^{ B}{}_{\tilde B}
	\tilde{\mathbb{B}}^{\tilde a\tilde A\tilde b\tilde B}\,({\mathsf{W}}{}_{, b})_{|{B}}\\
	&~~~+\frac{1}{2}J_{\Xi}(\accentset{\Xi}{F}^{-1})^A{}_{\tilde A}(\accentset{\xi}{F}^{-1})^{ b}{}_{\tilde b|\tilde B}
	\tilde{\mathbb{B}}^{\tilde a\tilde A\tilde b\tilde B}\,{\mathsf{W}}{}_{, b}\bigg]_{|A}\,.
\end{split}
\end{equation}
Using the shifter map, we try to write \eqref{key-1st-tr} as the in-plane governing equation of the physical plate \eqref{key-1-one}. Therefore, the referential mass density of the physical plate is given by $\rho_0=J_{\Xi}\tilde{\rho}_0$,\footnote{Conservation of mass for the physical and virtual plates implies that $\rho_0=\varrho\mathring{J}$ and $\tilde{\rho}_0=\tilde{\varrho}\mathring{\tilde{J}}$. Noting that $\mathring{\tilde{\varphi}}=id$, and $J_{\xi}=\mathring{\tilde{J}}J_{\Xi}\mathring{J}^{-1}$, the spatial mass density of the cloak is given by $\varrho=J_{\xi}\tilde{\rho}_0$.} and recalling that the shifter map is covariantly constant, one obtains
\begin{subequations}
	\begin{equation}\label{key-trans-1}
	{\mathbb{A}}^{ a A b B}=J_{\Xi}(\mathsf{s}^{-1})^a{}_{\tilde a}(\accentset{\Xi}{F}^{-1})^A{}_{\tilde A}(\mathsf{s}^{-1})^b{}_{\tilde b}(\accentset{\Xi}{F}^{-1})^B{}_{\tilde B}\tilde{\mathbb{A}}^{ \tilde a \tilde A \tilde b \tilde B}-\mathring{P}^{cA}(\mathring{F}^{-1})^B{}_cg^{ab}\,,
	\end{equation}
    \begin{equation}\label{key-b-vir-phy}
     {\mathbb{B}}^{ a A b B}=J_{\Xi}(\mathsf{s}^{-1})^a{}_{\tilde a}(\accentset{\Xi}{F}^{-1})^A{}_{\tilde A}(\accentset{\xi}{F}^{-1})^b{}_{\tilde b}(\accentset{\Xi}{F}^{-1})^B{}_{\tilde B}\tilde{\mathbb{B}}^{ \tilde a \tilde A \tilde b \tilde B}\,,\quad~\qquad\qquad\qquad
    \end{equation}
    \begin{equation}\label{key-bod-forc}
    \rho_0\mathring{\mathfrak B}^\perp=\frac{1}{2}J_{\Xi}(\mathsf{s}^{-1})^a{}_{\tilde a}(\accentset{\xi}{F}^{-1})^b{}_{\tilde b|\tilde B|\tilde A}\tilde{\mathbb{B}}^{ \tilde a \tilde A \tilde b \tilde B}g_{ab}\,,\qquad\qquad\quad~\qquad\qquad\qquad
    \end{equation}
    \begin{equation}\label{key-bod-mom}
    \rho_0\mathring{\mathfrak L}^b=-\frac{1}{2}J_{\Xi}(\mathsf{s}^{-1})^a{}_{\tilde a}(\accentset{\xi}{F}^{-1})^b{}_{\tilde c}\mathring{\tilde{F} }^{\tilde c}{}_{\tilde A}(\accentset{\xi}{F}^{-1})^c{}_{\tilde b|\tilde B}\tilde{\mathbb{B}}^{ \tilde a \tilde A \tilde b \tilde B}g_{ac}\,,\qquad\qquad\qquad
    \end{equation}
    \begin{equation}\label{key-bod-cons}
    (\mathring{\mathfrak B}^\top)^b=0\,,\qquad\qquad\qquad\qquad\qquad\qquad\qquad\qquad\qquad\qquad\qquad\qquad\quad~\, 
    \end{equation}
\end{subequations}
where in deriving \eqref{key-bod-forc}, the Piola identity $\left[J_{\Xi}(\accentset{\Xi}{F}^{-1})^A{}_{\tilde A}\right]_{|A}=0$, and the fact that the virtual plate has uniform elastic parameters were used.

\begin{remark}
Note that in this case we have two sets of governing equations that need to be simultaneously transformed under a cloaking map: the in-plane and the out-of-plane governing equations (unlike the previous case where the out-of-plane equilibrium equation was the only non-trivial governing equation). Note also that similar to 3D elasticity, the in-plane governing equations are transformed using a Piola transformation. This, in turn, implies that the density is transformed as $\rho_0=J_{\Xi}\tilde{\rho}_0$. Now because density must be transformed the same way for the in-plane and the out-of-plane governing equations, the scalar field $k$ introduced in \S\ref{four4.1} is equal to the Jacobian of the cloaking map $J_{\Xi}$, i.e., $k(X)=J_{\Xi}(X)$.	
\end{remark}
Similarly, we write \eqref{key-main}$_2$ as the out-of-plane governing equation of the physical plate \eqref{key-2-two} up to the Jacobian of the (referential) cloaking map $J_{\Xi}$. Hence, one finds the initial pre-stress, the tangential body force, and the flexural rigidity of the physical plate as\footnote{Note that
\begin{align*}
& (\mathring{F}^{-1})^A{}_{a|B}=\frac{\partial}{\partial X^B}\left[(\mathring{F}^{-1})^A{}_{a}\right]+\Gamma^A{}_{CB}(\mathring{F}^{-1})^C{}_{a}-\gamma^b{}_{ac}(\mathring{F}^{-1})^A{}_{b}\mathring{F}^c{}_B\,. \\
& (\accentset{\xi}{F}^{-1})^a{}_{\tilde a|\tilde{B}}=(\accentset{\xi}{F}^{-1})^a{}_{\tilde a|\tilde{b}}\mathring{\tilde F}^{\tilde b}{}_{\tilde B}=\mathring{\tilde F}^{\tilde b}{}_{\tilde B}\left(\frac{\partial}{\partial\tilde{x}^{\tilde b}}\left[(\accentset{\xi}{F}^{-1})^{ a}{}_{\tilde a}\right]+\gamma^a{}_{cb}(\accentset{\xi}{F}^{-1})^{ b}{}_{\tilde b}(\accentset{\xi}{F}^{-1})^{ c}{}_{\tilde a}-\tilde{\gamma}^{\tilde c}{}_{\tilde a\tilde b}(\accentset{\xi}{F}^{-1})^{ a}{}_{\tilde c}\right)\,,
\end{align*}
where $\tilde{\gamma}^{\tilde c}{}_{\tilde a\tilde b}$ are the (induced) Christoffel symbols corresponding to the virtual plate in its current configuration.
}
\begin{subequations}
	\begin{equation}\label{key-tr-2}
    \mathbb{C}^{aAbB}=J_{\Xi}(\accentset{\xi}{F}^{-1})^a{}_{\tilde a}(\accentset{\Xi}{F}^{-1})^A{}_{\tilde A}(\accentset{\xi}{F}^{-1})^b{}_{\tilde b}(\accentset{\Xi}{F}^{-1})^B{}_{\tilde B}\tilde{\mathbb{C}}^{\tilde a\tilde A\tilde b\tilde B}\,,\qquad\qquad\qquad\qquad\quad\quad\,~~\,\,
    \end{equation}
    \begin{equation}\label{key-tan-bod}
    \rho_0(\mathring{\mathfrak{B}}^\top)^b=\frac{1}{2}J_{\Xi}(\mathring{\tilde{F}}^{-1})^{\tilde B}{}_{\tilde a}(\accentset{\xi}{F}^{-1})^b{}_{\tilde b|\tilde C|\tilde A|\tilde B}\tilde{\mathbb{C}}^{\tilde a\tilde A\tilde b\tilde C}\,,\qquad\qquad\qquad\qquad\qquad\qquad\quad\quad~~~~~~~
    \end{equation}
    \begin{equation}\label{key-pre-stress}
    \begin{split}
    \mathring{P}^{bA}=&\frac{1}{2}\Big[(\mathring{F}^{-1})^B{}_{a|B}\mathbb{C}^{aCbA}{}_{|C}+(\mathring{F}^{-1})^B{}_a\mathbb{C}^{aCbA}{}_{|C|B}\Big]\\&-\frac{1}{2}J_{\Xi}(\mathring{\tilde F}^{-1})^{\tilde{B}}{}_{\tilde a}\tilde{\mathbb{C}}^{\tilde{a}\tilde{A}\tilde{b}\tilde{C}}\Big[(\accentset{\xi}{F}^{-1})^{ b}{}_{\tilde b|\tilde A}(\accentset{\Xi}{F}^{-1})^{ A}{}_{\tilde C|\tilde B}+(\accentset{\xi}{F}^{-1})^{ b}{}_{\tilde b|\tilde C}(\accentset{\Xi}{F}^{-1})^{ A}{}_{\tilde A|\tilde B}\\&+(\accentset{\xi}{F}^{-1})^{ b}{}_{\tilde b|\tilde B}(\accentset{\Xi}{F}^{-1})^{ A}{}_{\tilde C|\tilde A}+(\accentset{\xi}{F}^{-1})^{ b}{}_{\tilde b|\tilde C|\tilde B}(\accentset{\Xi}{F}^{-1})^{ A}{}_{\tilde A}+(\accentset{\xi}{F}^{-1})^{ b}{}_{\tilde b|\tilde A|\tilde B}(\accentset{\Xi}{F}^{-1})^{ A}{}_{\tilde C}\\&+(\accentset{\xi}{F}^{-1})^{ b}{}_{\tilde b|\tilde C|\tilde A}(\accentset{\Xi}{F}^{-1})^{ A}{}_{\tilde B}+(\accentset{\Xi}{F}^{-1})^{ A}{}_{\tilde C|\tilde A|\tilde B}(\accentset{\xi}{F}^{-1})^{ b}{}_{\tilde b}\Big]\,,\qquad\qquad\qquad\quad~~~~~\,\,\,\,~~~
    \end{split}
    \end{equation}
\end{subequations}
along with the following cloaking compatibility equations\footnote{One starts from the governing equations of the virtual plate and substitutes the derivatives with their corresponding transformed derivatives and compares the coefficients of the different derivatives in the transformed governing equations with those in the physical plate. This overdetermined system of equations gives all the transformed fields, and a set of cloaking compatibility equations.}

\begin{subequations}
\begin{align}\label{key-constraint-11}
	& (\mathring{F}^{-1})^B{}_{a|B}\mathbb{C}^{aAbC}+(\mathring{F}^{-1})^B{}_a\mathbb{C}^{aAbC}{}_{|B}
	+(\mathring{F}^{-1})^A{}_a\mathbb{C}^{aBbC}{}_{|B} \nonumber\\
	&~~~~~=J_{\Xi}(\mathring{\tilde F}^{-1})^{\tilde{B}}{}_{\tilde a}\tilde{\mathbb{C}}^{\tilde{a}\tilde{A}\tilde{b}\tilde{C}}
	\Big[(\accentset{\xi}{F}^{-1})^{ b}{}_{\tilde b|\tilde B}(\accentset{\Xi}{F}^{-1})^{ C}{}_{\tilde C}
	(\accentset{\Xi}{F}^{-1})^{ A}{}_{\tilde A}+(\accentset{\Xi}{F}^{-1})^{ C}{}_{\tilde C|\tilde B}
	(\accentset{\xi}{F}^{-1})^{ b}{}_{\tilde b}(\accentset{\Xi}{F}^{-1})^{ A}{}_{\tilde A} \nonumber\\
	&~~~~~+(\accentset{\Xi}{F}^{-1})^{ A}{}_{\tilde A|\tilde B}(\accentset{\xi}{F}^{-1})^{ b}{}_{\tilde b}
	(\accentset{\Xi}{F}^{-1})^{ C}{}_{\tilde C}+(\accentset{\xi}{F}^{-1})^{ b}{}_{\tilde b|\tilde C}
	(\accentset{\Xi}{F}^{-1})^{ C}{}_{\tilde A}(\accentset{\Xi}{F}^{-1})^{ A}{}_{\tilde B} \nonumber\\
	&~~~~~+(\accentset{\xi}{F}^{-1})^{ b}{}_{\tilde b|\tilde A}(\accentset{\Xi}{F}^{-1})^{ A}{}_{\tilde B}
	(\accentset{\Xi}{F}^{-1})^{ C}{}_{\tilde C}+(\accentset{\Xi}{F}^{-1})^{ C}{}_{\tilde C|\tilde A}
	(\accentset{\Xi}{F}^{-1})^{ A}{}_{\tilde B}(\accentset{\xi}{F}^{-1})^{ b}{}_{\tilde b}\Big]\,, \\
	\label{key-cons2}
	& (\mathring{F}^{-1})^B{}_{a|B}\mathbb{B}^{bCaA}{}_{|A}
	+(\mathring{F}^{-1})^B{}_{a}\mathbb{B}^{bCaA}{}_{|A|B}+2\left[\rho_0\left(\mathring{\mathfrak L}^ag^{bc}
	-\mathring{\mathfrak L}^bg^{ac}\right)(\mathring{F}^{-1})^C{}_{a}
	(\mathring{F}^{-1})^A{}_{c}\right]_{|A} \nonumber\\
	&~~~~~=J_{\Xi}(\mathring{\tilde{F}}^{-1})^{\tilde B}{}_{\tilde a}(\mathsf{s}^{-1})^b{}_{\tilde b}
	\tilde{\mathbb{B}}^{\tilde b\tilde C\tilde a\tilde A}(\accentset{\Xi}{F}^{-1})^C{}_{\tilde  C|\tilde A|\tilde B}\,, \\
	\label{key-cons3}
	& (\mathring{F}^{-1})^B{}_{a|B}\mathbb{B}^{bCaA}+(\mathring{F}^{-1})^A{}_a\mathbb{B}^{bCaB}{}_{|B}
	+(\mathring{F}^{-1})^B{}_a\mathbb{B}^{bCaA}{}_{|B}+2\rho_0\left(\mathring{\mathfrak L}^ag^{bc}
	-\mathring{\mathfrak L}^bg^{ac}\right)(\mathring{F}^{-1})^C{}_a(\mathring{F}^{-1})^A{}_c \nonumber\\
	&~~~~~=J_{\Xi}(\mathring{\tilde{F}}^{-1})^{\tilde B}{}_{\tilde a}(\mathsf{s}^{-1})^b{}_{\tilde b}
	\tilde{\mathbb{B}}^{\tilde b\tilde C\tilde a\tilde A}\Big[(\accentset{\Xi}{F}^{-1})^C{}_{\tilde C|\tilde A}
	(\accentset{\Xi}{F}^{-1})^A{}_{\tilde B}+(\accentset{\Xi}{F}^{-1})^C{}_{\tilde C|\tilde B}
	(\accentset{\Xi}{F}^{-1})^A{}_{\tilde A} \nonumber\\
	&~~~~~+(\accentset{\Xi}{F}^{-1})^C{}_{\tilde C}
	(\accentset{\Xi}{F}^{-1})^A{}_{\tilde A|\tilde B}\Big]\,,
\end{align}
\end{subequations}
and the ones given by \eqref{key-bod-cons} and \eqref{key-tan-bod}. Notice that $\pmb{\mathbb C}$ already possesses the major symmetries. 
Recalling that the elastic parameters of the virtual plate satisfy \eqref{key-ang-vir}, the relations $\mathbb{C}^{[aAcB}\mathring{F}^{b]}{}_A=0$, and $\mathbb{B}^{cB[aA}\mathring{F}^{b]}{}_A=0$, already hold, i.e., 
\begin{subequations}
\begin{equation}
\begin{split}
\mathbb{B}^{cB[aA}\mathring{F}^{b]}{}_A&=J_{\Xi}(\mathsf{s}^{-1})^c{}_{\tilde c}(\accentset{\Xi}{F}^{-1})^B{}_{\tilde B}(\accentset{\xi}{F}^{-1})^{[a}{}_{\tilde a}(\accentset{\Xi}{F}^{-1})^A{}_{\tilde A}\mathring{F}^{b]}{}_A\tilde{\mathbb{B}}^{ \tilde c \tilde B \tilde a \tilde A}\\&=J_{\Xi}(\mathsf{s}^{-1})^c{}_{\tilde c}(\accentset{\Xi}{F}^{-1})^B{}_{\tilde B}(\accentset{\xi}{F}^{-1})^{[a}{}_{\tilde a}(\accentset{\xi}{F}^{-1})^{b]}{}_{\tilde b}\mathring{\tilde F}^{\tilde b}{}_{\tilde A}\tilde{\mathbb{B}}^{ \tilde c \tilde B \tilde a \tilde A}\\&=J_{\Xi}(\mathsf{s}^{-1})^c{}_{\tilde c}(\accentset{\Xi}{F}^{-1})^B{}_{\tilde B}(\accentset{\xi}{F}^{-1})^{[b}{}_{\tilde a}(\accentset{\xi}{F}^{-1})^{a]}{}_{\tilde b}\mathring{\tilde F}^{\tilde b}{}_{\tilde A}\tilde{\mathbb{B}}^{ \tilde c \tilde B \tilde a \tilde A}=0\,,\,
\end{split}
\end{equation}
\begin{equation}
\begin{split}
\mathbb{C}^{[aAcB}\mathring{F}^{b]}{}_A&=J_{\Xi}(\accentset{\xi}{F}^{-1})^c{}_{\tilde c}(\accentset{\Xi}{F}^{-1})^B{}_{\tilde B}(\accentset{\xi}{F}^{-1})^{[a}{}_{\tilde a}(\accentset{\Xi}{F}^{-1})^A{}_{\tilde A}\mathring{F}^{b]}{}_A\tilde{\mathbb{C}}^{\tilde a\tilde A\tilde c\tilde B}\\&=J_{\Xi}(\accentset{\xi}{F}^{-1})^c{}_{\tilde c}(\accentset{\Xi}{F}^{-1})^B{}_{\tilde B}(\accentset{\xi}{F}^{-1})^{[a}{}_{\tilde a}(\accentset{\xi}{F}^{-1})^{b]}{}_{\tilde b}\mathring{\tilde F}^{\tilde b}{}_{\tilde A}\tilde{\mathbb{C}}^{\tilde a\tilde A\tilde c\tilde B}\\&=J_{\Xi}(\accentset{\xi}{F}^{-1})^c{}_{\tilde c}(\accentset{\Xi}{F}^{-1})^B{}_{\tilde B}(\accentset{\xi}{F}^{-1})^{[b}{}_{\tilde a}(\accentset{\xi}{F}^{-1})^{a]}{}_{\tilde b}\mathring{\tilde F}^{\tilde b}{}_{\tilde A}\tilde{\mathbb{C}}^{\tilde a\tilde A\tilde c\tilde B}=0\,.
\end{split}
\end{equation}
\end{subequations}
On the other hand, $\mathbb{A}^{[aAcB}\mathring{F}^{b]}{}_A=0$, and $\mathbb{B}^{[aAcB}\mathring{F}^{b]}{}_A=0$, respectively, imply that  
\begin{subequations}
\begin{equation}
\begin{split}
\mathbb{A}^{[aAcB}\mathring{F}^{b]}{}_A&=J_{\Xi}(\mathsf{s}^{-1})^c{}_{\tilde c}(\accentset{\Xi}{F}^{-1})^B{}_{\tilde B}(\mathsf{s}^{-1})^{[a}{}_{\tilde a}(\accentset{\Xi}{F}^{-1})^A{}_{\tilde A}\mathring{F}^{b]}{}_A\tilde{\mathbb{A}}^{ \tilde a \tilde A \tilde c \tilde B}-\mathring{P}^{dA}(\mathring{F}^{-1})^B{}_dg^{[ac}\mathring{F}^{b]}{}_A\\&=J_{\Xi}(\mathsf{s}^{-1})^c{}_{\tilde c}(\accentset{\Xi}{F}^{-1})^B{}_{\tilde B}(\mathsf{s}^{-1})^{[a}{}_{\tilde a}(\accentset{\xi}{F}^{-1})^{b]}{}_{\tilde b}\mathring{\tilde F}^{\tilde b}{}_{\tilde A}\tilde{\mathbb{A}}^{ \tilde a \tilde A \tilde c \tilde B}-\mathring{P}^{dA}(\mathring{F}^{-1})^B{}_dg^{[ac}\mathring{F}^{b]}{}_A=0\,,
\end{split}
\end{equation}
\begin{equation}
\begin{split}
\mathbb{B}^{[aAcB}\mathring{F}^{b]}{}_A&=J_{\Xi}(\accentset{\xi}{F}^{-1})^c{}_{\tilde c}(\accentset{\Xi}{F}^{-1})^B{}_{\tilde B}(\mathsf{s}^{-1})^{[a}{}_{\tilde a}(\accentset{\Xi}{F}^{-1})^A{}_{\tilde A}\mathring{F}^{b]}{}_A\tilde{\mathbb{B}}^{ \tilde a \tilde A \tilde c \tilde B}\\&=J_{\Xi}(\accentset{\xi}{F}^{-1})^c{}_{\tilde c}(\accentset{\Xi}{F}^{-1})^B{}_{\tilde B}(\mathsf{s}^{-1})^{[a}{}_{\tilde a}(\accentset{\xi}{F}^{-1})^{b]}{}_{\tilde b}\mathring{\tilde F}^{\tilde b}{}_{\tilde A}\tilde{\mathbb{B}}^{ \tilde a \tilde A \tilde c \tilde B}=0\,.\qquad\qquad\qquad\qquad\quad\quad
\end{split}
\end{equation}
\end{subequations}
Pushing forward theses expressions to the current configuration, one obtains
\begin{subequations}
\begin{equation}\label{key-1-non}
\begin{split}
	\mathbbm{c}^{[ab]cd} &=\frac{1}{\mathring{J}}\mathbb{A}^{[aAcB}\mathring{F}^{b]}{}_A\mathring{F}^{d}{}_B \\
	& =\frac{J_{\Xi}}{\mathring{J}}(\mathsf{s}^{-1})^c{}_{\tilde c}(\accentset{\Xi}{F}^{-1})^B{}_{\tilde B}
	\mathring{F}^{d}{}_B(\mathsf{s}^{-1})^{[a}{}_{\tilde a}(\accentset{\xi}{F}^{-1})^{b]}{}_{\tilde b}
	\mathring{\tilde F}^{\tilde b}{}_{\tilde A}\tilde{\mathbb{A}}^{ \tilde a \tilde A \tilde c \tilde B}\\
	&-\frac{1}{\mathring{J}}\mathring{P}^{eA}(\mathring{F}^{-1})^B{}_e\mathring{F}^{d}{}_Bg^{[ac}
	\mathring{F}^{b]}{}_A\\&=J_{\xi}(\mathsf{s}^{-1})^c{}_{\tilde c}
	(\accentset{\xi}{F}^{-1})^d{}_{\tilde d}(\mathsf{s}^{-1})^{[a}{}_{\tilde a}
	(\accentset{\xi}{F}^{-1})^{b]}{}_{\tilde b}\frac{1}{\mathring{\tilde J}}\mathring{\tilde F}^{\tilde b}{}_{\tilde A}
	\mathring{\tilde F}^{\tilde d}{}_{\tilde B}\tilde{\mathbb{A}}^{ \tilde a \tilde A \tilde c \tilde B}\\
	&-\frac{1}{\mathring{J}}\mathring{P}^{eA}(\mathring{F}^{-1})^B{}_e\mathring{F}^{d}{}_B
	g^{[ac}\mathring{F}^{b]}{}_A\\
	&=J_{\xi}(\mathsf{s}^{-1})^c{}_{\tilde c}(\accentset{\xi}{F}^{-1})^d{}_{\tilde d}
	(\mathsf{s}^{-1})^{[a}{}_{\tilde a}(\accentset{\xi}{F}^{-1})^{b]}{}_{\tilde b}
	\tilde{\mathbbm{c}}^{ \tilde a \tilde b \tilde c \tilde d}-\mathring{\sigma}^{[bd}g^{ca]}=0\,,
\end{split}
\end{equation}
\begin{equation}\label{key-2-non}
\begin{split}
	\mathbbm{b}^{[ab]cd} &=\frac{1}{\mathring J}\mathbb{B}^{[aAcB}\mathring{F}^{b]}{}_A
	\mathring{F}^d{}_B \\
	&=\frac{J_{\Xi}}{\mathring J}(\accentset{\xi}{F}^{-1})^c{}_{\tilde c}
	(\accentset{\Xi}{F}^{-1})^B{}_{\tilde B}\mathring{ F}^{ d}{}_{ B}(\mathsf{s}^{-1})^{[a}{}_{\tilde a}
	(\accentset{\xi}{F}^{-1})^{b]}{}_{\tilde b}\mathring{\tilde F}^{\tilde b}{}_{\tilde A}\tilde{\mathbb{B}}^{ \tilde a 
	\tilde A \tilde c \tilde B}\\
	&=J_{\xi}(\accentset{\xi}{F}^{-1})^c{}_{\tilde c}(\accentset{\xi}{F}^{-1})^d{}_{\tilde d}
	(\mathsf{s}^{-1})^{[a}{}_{\tilde a}(\accentset{\xi}{F}^{-1})^{b]}{}_{\tilde b}\frac{1}{\mathring{\tilde J}}
	\mathring{\tilde F}^{\tilde b}{}_{\tilde A}\mathring{\tilde F}^{\tilde d}{}_{\tilde B}\tilde{\mathbb{B}}^{ \tilde a 
	\tilde A \tilde c \tilde B}\\&=J_{\xi}(\accentset{\xi}{F}^{-1})^c{}_{\tilde c}
	(\accentset{\xi}{F}^{-1})^d{}_{\tilde d}(\mathsf{s}^{-1})^{[a}{}_{\tilde a}
	(\accentset{\xi}{F}^{-1})^{b]}{}_{\tilde b}\tilde{\mathbbm{b}}^{ \tilde a \tilde b \tilde c \tilde d}=0\,.\qquad\qquad
\end{split}
\end{equation}
\end{subequations}
Additionally, the initial body forces and the pre-stress need to satisfy the following balance of linear and angular momentum in the finitely deformed configuration
\begin{subequations}\label{key-all}
	\begin{equation}\label{key-key}
	\mathring{P}^{aA}{}_{|A}+\rho_0(\mathring{\mathfrak{B}}^\top)^a=0\,,\qquad\qquad
	\end{equation}
	\begin{equation}\label{key-two-sat}
	\rho_0\mathring{\mathfrak B}^\perp+\left[\rho_0\mathring{\mathfrak{L}}^a(\mathring{F}^{-1})^A{}_a\right]_{|A}=0\,,\qquad\qquad\qquad\quad~
	\end{equation}
	\begin{equation}\label{key-three-sat}
	\mathring{P}^{[aA}\mathring{F}^{b]}{}_A=0\,.
	\end{equation}
\end{subequations}
Notice that if \eqref{key-three-sat} holds (or, equivalently, $\mathring{\boldsymbol{\sigma}}$ is symmetric), then $\pmb{\mathbb A}$ possesses the major symmetries 
(cf. \eqref{key-trans-1}). The stress and couple-stress are transformed as
\begin{equation}
\begin{split}
\delta P^{aA}=&J_{\Xi}(\mathsf{s}^{-1})^a{}_{\tilde a}(\accentset{\Xi}{F}^{-1})^A{}_{\tilde A}\delta\tilde{P}^{\tilde a\tilde A}\\&+\frac{1}{2}J_{\Xi}(\mathsf{s}^{-1})^a{}_{\tilde a}(\accentset{\Xi}{F}^{-1})^A{}_{\tilde A}(\accentset{\xi}{F}^{-1})^b{}_{\tilde b}(\accentset{\Xi}{F}^{-1})^B{}_{\tilde B}\accentset{\xi}{F}^{\tilde c}{}_{b|B}\tilde{\mathbb{B}}^{\tilde a\tilde A\tilde b\tilde B}\widetilde{\mathsf W}_{,\tilde c}\,,\\
\delta\mathsf{M}^{aA}=&J_{\Xi}(\accentset{\xi}{F}^{-1})^a{}_{\tilde a}(\accentset{\Xi}{F}^{-1})^A{}_{\tilde A}\delta\tilde{\mathsf M}^{\tilde a\tilde A}\\&+\frac{1}{2}J_{\Xi}(\accentset{\xi}{F}^{-1})^a{}_{\tilde a}(\accentset{\Xi}{F}^{-1})^A{}_{\tilde A}(\accentset{\xi}{F}^{-1})^b{}_{\tilde b}(\accentset{\Xi}{F}^{-1})^B{}_{\tilde B}\accentset{\xi}{F}^{\tilde c}{}_{b|B}\tilde{\mathbb{C}}^{\tilde a\tilde A\tilde b\tilde B}\widetilde{\mathsf W}_{,\tilde c}\,.
\end{split}
\end{equation}
Using \eqref{key-bound-2}, one needs to have $\widetilde{\mathsf W}_{,\tilde a}=\mathsf{W}_{,a}(\mathsf{s}^{-1})^a{}_{\tilde a}$ on the boundary of the cloak $\partial_o\mathcal{C}$, which implies that $T\xi|_{\partial_o\mathcal{C}}=\accentset{\xi}{\mathbf F}|_{\partial_o\mathcal{C}}=id$. Additionally, the boundary surface traction and moment in the physical and virtual plates need to be identical on the boundary of the cloak, i.e., $(\delta\tilde{T}^\top)^{\tilde a}=\mathsf{s}^{\tilde a}{}_{a}(\delta T^\top)^a$, $\delta\tilde{T}^\perp=\delta{T}^\perp$, and $\delta\tilde{\mathsf{m}}^{\tilde a}=\mathsf{s}^{\tilde a}{}_a\delta{\mathsf{m}}^a$, on $\partial_o\mathcal{C}$ (see \eqref{key-bound-2}). Therefore, one needs to impose the following constraints on the outer boundary of the cloak: $\accentset{\xi}{F}^{\tilde a}{}_{a|A}\Big|_{\partial_o\mathcal{C}}=0$, $\accentset{\xi}{F}^{\tilde a}{}_{a|A|B}\Big|_{\partial_o\mathcal{C}}=0$, $T\Xi|_{\partial_o\mathcal C}=\accentset{\Xi}{\mathbf F}|_{\partial_o\mathcal C}=id$ (and thus, $\mathring{\mathbf{F}}|_{\partial_o\mathcal C}=id$, given that $\accentset{\xi}{\mathbf F}|_{\partial_o\mathcal{C}}=id$ and $\accentset{\xi}{\mathbf F}=\mathring{\tilde{\mathbf{F}}}\accentset{\Xi}{\mathbf F}\mathring{\mathbf F}^{-1}$). Similarly, knowing that the hole surface $\partial\tilde{\mathcal{E}}$ in the virtual plate is traction-free, the hole inner surface in the physical plate $\partial\mathcal{E}$ will be traction-free as well if one requires that $\accentset{\xi}{F}^{\tilde a}{}_{a|A}\Big|_{\partial\mathcal{E}}=0$, and $\accentset{\xi}{F}^{\tilde a}{}_{a|A|B}\Big|_{\partial\mathcal{E}}=0$.\footnote{Note that $\accentset{\xi}{F}^{\tilde a}{}_{a|A}\Big|_{\partial_o\mathcal{C}}=0$, implies that $\mathring{\boldsymbol{\mathfrak{L}}}|_{\partial_o\mathcal{C}}=\mathbf{0}$, and $\delta\boldsymbol{\mathfrak{L}}|_{\partial_o\mathcal{C}}=\mathbf{0}$, (see \eqref{key-bound-2} and \eqref{key-bod-mom}), and thus
\begin{equation*}
\begin{split}
\delta T^\perp&=-(\mathring{F}^{-1})^A{}_a\delta\mathsf{M}^{aB}{}_{|B}\mathsf{T}_{A}=-\left[J_{\Xi}(\accentset{\Xi}{F}^{-1})^B{}_{\tilde B}(\accentset{\xi}{F}^{-1})^a{}_{\tilde a}\delta\tilde{\mathsf M}^{\tilde a\tilde B}\right]_{|B}(\mathring{F}^{-1})^A{}_a\mathsf{T}_A\\&=-J_{\Xi}(\accentset{\Xi}{F}^{-1})^B{}_{\tilde B}(\accentset{\xi}{F}^{-1})^a{}_{\tilde a}\delta\tilde{\mathsf M}^{\tilde a\tilde B}{}_{|B}(\mathring{F}^{-1})^A{}_a\mathsf{T}_A=-(\mathring{\tilde F}^{-1})^{\tilde A}{}_{\tilde a}\delta\mathsf{M}^{\tilde a\tilde B}{}_{|\tilde B}\tilde{\mathsf T}_{\tilde A}=\delta \tilde{T}^\perp\,,
\end{split}\qquad \mathrm{on}\quad \partial_o\mathcal{C}\,,
\end{equation*}
where $\left[J_{\Xi}(\accentset{\Xi}{F}^{-1})^B{}_{\tilde B}\right]_{|B}=0$, and the fact that $\Xi$ (and $\xi$) fixes the boundary of the cloak $\partial_o\mathcal{C}$ to the third order were used.
} Furthermore, the hole must be traction-free in the physical plate in its finitely-deformed configuration, i.e., from \eqref{key-shell-hole}, one needs to have $(\mathring{{T}}^\top)^a|_{\partial\mathcal{E}}=\left(\mathring{P}^{aA}\mathsf{T}_A\right)\Big|_{\partial\mathcal{E}}={0}$, and $\mathring{T}^\perp|_{\partial\mathcal{E}}=-\left[(\mathring{F}^{-1})^A{}_a\rho\mathring{\mathfrak{L}}^a\mathsf{T}_{A}\right]\Big|_{\partial\mathcal{E}}=0$.

\begin{remark}\label{key-consq}
Upon using \eqref{key-tr-2} and the Piola identity, the cloaking compatibility equations \eqref{key-constraint-11} is simplified to read
\begin{equation}
\tilde{\mathbb{C}}^{\tilde a\tilde A\tilde b\tilde C}(\accentset{\xi}{F}^{-1})^b{}_{\tilde b}(\accentset{\Xi}{F}^{-1})^C{}_{\tilde C}(\mathring{F}^{-1})^A{}_a(\accentset{\xi}{F}^{-1})^a{}_{\tilde a|\tilde A}=\tilde{\mathbb C}^{\tilde a \tilde A\tilde b\tilde C}(\accentset{\xi}{F}^{-1})^c{}_{\tilde a}(\accentset{\Xi}{F}^{-1})^C{}_{\tilde A}(\mathring{F}^{-1})^A{}_c(\accentset{\xi}{F}^{-1})^b{}_{\tilde b|\tilde C}\,.
\end{equation}
\end{remark}

\begin{remark}
Note that for an isotropic (virtual) plate, the elastic constant $\tilde{\pmb{\mathbb{B}}}$ is given (with a slight abuse of notation) by
\begin{equation}\label{key-bb}
\tilde{\mathbb{B}}^{ \tilde a \tilde A \tilde b \tilde B}=\tilde{\mathsf{b}}_1\tilde{G}^{\tilde a \tilde A}\tilde{G}^{\tilde b\tilde B}+\tilde{\mathsf{b}}_2(\tilde{G}^{\tilde a\tilde b}\tilde{G}^{\tilde A\tilde B}+\tilde{G}^{\tilde a\tilde B}\tilde{G}^{\tilde b\tilde A})\,,
\end{equation}
for some scalars $\tilde{\mathsf{b}}_1$ and $\tilde{\mathsf{b}}_2$. This is a consequence of the fact that the most general form of a fourth-order isotropic tensor is $a_1\delta_{ij}\delta_{kl}+a_2\delta_{ik}\delta_{jl}+a_3\delta_{il}\delta_{jk}$ (for some scalars $a_i$, $i=1,2,3$) and the minor symmetries of $\tilde{\pmb{\mathbb{B}}}$ dictated by \eqref{key-ang-vir}.
\end{remark}
\begin{remark}
When restricting to the in-plane deformations, we recover our result \citep[\S4.4]{yavari2018nonlinear} for the elastodynamic cloaking of a cylindrical hole in the context of the small-on-large theory of (classical) elasticity in $3$D. For in-plane deformations, $\mathsf{W}=0$, and for a classical solid $\pmb{\mathbb B}$ (and $\pmb{\mathbb C}$) vanishes. Therefore, the out-of-plane equilibrium equation \eqref{key-2-two} is trivially satisfied, and the in-plane equilibrium equation \eqref{key-1-one} would be the only non-trivial governing equation. Note that in this case, the pre-stress is determined such that the (only non-trivial) balance of angular momentum \eqref{key-1-non} is satisfied. We should emphasis that in \citep{yavari2018nonlinear}, the variation of the body force is assumed to be independent of that of the motion. Therefore, the linearized equations involve the load increment $\delta\boldsymbol{\mathfrak B}$ independently of the infinitesimal deformation $\delta\varphi$ (see \citep[p.237]{MaHu1983}).
\end{remark}
%

\begin{remark}
Note that \eqref{key-two-sat} is trivially satisfied. To see this, using \eqref{key-bod-forc} and \eqref{key-bod-mom}, the expression in \eqref{key-two-sat} is simplified to read
\begin{equation}
\begin{split}
	\frac{1}{2}&J_{\Xi}(\mathsf{s}^{-1})^a{}_{\tilde a}(\accentset{\xi}{F}^{-1})^b{}_{\tilde b|\tilde B|\tilde A}\tilde{\mathbb{B}}^{ \tilde a \tilde A \tilde b \tilde B}g_{ab}-\frac{1}{2}\tilde{\mathbb{B}}^{ \tilde a \tilde A \tilde b \tilde B}g_{ac}(\mathsf{s}^{-1})^a{}_{\tilde a}\mathring{\tilde{F} }^{\tilde c}{}_{\tilde A}\left[J_{\Xi}(\accentset{\xi}{F}^{-1})^b{}_{\tilde c}(\accentset{\xi}{F}^{-1})^c{}_{\tilde b|\tilde B}(\mathring{F}^{-1})^A{}_b\right]_{|A}\\&=-\frac{1}{2}\tilde{\mathbb{B}}^{ \tilde a \tilde A \tilde b \tilde B}g_{ac}(\mathsf{s}^{-1})^a{}_{\tilde a}\mathring{\tilde{F} }^{\tilde c}{}_{\tilde A}(\accentset{\xi}{F}^{-1})^c{}_{\tilde b|\tilde B}\left[J_{\Xi}(\accentset{\xi}{F}^{-1})^b{}_{\tilde c}(\mathring{F}^{-1})^A{}_b\right]_{|A}\\&=-\frac{1}{2}\tilde{\mathbb{B}}^{ \tilde a \tilde A \tilde b \tilde B}g_{ac}(\mathsf{s}^{-1})^a{}_{\tilde a}(\accentset{\xi}{F}^{-1})^c{}_{\tilde b|\tilde B}\left[J_{\Xi}(\accentset{\Xi}{F}^{-1})^A{}_{\tilde A}\right]_{|A}=0\,,
	\end{split}
	\end{equation}
	where the relation $\accentset{\xi}{\mathbf F}=\mathring{\tilde{\mathbf{F}}}\accentset{\Xi}{\mathbf F}\mathring{\mathbf F}^{-1}$, and the Piola identity were used.
\end{remark}

\begin{remark}
If $\tilde{\pmb{\mathbb B}}$ vanishes for the virtual plate, which is the case if one assumes the Saint Venant-Kirchhoff energy function (cf. \eqref{saint}), then so does the tensor $\pmb{\mathbb B}$ for the physical plate, i.e., $\pmb{\mathbb B}=\mathbf{0}$. Moreover, if $\tilde{\pmb{\mathbb B}}=\mathbf{0}$, then $\mathring{\boldsymbol{\mathfrak B}}^\perp=\mathbf{0}$, $\mathring{\boldsymbol{\mathfrak L}}=\mathbf{0}$, and the cloaking compatibility equations \eqref{key-cons2} and \eqref{key-cons3} are trivially satisfied.  
\end{remark}

\begin{remark}
The second-order change (variation) in the energy (density) of the physical plate is given by
\begin{equation}
\delta^2 W=\frac{1}{2}\frac{\partial^2 W}{\partial C_{AB}\partial C_{CD}}\delta C_{AB}\delta C_{CD}+\frac{\partial^2 W}{\partial C_{AB}\partial \Theta_{CD}}\delta C_{AB}\delta \Theta_{CD}+\frac{1}{2}\frac{\partial^2 W}{\partial \Theta_{AB}\partial \Theta_{CD}}\delta \Theta_{AB}\delta \Theta_{CD} \,.
\end{equation}
Therefore, one obtains
\begin{equation}\label{key-11}
\delta^2 W=\frac{1}{2}\mathbb{A}^{aAbB}\mathsf{U}_{a|A}\mathsf{U}_{b|B}+\frac{1}{2}\mathbb{B}^{aAbB}\mathsf{U}_{a|A}(\mathsf{W}_{,b})_{|B}+\frac{1}{4}\mathbb{C}^{aAbB} (\mathsf{W}_{,a})_{|A}(\mathsf{W}_{,b})_{|B}\,.
\end{equation}
Similarly, we note that for the virtual plate
\begin{equation}\label{key-vvir}
\delta^2 \tilde{W}=\frac{1}{2}\tilde{\mathbb{A}}^{\tilde a\tilde A\tilde b\tilde B}\tilde{\mathsf{U}}_{\tilde a|\tilde A}\tilde{\mathsf{U}}_{\tilde b|\tilde B}+\frac{1}{2}\tilde{\mathbb{B}}^{\tilde a\tilde A\tilde b\tilde B}\tilde{\mathsf{U}}_{\tilde a|\tilde A}(\widetilde{\mathsf{W}}_{,\tilde b})_{|\tilde B}+\frac{1}{4}\tilde{\mathbb{C}}^{\tilde a\tilde A\tilde b\tilde B} (\widetilde{\mathsf{W}}_{,\tilde a})_{|\tilde A}(\widetilde{\mathsf{W}}_{,\tilde b})_{|\tilde B}\,.
\end{equation}
Using \eqref{key-grad-disp}, \eqref{disp-grad-both}, \eqref{key-trans-1}, \eqref{key-b-vir-phy}, \eqref{key-tr-2}, and \eqref{key-vvir}, one rewrites \eqref{key-11} as
\begin{equation}
\begin{split}
\delta^2 W=&J_{\Xi}\delta^2 \tilde{W}-\frac{1}{2}\mathring{P}^{cA}(\mathring{F}^{-1})^B{}_cg^{ab}\mathsf{s}^{\tilde a}{}_{a}\mathsf{s}^{\tilde b}{}_b\accentset{\Xi}{F}^{\tilde A}{}_A\accentset{\Xi}{F}^{\tilde B}{}_B\tilde{\mathsf U}_{\tilde a|\tilde A}\tilde{\mathsf U}_{\tilde b|\tilde B}\\&+\frac{1}{2}J_{\Xi}(\accentset{\xi}{F}^{-1})^b{}_{\tilde b}(\accentset{\Xi}{F}^{-1})^B{}_{\tilde B}\accentset{\xi}{F}^{\tilde c}{}_{b|B}\tilde{\mathbb B}^{\tilde a\tilde A\tilde b\tilde B}\tilde{\mathsf U}_{\tilde a|\tilde A}\widetilde{\mathsf W}_{,\tilde c}\\&+\frac{1}{2}J_{\Xi}(\accentset{\xi}{F}^{-1})^a{}_{\tilde a}(\accentset{\Xi}{F}^{-1})^A{}_{\tilde A}\accentset{\xi}{F}^{\tilde c}{}_{a|A}\tilde{\mathbb C}^{\tilde a\tilde A\tilde b\tilde B}\widetilde{\mathsf W}_{,\tilde c}(\widetilde{\mathsf W}_{,\tilde b})_{|\tilde B}\\&+\frac{1}{4}J_{\Xi}(\accentset{\xi}{F}^{-1})^a{}_{\tilde a}(\accentset{\Xi}{F}^{-1})^A{}_{\tilde A}(\accentset{\xi}{F}^{-1})^b{}_{\tilde b}(\accentset{\Xi}{F}^{-1})^B{}_{\tilde B}\accentset{\xi}{F}^{\tilde c}{}_{a|A}\accentset{\xi}{F}^{\tilde d}{}_{b|B}\tilde{\mathbb C}^{\tilde a\tilde A\tilde b\tilde B}\widetilde{\mathsf W}_{,\tilde c}\widetilde{\mathsf W}_{,\tilde d}\,.
\end{split}
\end{equation}
Hence, the positive-definiteness of the second order variation of the energy of the physical plate involves that of the virtual plate (i.e., $\delta^2 \tilde{W}$), along with the elastic parameters and (in-plane and out-of-plane) displacements of the virtual plate, the first and the second derivatives of the cloaking map, and the pre-stress. This is in contrast to transformation cloaking in classical (and generalized Cosserat) $3$D elasticity, where $\delta^2W=J_{\Xi}\delta^2\tilde{W}$ (see \citep{yavari2018nonlinear}). Thus, one cannot simply conclude that (the second order variation of) the energy density is positive-definite in the physical problem if and only if it is positive-definite in the virtual problem.
\end{remark}

%

\subsubsection{A circular cloak in the presence of in-plane and out-of-plane displacements}

Let us consider the cylindrical cloak example in the presence of the initial stress $\mathring{\mathbf P}$, the initial body force $\mathring{\boldsymbol{\mathfrak B}}$, and the initial body moment $\mathring{\boldsymbol{\mathfrak L}}$. 
The cloaking transformation $\xi$ maps a pre-stressed cylindrical annulus (in the physical plate) with inner and outer radii $r_i$ and $r_o$, respectively, to a cylindrical annulus (in the virtual plate) with inner and outer radii $\epsilon$ and $r_o$, respectively. Let, in polar coordinates, $(\tilde{r},\tilde{\theta})=\xi(r,\theta)=(f(r),\theta)$. Therefore 
\begin{equation}
\accentset{\xi}{\mathbf F}=\begin{bmatrix}
f'(r) & 0 \\
0 & 1     
\end{bmatrix}\,,
\end{equation}
where $f(r_o)=r_o$ and $f(r_i)=\epsilon$.
Let $(\tilde{R},\tilde{\Theta})=\Xi(R,\Theta)$, $(r,\theta)=\varphi(R,\Theta)$, and $(\tilde{r},\tilde{\theta})=\tilde{\varphi}(\tilde{R},\tilde{\Theta})$ such that $\mathring{\tilde\varphi}=id$ (and thus, $\mathring{\tilde{\mathbf{F}}}=id$). We assume that the physical plate is finitely deformed such that $(r,\theta)=\varphi(R,\Theta)=(\psi(R),\Theta)$. Therefore 
\begin{equation}
\mathring{\mathbf F}=\begin{bmatrix}
\psi'(R) & 0 \\
0 & 1     
\end{bmatrix}\,.
\end{equation}
Using $\accentset{\xi}{\mathbf F}=\mathring{\tilde{\mathbf{F}}}\accentset{\Xi}{\mathbf F}\mathring{\mathbf F}^{-1}$, one has
\begin{equation}
\accentset{\Xi}{\mathbf F}=\begin{bmatrix}
\psi'(R)f'(\psi(R)) & 0 \\
0 & 1     
\end{bmatrix}\,,
\end{equation}
and thus, $J_{\Xi}=\psi'(R)f(\psi(R))f'(\psi(R))/R$. The referential mass density of the cloak is given by
\begin{equation}
	\rho_0(R)=\psi'(R)f'(\psi(R))\frac{f(\psi(R))}{R}\tilde{\rho}_0\,,\quad R_i\leq R\leq R_o\,.
\end{equation}
Note that $\boldsymbol{\mathsf s}=\mathrm{diag}\left(1,{r}/{f(r)}\right)$.
Assuming the Saint Venant-Kirchhoff constitutive equation \eqref{saint} for the virtual plate, $\tilde{\pmb{\mathbb{C}}}$ is given by \eqref{key-vir-rig}, $\tilde{\pmb{\mathbb{B}}}=\mathbf{0}$ (and thus, ${\pmb{\mathbb{B}}}=\mathbf{0}$, see \eqref{key-b-vir-phy}, which using \eqref{key-bod-forc} and \eqref{key-bod-mom} implies that $\mathring{\boldsymbol{\mathfrak B}}^\perp=\mathbf{0}$, and $\mathring{\boldsymbol{\mathfrak L}}=\mathbf{0}$), and $\tilde{\pmb{\mathbb{A}}}$ is obtained as\footnote{Note that in the case of a general isotropic energy function for the virtual plate ${\pmb{\mathbb{B}}}$, $\mathring{\boldsymbol{\mathfrak L}}$, and $\mathring{\boldsymbol{\mathfrak B}}^\perp$ do not vanish and their expressions are given in Remark. \ref{key-rem}.}
\begin{equation}
	\tilde{\mathbb{A}}^{\tilde{a}\tilde{A}\tilde{b}\tilde{C}}
	=\frac{Eh}{2(1+\nu)}\mathring{\tilde F}^{\tilde a}{}_{\tilde M}\mathring{\tilde F}^{\tilde b}{}_{\tilde N}
	\left(\tilde{G}^{\tilde{A}\tilde{N}}\tilde{G}^{\tilde{C}\tilde{M}}+\tilde{G}^{\tilde{A}\tilde{C}}\tilde{G}^{\tilde{M}\tilde{N}}
	+\frac{2\nu}{1-\nu}\tilde{G}^{\tilde{A}\tilde{M}}\tilde{G}^{\tilde{C}\tilde{N}}\right)\,.
\end{equation}
Thus 
\begin{equation}\label{key-vir-rig1}
\hat{\tilde{\pmb{\mathbb{A}}}}=\left[\hat{\tilde{\mathbb{A}}}^{\tilde{a}\tilde{A}\tilde{b}\tilde{B}}\right]=\frac{Eh}{2(1+\nu)}\begin{bmatrix}
\begin{bmatrix}
\frac{2 }{1-\nu} & 0 \\
0 & \frac{2\nu }{1-\nu}  \\
\end{bmatrix} & \begin{bmatrix}
0 & 1 \\
1  & 0 \\
\end{bmatrix} \\
\begin{bmatrix}
0 & 1  \\
1 & 0 \\
\end{bmatrix} & \begin{bmatrix}
\frac{2\nu }{1-\nu}  & 0 \\
0 & \frac{2 }{1-\nu} \\
\end{bmatrix} \\
\end{bmatrix}\,.
\end{equation}
Using \eqref{key-tr-2}, the flexural rigidity tensor of the physical plate is given by
\begin{equation}
\begin{split}
\hat{{\pmb{\mathbb{C}}}}&=\left[\hat{{\mathbb{C}}}^{{a}{A}{b}{B}}\right]\\&=\frac{Eh^3}{12(1+\nu)}\begin{bmatrix}
\begin{bmatrix}
\frac{2 }{1-\nu}\frac{f(\psi(R))}{R\psi'(R){f'}^3(\psi(R))} & 0 \\
0 & \frac{2\nu }{1-\nu}\frac{\psi(R)}{f(\psi(R))f'(\psi(R))}  \\
\end{bmatrix} & \begin{bmatrix}
0 & \frac{R\psi'(R)}{f(\psi(R))f'(\psi(R))} \\
\frac{\psi(R)}{f(\psi(R))f'(\psi(R))}  & 0 \\
\end{bmatrix} \\
\begin{bmatrix}
0 & \frac{\psi(R)}{f(\psi(R))f'(\psi(R))}  \\
\frac{\psi^2(R)}{Rf(\psi(R))f'(\psi(R))\psi'(R)} & 0 \\
\end{bmatrix} & \begin{bmatrix}
\frac{2\nu }{1-\nu}\frac{\psi(R)}{f(\psi(R)){f'}(\psi(R))}  & 0 \\
0 & \frac{2 }{1-\nu}\frac{Rf'(\psi(R))\psi'(R)\psi^2(R)}{f^3(\psi(R))} \\
\end{bmatrix} \\
\end{bmatrix}\,.
\end{split}
\end{equation}
From \eqref{key-trans-1}, the first elasticity tensor of the physical plate is obtained as
\begin{equation}
\begin{split}
	\hat{{\pmb{\mathbb{A}}}}=\left[\hat{{\mathbb{A}}}^{{a}{A}{b}{B}}\right]=&\frac{Eh}{2(1+\nu)}\begin{bmatrix}
		\begin{bmatrix}
			\frac{2 }{1-\nu}\frac{f(\psi(R))}{Rf'(\psi(R))\psi'(R)} & 0 \\
			0 & \frac{2\nu }{1-\nu}  \\
		\end{bmatrix} & \begin{bmatrix}
			0 & \frac{Rf'(\psi(R))\psi'(R)}{f(\psi(R))} \\
			1  & 0 \\
		\end{bmatrix} \\
		\begin{bmatrix}
			0 & 1  \\
			\frac{f(\psi(R))}{Rf'(\psi(R))\psi'(R)} & 0 \\
		\end{bmatrix} & \begin{bmatrix}
			\frac{2\nu }{1-\nu}  & 0 \\
			0 & \frac{2 }{1-\nu}\frac{Rf'(\psi(R))\psi'(R)}{f(\psi(R))} \\
		\end{bmatrix} \\
	\end{bmatrix}\\&
	-\begin{bmatrix}
	\begin{bmatrix}
	\frac{1}{\psi'(R)}\hat{\mathring{P}}^{rR} & \frac{R}{\psi(R)}\hat{\mathring{P}}^{\theta R} \\
	0 & 0  \\
	\end{bmatrix} & \begin{bmatrix}
	\frac{1}{\psi'(R)}\hat{\mathring{P}}^{r\Theta} & \frac{R}{\psi(R)}\hat{\mathring{P}}^{\theta\Theta} \\
	0  & 0 \\
	\end{bmatrix} \\
	\\
	\begin{bmatrix}
	0 & 0  \\
	\frac{1}{\psi'(R)}\hat{\mathring{P}}^{rR} & \frac{R}{\psi(R)}\hat{\mathring{P}}^{\theta R} \\
	\end{bmatrix} & \begin{bmatrix}
	0  & 0 \\
	\frac{1}{\psi'(R)}\hat{\mathring{P}}^{r\Theta} & \frac{R}{\psi(R)}\hat{\mathring{P}}^{\theta\Theta} \\
	\end{bmatrix} \\
	\end{bmatrix}\,,
\end{split}	
\end{equation}
where utilizing \eqref{key-pre-stress}, the pre-stress is calculated as
\begin{equation}
	\begin{split}
		\hat{\mathring{P}}^{r\Theta}=&\hat{\mathring{P}}^{\theta R}=0\,,\\
\hat{\mathring{P}}^{rR}=&\frac{E h^3}{12 \left(\nu ^2-1\right) R \psi (R) f^3(\psi (R)) {f'}^5(\psi (R))} \bigg\{\psi (R) f(\psi (R)) {f'}^5(\psi (R)) \left[(\nu -2) R \psi '(R)+2 \psi
	(R)\right]\\&+R \psi^2 (R) \psi '(R) {f'}^6(\psi (R))+f^2(\psi (R)) {f'}^3(\psi (R)) \Big[\nu  R \psi (R)
	\psi '(R) f''(\psi (R))\\&+f'(\psi (R)) \left\{-(\nu -1) R \psi '(R)-2 \psi (R)\right\}\Big]-\psi (R)
	f^3(\psi (R)) {f'}^2(\psi (R)) f''(\psi (R))\\&+\psi (R) f^4(\psi (R)) \left[3 {f''}^2(\psi (R))-f^{(3)}(\psi
	(R)) f'(\psi (R))\right]\bigg\}\,,\\
\hat{\mathring{P}}^{\theta\Theta}=&\frac{E h^3}{12 \left(\nu ^2-1\right) R \psi (R) f^4(\psi (R)) {f'}^3(\psi (R))} \bigg\{-3 R \psi^3 (R) \psi '(R) {f'}^5(\psi (R))\\&+\psi^2 (R) f(\psi (R)) {f'}^3(\psi (R))
	\left[R \psi (R) \psi '(R) f''(\psi (R))+f'(\psi (R)) \left\{\psi (R)-2 (\nu -2) R \psi
	'(R)\right\}\right]\\&+\psi (R) f^2(\psi (R)) {f'}^2(\psi (R)) \left[f'(\psi (R)) \left\{\nu  R \psi '(R)+(\nu
	-2) \psi (R)\right\}-\nu  R \psi (R) \psi '(R) f''(\psi (R))\right]\\&+f^3(\psi (R)) \Big[\nu  R \psi^2
	(R) f^{(3)}(\psi (R)) \psi '(R) f'(\psi (R))+\Big(\left(\psi (R)-R \psi '(R)\right) f'(\psi (R))\\&-2 R
	\psi (R) \psi '(R) f''(\psi (R))\Big) \left\{\nu  \psi (R) f''(\psi (R))+(1-\nu ) f'(\psi
	(R))\right\}\Big]\bigg\}\,.\\
	\end{split}
\end{equation}
Using \eqref{key-tan-bod}, one finds the components of the tangential body force as $(\mathring{\mathfrak{B}}^\top)^{\theta}=0$, and
\begin{equation}\label{key-cons-further}
\begin{split}
(\mathring{\mathfrak{B}}^\top)^{r}=& \frac{E h^3 }{12\tilde{\rho}_0 \left(\nu ^2-1\right)  f^4(\psi (R))
	{f'}^7(\psi (R))} 
	\bigg\{3 \psi (R) {f'}^7(\psi (R))-3 f(\psi (R)) {f'}^6(\psi (R))\\&-3 f^2(\psi (R))
	{f'}^4(\psi (R)) f''(\psi (R))\\&+2 f^3(\psi (R)) {f'}^2(\psi (R)) \left[f^{(3)}(\psi (R)) f'(\psi (R))-3
	{f''}^2(\psi (R))\right]\\&+f^4(\psi (R)) \left[15 {f''}^3(\psi (R))+f^{(4)}(\psi (R)) {f'}^2(\psi (R))-10
	f^{(3)}(\psi (R)) f'(\psi (R)) f''(\psi (R))\right]\bigg\}\,.\\
\end{split}
\end{equation}
However, note that \eqref{key-bod-cons} implies that $(\mathring{\mathfrak{B}}^\top)^{r}=0$, and thus, \eqref{key-cons-further} can be viewed as a constraint.
Notice that \eqref{key-three-sat} is trivially satisfied. The cloaking compatibility equations \eqref{key-constraint-11} gives the following ODE
\begin{equation}\label{jimbo}
\begin{split}
f'(\psi (R)) &\left[f(\psi (R))-\psi (R) f'(\psi (R))\right] \left[\psi (R) f'(\psi (R))+(\nu -1) f(\psi
(R))\right]\\&-\nu  \psi (R) f^2(\psi (R)) f''(\psi (R))=0\,.
\end{split}
\end{equation}
Recalling that $r=\psi(R)$, we may rewrite \eqref{jimbo} as
\begin{equation}\label{key-jim}
	f'(r) \left[f(r)-r f'(r)\right] \left[r f'(r)+(\nu -1) f(r)\right]-\nu  r f^2(r) f''(r)=0\,.
\end{equation}
Noting that \eqref{key-jim} is a second-order ODE and the cloaking transformation $\xi$ needs to satisfy $f(r_o)=r_o$, $f(r_i)=\epsilon$, and $f'(r_o)=1$, one concludes that cloaking is not possible. The balance of linear momentum in the finitely deformed configuration \eqref{key-key} is simplified to read
\begin{equation}\label{key-17}
\begin{split}
3& R \psi (R)^3 \psi '(R) \left(\psi (R)-R \psi '(R)\right) f'(\psi (R))^5+\psi (R)^2 f(\psi (R)) f'(\psi (R))^3 \Big[R^2 \psi (R) \psi ''(R) f'(\psi (R))\\&+\left(R \psi '(R)-\psi (R)\right) \left\{R \psi (R) \psi '(R) f''(\psi (R))+f'(\psi (R)) \left[(5-2 \nu
) R \psi '(R)+\psi (R)\right]\right\}\Big]\\&+\psi (R) f(\psi (R))^2 f'(\psi (R))^2 \Big[(\nu -2) R^2 \psi (R) \psi ''(R) f'(\psi (R))\\&+\left(\psi (R)-R \psi '(R)\right) \left\{\nu  R \psi (R) \psi '(R) f''(\psi (R))+f'(\psi (R)) \left[(2-\nu ) \psi (R)-(\nu
-1) R \psi '(R)\right]\right\}\Big]\\&+f(\psi (R))^3 \bigg((\nu -1) R^2 \psi '(R)^2 f'(\psi (R))^2-(\nu -1) R \psi (R) f'(\psi (R)) \Big[f'(\psi (R)) \left(R \psi ''(R)+2 \psi '(R)\right)\\&-R \psi '(R)^2 f''(\psi (R))\Big]+\psi (R)^2 \Big[-2 \nu  R^2 \psi
'(R)^2 f''(\psi (R))^2+(\nu -1) f'(\psi (R))^2\\&+R f'(\psi (R)) \left(\nu  R f^{(3)}(\psi (R)) \psi '(R)^2+f''(\psi (R)) \left[\nu  R \psi ''(R)+\psi '(R)\right]\right)\Big]\\&-\nu  \psi (R)^3 \left(f'(\psi (R)) \left(R f^{(3)}(\psi (R)) \psi '(R)+f''(\psi
(R))\right)-2 R \psi '(R) f''(\psi (R))^2\right)\bigg)=0\,.
\end{split}
\end{equation}
Provided that $\nu\neq0$, one may use \eqref{key-jim} to obtain expressions for $f''(\psi(R))$ and $f^{(3)}(\psi(R))$ in terms of $\psi(R)$, $\psi'(R)$, $f(\psi(R))$, and $f'(\psi(R))$. Plugging these expressions into \eqref{key-17}, it is straightforward to verify that \eqref{key-17} is trivially satisfied. Therefore, the balance of linear and angular momenta (cf. \eqref{key-all}) for the physical plate in its finitely deformed configuration are satisfied as long as the cloaking map satisfies the constraint given by \eqref{key-jim}.

\begin{remark}\label{key-rem}
Assuming that $\tilde{\pmb{\mathbb B}}$ is given by \eqref{key-bb}, one obtains $\pmb{\mathbb B}$, the initial normal body force $\mathring{ \mathfrak B}^\perp$, and the initial body moment $\mathring{ \boldsymbol{\mathfrak{L}}}$ of the physical plate using \eqref{key-b-vir-phy}, \eqref{key-bod-forc}, and \eqref{key-bod-mom}, respectively, as 
\begin{equation}
\hat{{\pmb{\mathbb{B}}}}=\left[\hat{{\mathbb{B}}}^{{a}{A}{b}{B}}\right]=\begin{bmatrix}
\begin{bmatrix}
\frac{\big(\tilde{\mathsf{b}}_1+2\tilde{\mathsf{b}}_2\big)f(\psi(R)) }{R{f'}^2(\psi(R))\psi'(R)} & 0 \\
0 & \frac{\tilde{\mathsf{b}}_1\psi(R) }{f(\psi(R))}  \\
\end{bmatrix} & \begin{bmatrix}
0 & \frac{\tilde{\mathsf{b}}_2R\psi'(R)}{f(\psi(R))} \\
\frac{\tilde{\mathsf{b}}_2\psi(R)}{f(\psi(R))}  & 0 \\
\end{bmatrix} \\
\begin{bmatrix}
0 & \frac{\tilde{\mathsf b}_2}{f'(\psi(R))}  \\
\frac{\tilde{\mathsf b}_2\psi(R)}{Rf'(\psi(R))\psi'(R)} & 0 \\
\end{bmatrix} & \begin{bmatrix}
\frac{\tilde{\mathsf b}_1 }{f'(\psi(R))}  & 0 \\
0 & \frac{\big(\tilde{\mathsf{b}}_1+2\tilde{\mathsf{b}}_2\big)R\psi(R)f'(\psi(R))\psi'(R) }{f^2(\psi(R))} \\
\end{bmatrix} \\
\end{bmatrix}\,,
\end{equation}	
\begin{equation}
\begin{split}
\mathring{ \mathfrak B}^\perp=&-\frac{(\tilde{\mathsf b}_1+2 \tilde{\mathsf b}_2)}{2\tilde{\rho}_0 f^3(\psi (R)) {f'}^5(\psi (R))} \bigg[-\psi (R) {f'}^5(\psi (R))+f(\psi (R)) {f'}^4(\psi (R))\\&+2 f^2(\psi (R)) {f'}^2(\psi (R)) f''(\psi (R))+f^3(\psi (R)) \left(f^{(3)}(\psi (R)) f'(\psi (R))-3 {f''}^2(\psi (R))\right)\bigg]\,, 
\end{split}
\end{equation}
\begin{equation}
\mathring{\mathfrak{L}}^r=\frac{(\tilde{\mathsf b}_1+2 \tilde{\mathsf b}_2)}{2\tilde{\rho}_0 f^2(\psi (R)) {f'}^4(\psi (R))} \Big[f^2(\psi (R)) f''(\psi
	(R))+\psi (R) {f'}^3(\psi (R))-f(\psi (R)) {f'}^2(\psi
	(R))\Big]\,,
\end{equation}
and the circumferential component of the body moment vanishes, i.e., $\mathring{\mathfrak{L}}^{\theta}=0$.
\end{remark}

\begin{remark}[General cloaking transformations]
Next we show that if $\tilde{\mathsf b}_2>0$ and $\tilde{\mathsf b}_1+\tilde{\mathsf b}_2>0$, i.e., if the tensor $\tilde{\pmb{\mathbb B}}$ for the virtual plate is positive definite, then transformation cloaking would not be realizable even if one uses a general cloaking map $\xi$ for an arbitrary hole surrounded by a cloak (with an arbitrary shape). Without loss of generality, we use Cartesian coordinates, where the shifters and the metrics have trivial representations. Let us consider an arbitrary cloaking map $\xi$ such that
\begin{equation}
\accentset{\xi}{\mathbf F}^{-1}=\begin{bmatrix}
\mathsf{F}_{11}(x,y) & \mathsf{F}_{12}(x,y) \\
\mathsf{F}_{21}(x,y) & \mathsf{F}_{22}(x,y)     
\end{bmatrix}\,.
\end{equation}
Therefore, $J_{\xi}=\left[\mathsf{F}_{11}\mathsf{F}_{22}-\mathsf{F}_{12}\mathsf{F}_{21}\right]^{-1}$, and \eqref{key-2-non} is simplified to read
\begin{subequations}
\begin{align}
	\label{key-1q} 
	& \tilde{\mathsf{b}}_1 \mathsf{F}_{21}\left(\mathsf{F}_{11}^2+\mathsf{F}_{12}^2\right)
	+2 \tilde{\mathsf{b}}_2 \mathsf{F}_{11} \left(\mathsf{F}_{11} \mathsf{F}_{21}
	+\mathsf{F}_{12}\mathsf{F}_{22}\right)-\mathsf{F}_{12} \left(\tilde{\mathsf{b}}_1
	+2 \tilde{\mathsf{b}}_2\right) \left(\mathsf{F}_{11}^2+\mathsf{F}_{12}^2\right)=0\,, \\
	\label{key-2q}
	& \mathsf{F}_{21} \left(\mathsf{F}_{21}^2+\mathsf{F}_{22}^2\right) \left(\tilde{\mathsf{b}}_1
	+2\tilde{\mathsf{b}}_2\right)-2 \mathsf{F}_{22} \tilde{\mathsf{b}}_2 \left(\mathsf{F}_{11} \mathsf{F}_{21}
	+\mathsf{F}_{12} \mathsf{F}_{22}\right)-\mathsf{F}_{12} \tilde{\mathsf{b}}_1
	\left(\mathsf{F}_{21}^2+\mathsf{F}_{22}^2\right)=0\,, \\
	\label{key-3q}
	& \mathsf{F}_{11} \left(\mathsf{F}_{22}^2\tilde{\mathsf b}_2-\mathsf{F}_{12} 
	\mathsf{F}_{21}	(\tilde{\mathsf b}_1+\tilde{\mathsf b}_2)+\mathsf{F}_{21}^2 (\tilde{\mathsf b}_1
	+2 \tilde{\mathsf b}_2)\right) \nonumber\\
	&+\mathsf{F}_{12} \mathsf{F}_{22} \left(\mathsf{F}_{21} (\tilde{\mathsf b}_1
	+\tilde{\mathsf b}_2)-\mathsf{F}_{12} (\tilde{\mathsf b}_1+2\tilde{\mathsf b}_2)\right)
	-\mathsf{F}_{11}^2 \mathsf{F}_{22} \tilde{\mathsf b}_2=0\,.
\end{align}
\end{subequations}
Provided that $\mathsf{F}_{12}^2\tilde{\mathsf b}_1+\mathsf{F}_{11}^2(\tilde{\mathsf{b}}_1+2\tilde{\mathsf b}_2)\neq 0$, from \eqref{key-1q} one obtains
\begin{equation}\label{key-x}
\mathsf{F}_{21}=\mathsf{F}_{12}\frac{(\mathsf{F}_{11}^2+\mathsf{F}_{12}^2)(\tilde{\mathsf b}_1+2\tilde{\mathsf b}_2)-2\mathsf{F}_{11}\mathsf{F}_{22}\tilde{\mathsf b}_2}{\mathsf{F}_{12}^2\tilde{\mathsf b}_1+\mathsf{F}_{11}^2(\tilde{\mathsf{b}}_1+2\tilde{\mathsf b}_2)}\,.
\end{equation}
Substituting for $\mathsf{F}_{21}$ into \eqref{key-3q}, one has
\begin{equation}\label{key-ye}
\begin{split}
\frac{(\mathsf{F}_{11}^2+\mathsf{F}_{12}^2)(\mathsf{F}_{12}^2-\mathsf{F}_{11}\mathsf{F}_{22})\tilde{\mathsf b}_2(\tilde{\mathsf b}_1+2\tilde{\mathsf{b}}_2)}{\left[\mathsf{F}_{12}^2\tilde{\mathsf b}_1+\mathsf{F}_{11}^2(\tilde{\mathsf{b}}_1+2\tilde{\mathsf b}_2)\right]^2}\bigg[&\mathsf{F}_{12}^2\mathsf{F}_{22}\tilde{\mathsf b}_1+\mathsf{F}_{11}^2(\mathsf{F}_{11}-\mathsf{F}_{22})(\tilde{\mathsf b}_1+2\tilde{\mathsf b}_2)\\&+\mathsf{F}_{11}\mathsf{F}_{12}^2(3\tilde{\mathsf b}_1+4\tilde{\mathsf b}_2)\bigg]=0\,.
\end{split}
\end{equation}
Using \eqref{key-x}, the Jacobian of the cloaking map is simplified to read
\begin{equation}
J_{\xi}=\frac{\mathsf{F}_{12}^2\tilde{\mathsf b}_1+\mathsf{F}_{11}^2(\tilde{\mathsf{b}}_1+2\tilde{\mathsf b}_2)}{(\mathsf{F}_{11}^2+\mathsf{F}_{12}^2)(\mathsf{F}_{11}\mathsf{F}_{22}-\mathsf{F}_{12}^2)(\tilde{\mathsf b}_1+2\tilde{\mathsf{b}}_2)}\,.
\end{equation}
Knowing that $\tilde{\mathsf b}_2>0$, and the Jacobian of the cloaking map cannot be singular, \eqref{key-ye} implies that 
\begin{equation}
\mathsf{F}_{12}^2\mathsf{F}_{22}\tilde{\mathsf b}_1+\mathsf{F}_{11}^2(\mathsf{F}_{11}-\mathsf{F}_{22})(\tilde{\mathsf b}_1+2\tilde{\mathsf b}_2)+\mathsf{F}_{11}\mathsf{F}_{12}^2(3\tilde{\mathsf b}_1+4\tilde{\mathsf b}_2)=0\,,
\end{equation}
where as long as $\mathsf{F}_{12}^2\tilde{\mathsf b}_1-\mathsf{F}_{11}^2(\tilde{\mathsf{b}}_1+2\tilde{\mathsf b}_2)\neq 0$, gives
\begin{equation}\label{key-y}
\mathsf{F}_{22}=\mathsf{F}_{11}\frac{\mathsf{F}_{11}^2(\tilde{\mathsf b}_1+2\tilde{\mathsf b}_2)+\mathsf{F}_{12}^2(3\tilde{\mathsf b}_1+4\tilde{\mathsf b}_2)}{\mathsf{F}_{11}^2(\tilde{\mathsf{b}}_1+2\tilde{\mathsf b}_2)-\mathsf{F}_{12}^2\tilde{\mathsf b}_1}\,.
\end{equation}
Plugging \eqref{key-x} and \eqref{key-y} into \eqref{key-2q}, one obtains
\begin{equation}
\frac{\mathsf{F}_{12}(\mathsf{F}_{11}^2+\mathsf{F}_{12}^2)^4\tilde{\mathsf b}_2(\tilde{\mathsf b}_1+\tilde{\mathsf b}_2)(\tilde{\mathsf b}_1+2\tilde{\mathsf b}_2)^2}{\left[\mathsf{F}_{11}^2(\tilde{\mathsf{b}}_1+2\tilde{\mathsf b}_2)-\mathsf{F}_{12}^2\tilde{\mathsf b}_1\right]^3}=0\,,
\end{equation}
where recalling that $\tilde{\mathsf b}_1+\tilde{\mathsf b}_2>0$, implies that $\mathsf{F}_{12}=0$, and thus, $\mathsf{F}_{21}=0$ (cf. \eqref{key-x}), and from \eqref{key-y}, $\mathsf{F}_{11}=\mathsf{F}_{22}$. Given that $\accentset{\xi}{\mathbf F}|_{\partial_o\mathcal{C}}=id$, one concludes that $\xi$ must be the identity, i.e., cloaking is not possible.

Let us consider the case where $\mathsf{F}_{12}^2\tilde{\mathsf b}_1+\mathsf{F}_{11}^2(\tilde{\mathsf{b}}_1+2\tilde{\mathsf b}_2)= 0$, and thus
\begin{equation}\label{key-cont}
\tilde{\mathsf b}_1=-\frac{2\mathsf{F}_{11}^2}{\mathsf{F}_{11}^2+\mathsf{F}_{12}^2}\tilde{\mathsf b}_2\,. 
\end{equation}
Therefore, \eqref{key-1q} is simplified to read $2\tilde{\mathsf b}_2\mathsf{F}_{12}(\mathsf{F}_{12}^2-\mathsf{F}_{11}\mathsf{F}_{22})=0$, implying that $\mathsf{F}_{12}^2=\mathsf{F}_{11}\mathsf{F}_{22}$ (or $\mathsf{F}_{22}=\mathsf{F}_{12}^2/\mathsf{F}_{11}$).\footnote{Note that $\mathsf{F}_{12}\neq0$, because otherwise, $\tilde{\mathsf b}_1=-2\tilde{\mathsf b}_2$ (from \eqref{key-cont}), contradicting the positive definiteness of $\tilde{\pmb {\mathbb B}}$.} Substituting for $\tilde{\mathsf b}_1$ and $\mathsf{F}_{22}$ into \eqref{key-2q} and \eqref{key-3q}, one obtains
\begin{subequations}\label{key-qq}
\begin{align}
	& 2\tilde{\mathsf b}_2\frac{\mathsf{F}_{12}(\mathsf{F}_{12}-\mathsf{F}_{21})}
	{\mathsf{F}_{11}^2(\mathsf{F}_{11}^2+\mathsf{F}_{12}^2)}\left[\mathsf{F}_{12}^5
	+\mathsf{F}_{11}^2\mathsf{F}_{21}(\mathsf{F}_{11}^2+\mathsf{F}_{12}^2)
	+\mathsf{F}_{11}^2\mathsf{F}_{21}^2\mathsf{F}_{12}\right]=0\,, \\
	& \tilde{\mathsf b}_2\frac{\mathsf{F}_{12}(\mathsf{F}_{21}
	-\mathsf{F}_{12})}{\mathsf{F}_{11}(\mathsf{F}_{11}^2+\mathsf{F}_{12}^2)}\left[\mathsf{F}_{11}^4
	+\mathsf{F}_{12}^4+2\mathsf{F}_{11}^2\mathsf{F}_{12}\mathsf{F}_{21}\right]=0\,.
\end{align}
\end{subequations}
Note that $J_{\xi}=\left[\mathsf{F}_{11}\mathsf{F}_{22}-\mathsf{F}_{12}\mathsf{F}_{21}\right]^{-1}=\left[\mathsf{F}_{12}(\mathsf{F}_{12}-\mathsf{F}_{21})\right]^{-1}$, and thus, \eqref{key-qq} implies that
\begin{subequations}
\begin{align}
	& \mathsf{F}_{12}^5+\mathsf{F}_{11}^2\mathsf{F}_{21}(\mathsf{F}_{11}^2
	+\mathsf{F}_{12}^2)+\mathsf{F}_{11}^2\mathsf{F}_{21}^2\mathsf{F}_{12}=0\,, \\
	& \mathsf{F}_{11}^4+\mathsf{F}_{12}^4+2\mathsf{F}_{11}^2\mathsf{F}_{12}\mathsf{F}_{21}=0\,,
\end{align}
\end{subequations}
from which, one obtains $\mathsf{F}_{12}=-\mathsf{F}_{21}$ and $\mathsf{F}_{11}^2=-\mathsf{F}_{12}\mathsf{F}_{21}$. Thus, $\mathsf{F}_{11}^2=\mathsf{F}_{12}^2$, and using \eqref{key-cont}, one concludes that $\tilde{\mathsf b}_1+\tilde{\mathsf b}_2=0$, i.e., $\tilde{\pmb{\mathbb B}}$ is not positive definite, which is a contradiction.
Similarly, it is straightforward to show that assuming $\mathsf{F}_{12}^2\tilde{\mathsf b}_1-\mathsf{F}_{11}^2(\tilde{\mathsf{b}}_1+2\tilde{\mathsf b}_2)=0$, the Jacobian of the cloaking map is forced to be singular.
\end{remark}
\section{Concluding remarks}\label{five5}

In this paper, we formulated the problem of elastodynamic transformation cloaking in plates. In particular, we considered transformation cloaking in Kirchhoff-Love plates as well as elastic plates with both the in-plane and the out-of-plane displacements.
Using a Lagrangian field theory, the governing equations of nonlinear (and linearized) elastic shells (and plates) were derived by characterizing the geometry of a shell as an embedded hypersuface in the Euclidean space using the first and the second fundamental forms. The body forces and body moments were taken into consideration in the boundary-value problem of an elastic plate using the Lagrange--d'Alembert principle. A cloaking map transforms the boundary-value problem of an isotropic and homogeneous elastic plate (virtual problem) to that of an anisotropic and inhomogeneous elastic plate with a finite hole covered by a cloak (physical problem) that is designed such that the response of the virtual plate is mimicked outside the cloak. 

Cloaking in Kirchhoff-Love plates involves transforming the (out-of-plane) governing equation of the virtual plate to that of the physical plate up to an unknown scalar field via a cloaking map. In doing so, one obtains a set of constraints (cloaking compatibility equations) involving the cloaking transformation, the scalar field, and the elastic parameters of the virtual plate. In addition, on the boundary of the cloak and the hole there are some conditions that the cloaking transformation and the scalar field need to satisfy. In particular, the cloaking map needs to fix the outer boundary of the cloak up to the third order and the scalar field needs to be the identity up to the first order on the outer boundary of the cloak. In the example of a circular hole, we showed that cloaking a circular hole in Kirchhoff-Love plates is not possible for a generic radial cloaking map; the obstruction to transformation cloaking are the cloaking compatibility equations and the boundary conditions that the cloaking map needs to satisfy.

In the case of a hole with an arbitrary shape, the cloaking compatibility equations are a system of second-order nonlinear PDEs, and the balance of linear and angular momenta for the (physical) plate in its finitely-deformed configuration lead to fourth-order and third-order nonlinear PDEs. The complexity of this system of nonlinear PDEs makes studying the obstruction to cloaking for an arbitrary cloaking map very difficult. This is in contrast to $3$D elastodynamics and elastic plates with both the in-plane and the out-of-plane deformations, where the linearized balance of angular momentum is the obstruction to cloaking. The nature of the linearized balance of angular momentum usually allows one to analyze these equations for an arbitrary cloaking map and to prove obstruction to cloaking. Note that Kirchoff-Love plates can only have bending deformations, and in this case, the linearized balance of angular momentum only implies that the flexural rigidity tensor must have the minor symmetries. As the flexural rigidity tensor preserves its minor (and major) symmetries under the cloaking map, the linearized balance of angular momentum is trivially satisfied. Therefore, one only needs to analyze the cloaking compatibility equations and the balance of linear and angular momenta in the finitely-deformed configuration (of the physical plate) to study obstruction to transformation cloaking.

Next, we relaxed the pure bending assumption and formulated the transformation cloaking problem for an elastic plate in the presence of the in-plane and the out-of-plane displacements. The physical plate is initially stressed and is subjected to (in-plane and out-of-plane) body forces and moments in its finitely-deformed configuration. This problem involves transforming the in-plane governing equations using the Piola transformation given by the cloaking map as well as transforming the out-of-plane governing equation up to the Jacobian of the cloaking map. Assuming a general radial cloaking map, we showed that cloaking is not possible for a circular hole; the cloaking compatibility equations and the boundary conditions that the cloaking map needs to satisfy obstruct transformation cloaking, similar to the case of Kirchhoff-Love plates. We also showed that if for the virtual plate the elasticity tensor pertaining to the coupling between the in-plane and the out-of-plane displacements is positive-definite, then cloaking is not possible even if one uses a general cloaking map for a hole and a cloak with arbitrary shapes; the balance of angular momentum is the obstruction to cloaking similar to transformation cloaking in $3$D elastodynamics.

An extension of this work would be a further generalization of the present theory by assigning a set of deformable directors at each point of an elastic shell in its reference and current configurations. This will lead to a (normal and tangential) hyperstress in addition to the stress and the couple-stress of the present theory. One should note that the balance of linear (and angular) momentum and micro linear (and angular) momentum in this case are coupled in a way that makes transformation cloaking highly non-trivial. Moreover, one needs to derive a set of compatibility conditions for the normal and tangential components of the director gradient. This will be the subject of a future communication.

\section*{Acknowledgement}

This research was supported by ARO W911NF-18-1-0003 (Dr. David Stepp). A.G. benefited from discussions with Fabio Sozio, Arzhang Angoshtari, Amirhossein Tajdini, and Souhayl Sadik.

\bibliographystyle{abbrvnat}
\bibliography{ref}

\begin{appendices}

\section{Variations of some geometric objects}
\label{Geometry1}

In this appendix, we discuss the derivation of the variations of the right Cauchy-Green deformation tensor $\mathbf{C}$, the unit normal vector field (of the deformed shell) $\boldsymbol{\mathcal{N}}$, and $\mathbf{\Theta}$ used in obtaining the Euler-Lagrange equations in \S\ref{three3} (see also \citep{capovilla2002stresses,kadianakis2013kinematics,kadianakis2018inf}).

\paragraph{Lie derivative.}Let $\mathbf{w}:\mathcal{U}\rightarrow T\mathcal{S}$ be a ($C^1$) vector field, where $\mathcal{U} \subset \mathcal{S}$ is an open neighborhood. A curve $\alpha:I \rightarrow \mathcal{S}$, where $I$ is an open interval, is an \emph{integral curve} of $\mathbf{w}$ provided that $\frac{d \alpha(t)}{dt}=\mathbf{w}(\alpha(t)),~\forall~t \in I$.
Consider a time-dependent vector field $\mathbf{w}:\mathcal{S}\times I\rightarrow T\mathcal{S}$, where $I$ is some open interval. The collection of maps $\psi_{\tau,t}$ is the flow of $\mathbf{w}$ if for each $t$ and $x$, $\tau\mapsto\psi_{\tau,t}(x)$ is an integral curve of $\mathbf{w}_t$, i.e., $\frac{d}{d\tau}\psi_{\tau,t}(x)=\mathbf{w}(\psi_{\tau,t}(x),\tau)$,
and $\psi_{t,t}(x)=x$. Assume that $\mathbf{t}$ is a time-dependent tensor field on $\mathcal{S}$, i.e., $\mathbf{t}_t(x)=\mathbf{t}(x,t)$ is a tensor. The Lie derivative of $\mathbf{t}$ with respect to $\mathbf{w}$ is defined as $\mathbf{L}_{\mathbf{w}}\mathbf{t}=\frac{d}{d \tau}\psi_{\tau,t}^* \mathbf{t}_{\tau} \Big|_{\tau=t}$.
Note that $\psi_{\tau,t}$ maps $\mathbf{t}_t$ to $\mathbf{t}_{\tau}$. Therefore, to calculate the Lie derivative one drags $\mathbf{t}$ along the flow of $\mathbf{w}$ from $\tau$ to $t$ and then differentiates the Lie dragged tensor with respect to $\tau$.
The \emph{autonomous} Lie derivative of $\mathbf{t}$ with respect to $\mathbf{w}$ is defined as $\mathfrak{L}_{\mathbf{w}}\mathbf{t}=\frac{d}{d \tau}\psi_{\tau,t}^* \mathbf{t}_{t} \Big|_{\tau=t}$.
Hence, $\mathbf{L}_{\mathbf{w}}\mathbf{t}=\partial \mathbf{t}/\partial t+\mathfrak{L}_{\mathbf{w}}\mathbf{t}$.
The Lie derivative for a scalar $f$ is given by $\mathbf{L}_{\mathbf{w}}f=\partial f/\partial t+\mathbf{w}[f]$. In a coordinate chart $\{x^a\}$, this is written as, $\mathbf{L}_{\mathbf{w}}f=\frac{\partial f}{\partial t}+\frac{\partial f}{\partial x^a}w^a$.
For a vector $\mathbf{u}$, it can be shown that $\mathbf{L}_{\mathbf{w}}\mathbf{u}=\frac{\partial \mathbf{w}}{\partial t}+[\mathbf{w},\mathbf{u}]$.
If $\nabla$ is a torsion-free connection, then $[\mathbf{w},\mathbf{u}]=\nabla_{\mathbf{w}}\mathbf{u}-\nabla_{\mathbf{u}}\mathbf{w}$, and thus, $\mathbf{L}_{\mathbf{w}}\mathbf{u}=\frac{\partial \mathbf{w}}{\partial t}+\nabla_{\mathbf{w}}\mathbf{u}-\nabla_{\mathbf{u}}\mathbf{w}$.

The rate of deformation tensor for shells is defined as \citep{MaHu1983}
\begin{equation}
	2\mathbf{D}^\flat=\varphi^*_t\left[(\nabla^{\mathbf g}\mathbf{v}^\top)^\flat
	+[(\nabla^{\mathbf g}\mathbf{v}^\top)^\flat]^{\mathsf T}-2v^\perp\boldsymbol{\theta}\right]\,,
\end{equation}
where the spatial velocity is decomposed into the normal and tangential components as $\mathbf{v}=\mathbf{v}^\top+v^\perp\mathbf{n}$. In components
\begin{equation}
2 D_{AB}=(v^\top_{a|b}+v^\top_{b|a})F^a{}_AF^b{}_B-2v^\perp\Theta_{AB}\,.
\end{equation}
Note that
\begin{equation}
\mathbf{L}_{\mathbf{v}^\top}\mathbf{g}=(\nabla^{\mathbf g}\mathbf{v}^\top)^\flat+[(\nabla^{\mathbf g}\mathbf{v}^\top)^\flat]^{\mathsf T}\,.
\end{equation}
Therefore
\begin{equation}
2\mathbf{D}^\flat=\varphi^*_t(\mathbf{L}_{\mathbf{v}^\top}\mathbf{g})-2v^\perp\mathbf{\Theta}\,.
\end{equation}
Knowing that $\varphi_t^*(\mathbf{L}_{\mathbf{v}}\mathbf{g})=2\mathbf{D}^\flat$ (see, e.g., \citep{MaHu1983,simo1984stress}), one obtains 
\begin{equation}
\varphi_t^*(\mathbf{L}_{\mathbf{v}}\mathbf{g})=\varphi^*_t(\mathbf{L}_{\mathbf{v}^\top}\mathbf{g})-2v^\perp\mathbf{\Theta}\,.
\end{equation}
Thus, $\mathbf{L}_{\delta\varphi}\mathbf{g}=\mathbf{L}_{\delta\varphi^\top}\mathbf{g}-2\,\delta \varphi^\perp\,\boldsymbol{\theta}$, and hence, $\delta\mathbf{C}^\flat=\varphi^*_t(\mathbf{L}_{\delta\varphi}\mathbf{g})=\varphi^*_t\,\mathbf{L}_{\delta\varphi^\top}\mathbf{g}-2\,\delta\varphi^\perp\,\mathbf{\Theta}$. 
Also note that
\begin{equation}
\mathbf{L}_{\delta\varphi^\top}\mathbf{g}=\left[g_{cb}(\delta\varphi^\top)^c{}_{|a}+g_{ac}(\delta\varphi^\top)^c{}_{|b}\right]dx^a\otimes dx^b\,.
\end{equation}
Hence, in components, one obtains
\begin{equation}
\delta C_{AB}=F^a{}_A\,\delta{\varphi}^{\top}_{a|B}+F^b{}_B\,\delta\varphi^{\top}_{b|A}-2\,\delta\varphi^\perp\,\Theta_{AB}\,.
\end{equation}
Or
\begin{equation}
\delta C_{AB}=F^a{}_A\,g_{ac}(\delta{\varphi}^{\top})^c{}_{|B}+F^b{}_B\,g_{bc}(\delta\varphi^{\top})^c{}_{|A}-2\,\delta\varphi^\perp\,F^a{}_AF^b{}_B\theta_{ab}\,. 
\end{equation}
Therefore, \eqref{cauchy-var} is implied. 

The covariant derivative of $\mathbf{v}$ is computed as
\begin{equation}
\bar{\nabla}^{\bar{\mathbf g}}\mathbf{v}=\bar{\nabla}^{\bar{\mathbf g}}(\mathbf{v}^\top+v^\perp\mathbf{n})=\bar{\nabla}^{\bar{\mathbf g}}\mathbf{v}^\top+v^\perp\bar{\nabla}^{\bar{\mathbf g}}\mathbf{n}+\mathbf{d}v^\perp\otimes\mathbf{n}\,.
\end{equation}
Using the relations \eqref{sec-fun-form} and \eqref{2nd-fun} in the ambient space, one obtains
\begin{equation}\label{vel-appen}
\bar{\nabla}^{\bar{\mathbf g}}\mathbf{v}=(\nabla^{\mathbf g}\mathbf{v}^\top-v^\perp\boldsymbol{\theta})+(\mathbf{d}v^\perp+\boldsymbol{\theta}\cdot\mathbf{v}^\top)\otimes\mathbf{n}\,.
\end{equation}
In components
\begin{equation}
	\bar{\nabla}^{\bar{\mathbf g}}\mathbf{v}
	=\left[(v^\top)^a{}_{|b}-v^\perp\theta^a{}_b\right]\partial_a\otimes dx^b
	+\left[v^\perp{}_{,b}+\theta_{bc}(v^\top)^c\right]\mathbf{n}\otimes dx^b\,.
\end{equation}
Similarly, one can write
\begin{equation}
	\bar{\nabla}^{\bar{\mathbf g}}\delta\varphi
	=\left[(\delta\varphi^\top)^a{}_{|b}-\delta\varphi^\perp\theta^a{}_b\right]\partial_a\otimes dx^b
	+\left[\delta\varphi^\perp{}_{,b}+\theta_{bc}(\delta\varphi^\top)^c\right]\mathbf{n}\otimes dx^b\,.
\end{equation}
Note that for an arbitrary vector field $\mathbf W={\mathbf W}^\top+W^\perp\boldsymbol{\mathcal{N}}$ defined on a surface embedded in $\mathbb{R}^3$, the tangential and normal components of the covariant derivative with respect to the surface coordinates are similarly given by  
\begin{equation}
	(\bar{\nabla}^{\bar{\mathbf g}}\mathbf{W})^\top=\nabla^{\mathbf g}\mathbf{W}^\top-W^\perp\boldsymbol{\theta}\,,
	 \quad\quad 
	 (\bar{\nabla}^{\bar{\mathbf g}}\mathbf{W})^\perp=\mathbf{d}W^\perp+\boldsymbol{\theta}\cdot\mathbf{W}^\top\,.
\end{equation}
Therefore, $\bar{\nabla}^{\bar{\mathbf g}}\mathbf{W}=(\bar{\nabla}^{\bar{\mathbf g}}\mathbf{W})^\top+(\bar{\nabla}^{\bar{\mathbf g}}\mathbf{W})^\perp$, in components reads
\begin{equation}\label{surface-covarinat}
	(\bar{\nabla}^{\bar{\mathbf g}}\mathbf{W})^\top
	=\left[(W^\top)^a{}_{|b}-W^\perp\theta^a{}_b\right]\partial_a\otimes dx^b\,,\qquad 
	(\bar{\nabla}^{\bar{\mathbf g}}\mathbf{W})^\perp
	=\left[W^\perp{}_{,b}+\theta_{bc}(W^\top)^c\right]\boldsymbol{\mathcal{N}}\otimes dx^b\,.
\end{equation}
Thus, one can use \eqref{surface-covarinat}$_1$ to write the variation of the deformation gradient in components as
\begin{equation}\label{ff-key}
\delta F^a{}_A=(\delta\varphi^\top)^a{}_{|A}-\theta^a{}_bF^b{}_A\delta\varphi^\perp\,.
\end{equation}
Therefore, \eqref{key-ff} follows.
At any time $t$, the deformation map $\varphi_t:\mathcal{H}\to\mathcal{S}$ is a smooth embedding of the (undeformed) shell into the ambient space.  For each $X\in\mathcal{H}$, let $d\varphi_t(X):T_X\mathcal{H}\to T_{\varphi_t(X)}\mathcal{S}$ be the tangent of $\varphi_t$ at $X$.
The variation of the unit normal vector $\boldsymbol{\mathcal{N}}_{\epsilon}=\mathbf{n}_{\epsilon}\circ\varphi_{\epsilon,t}$ is defined as
\begin{equation}
\delta\boldsymbol{\mathcal{N}}=\frac{d}{d\epsilon}\boldsymbol{\mathcal{N}}_{\epsilon}\Big|_{\epsilon=0}=\bar{\nabla}_{\frac{\partial}{\partial\epsilon}}^{\bar{\mathbf g}}\boldsymbol{\mathcal{N}}_{\epsilon}\Big|_{\epsilon=0}=D_{\varphi_{\epsilon}(X,t)}\boldsymbol{\mathcal{N}}_{\epsilon}\Big|_{\epsilon=0}={\bar{\nabla}}^{\bar{\mathbf g} }_{\delta\varphi}\boldsymbol{\mathcal{N}}.
\end{equation}
In order to compute the variation, let $\mathbf{W}$ be a vector field in $\mathcal S$ tangent to $\varphi_t(\mathcal H)$  and note that
\begin{equation}
{\bar{\nabla}}^{\bar{\mathbf g} }_{\delta\varphi}\mathbf{W}=[\delta\varphi,\mathbf{W}]+{\bar{\nabla}}^{\bar{\mathbf g} }_{\mathbf{W}}\delta\varphi\,.
\end{equation}
From \eqref{vel-appen}, one obtains
\begin{equation}\label{app-deltaphi}
{\bar{\nabla}}^{\bar{\mathbf g} }_{\mathbf{W}}\delta\varphi=\left(\nabla^{\mathbf g}_{\mathbf{W}}{\delta\varphi^\top}-\delta\varphi^\perp\boldsymbol{\theta}\cdot\mathbf{W}\right)+\left((\mathbf{d}\,\delta\varphi^\perp)\cdot\mathbf{W}+\boldsymbol{\theta}\cdot\delta\varphi^\top\cdot\mathbf{W}\right)\mathbf{n}\,.
\end{equation}
Therefore
\begin{equation}
\bar{\mathbf g}\left(\boldsymbol{\mathcal{N}},{\bar{\nabla}}^{\bar{\mathbf g} }_{\mathbf{W}}\delta\varphi\right)=(\mathbf{d}\,\delta\varphi^\perp)\cdot\mathbf{W}+\boldsymbol{\theta}\cdot\delta\varphi^\top\cdot\mathbf{W}=-\bar{\mathbf g}\left({\bar{\nabla}}^{\bar{\mathbf g} }_{\delta\varphi}\boldsymbol{\mathcal{N}},\mathbf{W}\right),
\end{equation}
where the second equality is a consequence of the metric compatibility of $\bar{\nabla}^{\bar{\mathbf g}}$ \eqref{me-comp}. By arbitrariness of $\mathbf W$, we have
\begin{equation}\label{app-deltan}
\delta\boldsymbol{\mathcal N}={\bar{\nabla}}^{\bar{\mathbf g} }_{\delta\varphi}\boldsymbol{\mathcal{N}}=-\boldsymbol{\theta}\cdot\delta\varphi^\top-\mathbf{d}\,\delta\varphi^\perp\,.
\end{equation}
In components
\begin{equation}
\delta\boldsymbol{\mathcal N}=-\left[(\delta\varphi^\top)^b\theta^a{}_b+\delta\varphi^\perp{}_{,b}g^{ab}\right]\partial_a \,.
\end{equation}
Hence, \eqref{normal-var} and \eqref{norm-var} are followed.
Note that $\delta\mathbf{\Theta}^\flat=\varphi^*_t(\mathbf{L}_{\delta\varphi}\boldsymbol{\theta})$, where
$
\mathbf{L}_{\delta\varphi}\boldsymbol{\theta}=\mathbf{L}_{\delta\varphi^\top}\boldsymbol{\theta}-\delta\varphi^\perp\mathbf{III}+\mathrm{Hess}_{\delta\varphi^\perp}\,,
$
i.e.,
\begin{equation}\label{theta-variation}
\delta\mathbf{\Theta}^\flat=\varphi^*_t\mathbf{L}_{\delta\varphi^\top}\boldsymbol{\theta}-\delta\varphi^\perp\,\varphi_t^*\mathbf{III}\,+\varphi_t^*\mathrm{Hess}_{\delta\varphi^\perp},
\end{equation}
where for $\boldsymbol{x},\boldsymbol{y}\in\mathcal{X}(\varphi(\mathcal{H}))$, the third fundamental form of the deformed hypersurface $\mathbf{III}$ and the Hessian of $\delta\varphi^\perp$, i.e., $\mathrm{Hess}_{\delta\varphi^\perp}$ are given by \eqref{3rd-fundamental-form} and \eqref{Hessian}.
 The Lie derivative with respect to the tangential component of the variation field is given in components by (see, e.g., \citep[p.97]{MaHu1983})
\begin{equation}
\mathbf{L}_{\delta\varphi^\top}\boldsymbol{\theta}^\flat=\left[\theta_{ab|c}\,(\delta\varphi^\top)^c+\theta_{ac}(\delta\varphi^\top)^c{}_{|b}+\theta_{cb}(\delta\varphi^\top)^c{}_{|a}\right]dx^a\otimes dx^b\,.
\end{equation}
Therefore, in components, one can write the variation of $\mathbf{\Theta}^\flat$ as
\begin{equation}\label{sec-fun-form-var1}
\begin{split}
\delta\Theta_{AB}=&F^a{}_AF^b{}_B\,\theta_{ab|c}\,(\delta\varphi^\top)^c+F^a{}_A\theta_{ac}(\delta\varphi^\top)^c{}_{|B}+F^b{}_B\theta_{bc}(\delta\varphi^\top)^c{}_{|A}\\&-\delta\varphi^\perp F^a{}_AF^b{}_B\theta_{ac}\theta_{bd}\,g^{cd}+F^b{}_A\left(\frac{\partial\,\delta\varphi^\perp}{\partial\, x^b}\right)_{|B}\,.
\end{split}
\end{equation}
Using \eqref{ff-key} and the fact that $\Theta_{AB}=F^a{}_AF^b{}_B\theta_{ab}$, one obtains the variation of $\boldsymbol{\theta}^\flat$ as
\begin{equation}
\delta \theta_{ab}=\theta_{ab|c}(\delta\varphi^\top)^c+\delta\varphi^\perp\theta_{ac}\theta_{bd}g^{cd}+(\delta\varphi^\perp{}_{,a})_{|b}\,.
\end{equation}
Hence, \eqref{sec-fun-form-var} and \eqref{delta-teta-pla} are obtained.

\paragraph{Proof of the relation \eqref{theta-variation}.}
Here, we give a proof of the relation \eqref{theta-variation}, see also \citep{capovilla2002stresses,lenz2000stability,deserno2004notes}. For the sake of simplicity, we assume that the surface is embedded in three-dimensional Euclidean space. For this proof, we adopt a notation different from the rest of the paper. 
Let us consider an embedded surface denoted by $\Sigma$ in $\mathbb{R}^3$. The surface geometry is locally described by three functions $\mathbf{x}(x^1,x^2,x^3)=\mathbf{X}(\nu^\alpha)$ in the Cartesian coordinates $\{x^1,x^2,x^3\}$ such that $\{\nu^\alpha\}$, $\alpha=1,2$, is a local coordinate chart on the surface. Let us define two tangent vectors $\mathbf{e}_{\alpha}={\partial}\mathbf{X}/{\partial \nu^\alpha}$ on the surface. We note that the surface geometry is completely described by its induced metric $\eta_{\alpha\beta}$ and its induced second fundamental form $\Lambda_{\alpha\beta}$.\footnote{Note that $\boldsymbol{\eta}$ and $\mathbf{\Lambda}$, respectively, correspond to $\mathbf C$ and $\mathbf{\Theta}$ defined previously.} Note that $\eta_{\alpha\beta}=\mathbf{e}_{\alpha}\cdot\mathbf{e}_{\beta}$, where ``$\cdot$'' denotes the dot product in $\mathbb{R}^3$. Let us denote the surface covariant derivative with $\nabla_\alpha$. Then, one can write the Gauss-Weingarten equations as $\nabla_{\alpha}\mathbf{e}_{\beta}=\Lambda_{\alpha\beta}\mathbf{n}$, and $\nabla_{\alpha}\mathbf{n}=-\Lambda_{\alpha\beta}\mathbf{e}^{\beta}$, where $\mathbf n$ is the unit normal vector to the surface. Let us consider the deformation of the embedding functions of the surface $\mathbf{X}(\nu)\to\mathbf{X}(\nu)+\delta\mathbf{X}(\nu)$ such that the variation field $\delta\mathbf{X}$ is decomposed into the tangential and normal components as $\delta\mathbf{X}=(\psi^\top)^\alpha\mathbf{e}_{\alpha}+\psi^\perp\mathbf{n}$. Using the relation $\Lambda_{\alpha\beta}=\mathbf{n}\cdot\nabla_{\alpha}\mathbf{e}_{\beta}$, one may write
\begin{equation}
\delta \Lambda_{\alpha\beta}=\delta\mathbf{n}\cdot\nabla_{\alpha}\nabla_{\beta}\mathbf{X}+\mathbf{n}\cdot\nabla_{\alpha}\nabla_{\beta}\delta\mathbf{X}\,.
\end{equation}
Knowing that the variation of the unit normal vector is purely tangential, the first term vanishes, i.e., $\delta\mathbf{n}\cdot\nabla_{\alpha}\nabla_{\beta}\mathbf{X}=\delta\mathbf{n}\cdot\nabla_{\alpha}\mathbf{e}_{\beta}=\Lambda_{\alpha\beta}(\delta\mathbf{n})\cdot\mathbf{n}=0$. After some simplification and using the Codazzi-Mainardi equation $\nabla_{\alpha}\Lambda_{\beta\gamma}=\nabla_{\beta}\Lambda_{\alpha\gamma}$, one obtains
\begin{equation}\label{jj-key}
\delta \Lambda_{\alpha\beta}=\Lambda_{\beta\gamma}\nabla_{\alpha}(\psi^\top)^\gamma+\Lambda_{\alpha\gamma}\nabla_{\beta}(\psi^\top)^\gamma+(\psi^\top)^{\gamma}\nabla_{\gamma}\Lambda_{\alpha\beta}-\Lambda_{\alpha\gamma}\Lambda^{\gamma}{}_{\beta}\psi^\perp+\nabla_{\alpha}\nabla_{\beta}\psi^\perp\,.
\end{equation}
Notice that the first three terms correspond to the Lie derivative of the induced second fundamental form with respect to the tangential component of the variation field, and thus, \eqref{jj-key} can be rewritten as
\begin{equation}
\Lambda_{\alpha\beta}=(\mathbf{L}_{\psi^\top}\Lambda)_{\alpha\beta}-\Lambda_{\alpha\gamma}\Lambda^{\gamma}{}_{\beta}\psi^\perp+\nabla_{\alpha}\nabla_{\beta}\psi^\perp \,.
\end{equation}
Therefore, \eqref{theta-variation} follows.

\section{The Euler-Lagrange equations of elastic shells}
\label{EU-Lag}

In this appendix, we discuss the derivation of the Euler-Lagrange equations. Substituting \eqref{vel-var}, \eqref{cau-gr-var}, \eqref{theta-var}, and \eqref{norm-var} into \eqref{eu-lagrange}, one obtains
\begin{equation}
\begin{split}
\int_{t_0}^{t_1}\int_{\mathcal{H}}\Bigg\{&\rho(\mathfrak{B}^\top)_a(\delta\varphi^\top)^a+\rho\mathfrak{B}^\perp\cdot\delta\varphi^\perp+\frac{\partial\mathcal{L}}{\partial\dot{\varphi}}\cdot\frac{D\,\delta\varphi}{dt}-\rho\mathfrak{L}_b\left((\delta\varphi^\top)^a\theta^b{}_a+\frac{\partial\,\delta\varphi^\perp}{\partial x^a}g^{ab}\right)\\&+\frac{\partial\mathcal{L}}{\partial C_{AB}}\left(F^a{}_A\delta\varphi^\top_{a|B}+F^b{}_B\delta\varphi^\top_{b|A}-2\delta\varphi^\perp F^a{}_AF^b{}_B\theta_{ab}\right)\\&+\frac{\partial\mathcal{L}}{\partial\Theta_{AB}}\bigg[F^a{}_AF^b{}_B\,\theta_{ab|c}\,(\delta\varphi^\top)^c+F^a{}_A\theta_{ac}(\delta\varphi^\top)^c{}_{|B}+F^b{}_B\theta_{bc}(\delta\varphi^\top)^c{}_{|A}\\&-\delta\varphi^\perp F^a{}_AF^b{}_B\theta_{ac}\theta_{bd}\,g^{cd}+F^b{}_A\left(\frac{\partial\,\delta\varphi^\perp}{\partial\, x^b}\right)_{|B}\bigg]\Bigg\}dAdt=0\,.
\end{split}
\end{equation}
After some simplification, we have
\begin{equation}
\begin{split}
\int_{t_0}^{t_1}&\int_{\mathcal{H}}\Bigg\{\rho(\mathfrak{B}^\top)_a(\delta\varphi^\top)^a+\rho\mathfrak{B}^\perp\cdot\delta\varphi^\perp+\frac{d}{dt}\left(\frac{\partial\mathcal{L}}{\partial\dot{\varphi}}\cdot\,\delta\varphi\right)-\frac{d}{dt}\left(\frac{\partial\mathcal{L}}{\partial\dot{\varphi}}\right)\cdot\,\delta\varphi-\rho\mathfrak{L}_b(\delta\varphi^\top)^a\theta^b{}_a\\&-\left(\rho\mathfrak{L}_b\delta\varphi^\perp\, g^{ab}(F^{-1})^A{}_a\right)_{|A}+\left(\rho\mathfrak{L}_b g^{ab}(F^{-1})^A{}_a\right)_{|A}\delta\varphi^\perp\\&-2\frac{\partial\mathcal{L}}{\partial C_{AB}} F^a{}_AF^b{}_B\theta_{ab}\delta\varphi^\perp+2\left(\frac{\partial\mathcal{L}}{\partial C_{AB}}F^b{}_Bg_{cb}(\delta\varphi^\top)^c\right)_{|A}-2\left(\frac{\partial\mathcal{L}}{\partial C_{AB}}F^b{}_Bg_{cb}\right)_{|A}(\delta\varphi^\top)^c\\&+\frac{\partial\mathcal{L}}{\partial\Theta_{AB}}F^a{}_AF^b{}_B\,\theta_{ab|c}\,(\delta\varphi^\top)^c+2\left(\frac{\partial\mathcal{L}}{\partial\Theta_{AB}}F^a{}_A\theta_{ac}(\delta\varphi^\top)^c\right)_{|B}\\&-2\left(\frac{\partial\mathcal{L}}{\partial\Theta_{AB}}F^a{}_A\theta_{ac}\right)_{|B}(\delta\varphi^\top)^c-\frac{\partial\mathcal{L}}{\partial\Theta_{AB}} F^a{}_AF^b{}_B\theta_{ac}\theta_{bd}\,g^{cd}\delta\varphi^\perp\\&+\left(\frac{\partial\mathcal{L}}{\partial\Theta_{AB}}F^b{}_A\frac{\partial\,\delta\varphi^\perp}{\partial\, x^b}\right)_{|B}-\left[\left(\frac{\partial\mathcal{L}}{\partial\Theta_{AB}}F^b{}_A\right)_{|B}\delta\varphi^\perp(F^{-1})^D{}_b\right]_{|D}\\&+\left[\left(\frac{\partial\mathcal{L}}{\partial\Theta_{AB}}F^b{}_A\right)_{|B}(F^{-1})^D{}_b\right]_{|D}\delta\varphi^\perp \Bigg\}dAdt=0\,.
\end{split}
\end{equation}
This can further be simplified to read
\begin{equation}
\begin{split}
\int_{t_0}^{t_1}&\int_{\mathcal{H}}\Bigg\{\Bigg[\rho(\mathfrak{B}^\top)_a-\frac{d}{dt}\left(\frac{\partial\mathcal{L}}{\partial(\dot{\varphi}^\top)^a}\right)-\rho\mathfrak{L}_b\theta^b{}_a-2\left(\frac{\partial\mathcal{L}}{\partial C_{AB}}F^b{}_Bg_{ab}\right)_{|A}+\frac{\partial\mathcal{L}}{\partial\Theta_{AB}}F^c{}_AF^b{}_B\,\theta_{cb|a}\\&-2\left(\frac{\partial\mathcal{L}}{\partial\Theta_{AB}}F^b{}_A\theta_{ba}\right)_{|B}\Bigg](\delta\varphi^\top)^a+\Bigg[\rho\mathfrak{B}^\perp-\frac{d}{dt}\left(\frac{\partial\mathcal{L}}{\partial\dot{\varphi}^\perp}\right)+\left(\rho\mathfrak{L}_b g^{ab}(F^{-1})^A{}_a\right)_{|A}\\&-2\frac{\partial\mathcal{L}}{\partial C_{AB}} F^a{}_AF^b{}_B\theta_{ab}-\frac{\partial\mathcal{L}}{\partial\Theta_{AB}} F^a{}_AF^b{}_B\theta_{ac}\theta_{bd}\,g^{cd}+\left(\left(\frac{\partial\mathcal{L}}{\partial\Theta_{AB}}F^b{}_A\right)_{|B}(F^{-1})^D{}_b\right)_{|D}\Bigg]\delta\varphi^\perp\\&+\frac{d}{dt}\left(\frac{\partial\mathcal{L}}{\partial\dot{\varphi}}\cdot\,\delta\varphi\right)-\left(\rho\mathfrak{L}_b\delta\varphi^\perp\, g^{ab}(F^{-1})^A{}_a\right)_{|A}+2\left(\frac{\partial\mathcal{L}}{\partial C_{AB}}F^b{}_Bg_{cb}(\delta\varphi^\top)^c\right)_{|A}\\&+2\left(\frac{\partial\mathcal{L}}{\partial\Theta_{AB}}F^a{}_A\theta_{ac}(\delta\varphi^\top)^c\right)_{|B}+\left(\frac{\partial\mathcal{L}}{\partial\Theta_{AB}}\,\frac{\partial\,\delta\varphi^\perp}{\partial X^A}\right)_{|B}\\&-\left[\left(\frac{\partial\mathcal{L}}{\partial\Theta_{AB}}F^b{}_A\right)_{|B}\delta\varphi^\perp(F^{-1})^D{}_b\right]_{|D}\Bigg\}dAdt=0\,.
\end{split}
\end{equation} 
We assume that $\delta\varphi(X,t_0)=\delta\varphi(X,t_1)=0$. Using Stokes' theorem, we have
\begin{equation}\label{fin-eu}
\begin{split}
\int_{t_0}^{t_1}&\int_{\mathcal{H}}\Bigg\{\Bigg[\rho(\mathfrak{B}^\top)_a-\frac{d}{dt}\left(\frac{\partial\mathcal{L}}{\partial(\dot{\varphi}^\top)^a}\right)-\rho\mathfrak{L}_b\theta^b{}_a-2\left(\frac{\partial\mathcal{L}}{\partial C_{AB}}F^b{}_Bg_{ab}\right)_{|A}+\frac{\partial\mathcal{L}}{\partial\Theta_{AB}}F^c{}_AF^b{}_B\,\theta_{cb|a}\\&-2\left(\frac{\partial\mathcal{L}}{\partial\Theta_{AB}}F^b{}_A\theta_{ba}\right)_{|B}\Bigg](\delta\varphi^\top)^a+\Bigg[\rho\mathfrak{B}^\perp-\frac{d}{dt}\left(\frac{\partial\mathcal{L}}{\partial\dot{\varphi}^\perp}\right)+\left(\rho\mathfrak{L}_b g^{ab}(F^{-1})^A{}_a\right)_{|A}\\&-2\frac{\partial\mathcal{L}}{\partial C_{AB}} F^a{}_AF^b{}_B\theta_{ab}-\frac{\partial\mathcal{L}}{\partial\Theta_{AB}} F^a{}_AF^b{}_B\theta_{ac}\theta_{bd}\,g^{cd}\\
&+\left(\left(\frac{\partial\mathcal{L}}{\partial\Theta_{AB}}F^b{}_A\right)_{|B}(F^{-1})^D{}_b\right)_{|D}\Bigg]\delta\varphi^\perp\Bigg\}dAdt+\int_{t_0}^{t_1}\int_{\partial\mathcal{H}}\Bigg\{\frac{\partial\mathcal{L}}{\partial\Theta_{AB}}\,\frac{\partial\,\delta\varphi^\perp}{\partial X^A}\mathsf{T}_{B}\\&-\Bigg[\rho\mathfrak{L}_b g^{ab}(F^{-1})^A{}_a+\left(\frac{\partial\mathcal{L}}{\partial\Theta_{CB}}F^a{}_C\right)_{|B}(F^{-1})^A{}_a\Bigg]\mathsf{T}_{A}\delta\varphi^\perp\\&+2\Bigg[\frac{\partial\mathcal{L}}{\partial C_{AB}}F^a{}_Bg_{ac}+\frac{\partial\mathcal{L}}{\partial\Theta_{AB}}F^a{}_B\theta_{ac}\Bigg]\mathsf{T}_{A}(\delta\varphi^\top)^c\Bigg\}dL\,dt=0\,,
\end{split}
\end{equation}
where $\boldsymbol{\mathsf{T}}$ is the outward vector field normal to the boundary curve $\partial\mathcal{H}$. Knowing that $\delta\varphi^\top$, $\delta\varphi^\perp$, and $\mathbf{d}(\delta\varphi^\perp)$ are arbitrary, from \eqref{fin-eu} the Euler-Lagrange equations \eqref{eus-lag}, along with the boundary conditions \eqref{eu-bound} are obtained.

\end{appendices}

\end{document}